\documentclass[conference]{IEEEtran}
\IEEEoverridecommandlockouts

\usepackage{cite}
\usepackage{amsmath,amssymb,amsfonts}
\usepackage{graphicx}
\usepackage{textcomp}
\usepackage{xcolor}
\usepackage{float}
\usepackage{comment}
\usepackage{hyperref}
\usepackage{caption}
\def\BibTeX{{\rm B\kern-.05em{\sc i\kern-.025em b}\kern-.08em
    T\kern-.1667em\lower.7ex\hbox{E}\kern-.125emX}}
\usepackage{algorithm}

\usepackage{algpseudocode}
\usepackage{booktabs}
\usepackage{subcaption}
\usepackage{textpos}
\usepackage{amsthm}    
\newtheorem{definition}{Definition}
\newtheorem{example}{Example}
\newtheorem{assumption}{Assumption}

\newtheorem{theorem}{Theorem}
\newtheorem{lemma}{Lemma}

\newcommand{\norm}[1]{\left\lVert#1\right\rVert}
\newcommand{\ve}{\varepsilon}
\newcommand{\N}{\mathbb{N}}
\newcommand{\R}{\mathbb{R}}

\newcommand{\E}{\mathbb{E}}

\newcommand{\PR}{\mathbb{P}}
  
\usepackage{ulem}

\begin{document}

\title{Lower Bounds for Rényi Differential Privacy \\ in a Black-Box Setting}

\author{\IEEEauthorblockN{Tim Kutta}
\IEEEauthorblockA{\textit{}
\textit{Ruhr-University Bochum}\\ tim.kutta@rub.de}
\and
\IEEEauthorblockN{Önder Askin}
\IEEEauthorblockA{\textit{}
\textit{Ruhr-University Bochum}\\ oender.askin@rub.de}
\and
\IEEEauthorblockN{Martin Dunsche}
\IEEEauthorblockA{\textit{}
\textit{Ruhr-University Bochum}\\ martin.dunsche@rub.de}
}

\maketitle
\begin{abstract}
We present new methods for assessing the privacy guarantees of an algorithm with regard to Rényi Differential Privacy. To the best of our knowledge, this work is the first to address this problem in a black-box scenario, where only algorithmic outputs are available. To quantify privacy leakage, we devise a new estimator for the Rényi divergence of a pair of output distributions. This estimator is transformed into a  statistical lower bound that is proven to hold for large samples with high probability. Our method is applicable for a broad class of algorithms, including many well-known examples from the privacy literature. We demonstrate the effectiveness of our approach by experiments encompassing algorithms and privacy enhancing methods that have not been considered in related works.
\end{abstract}

\section{Introduction}

 Differential Privacy (DP) \cite{Dwork2006} has emerged as a standard concept to assess and mitigate the privacy leakage of algorithms that release data.
 Algorithms that satisfy DP process databases with random noise to mask individual users' contributions. DP provides robust and analytically stringent privacy guarantees and is employed where sensitive data is at stake \cite{Erlingsson2014, Blocki2016, Bolin2017, Abowd2018}. However, the rigorous requirements of DP preclude many otherwise useful privatization schemes, particularly those based on Gaussian noise. As a consequence, various relaxations of DP have been proposed to accommodate a wider range of algorithms, while still preserving privacy in a meaningful way. The most prominent of these are approximate DP \cite{Dwork2006_b} and Rényi Differential Privacy (RDP) \cite{Mironov2017}. 
Both variants broaden the class of privacy preserving mechanisms (they crucially allow for the use of Gaussian noise), while maintaining key features of DP such as stability under post-processing. RDP in particular has attracted growing interest as an analytical framework to closely track the privacy loss in iterative procedures \cite{Wang2019, Feldman2021, Chourasia2021}. Moreover, it is increasingly used to study privacy enhancing methods such as shuffling \cite{Girgis2021} or subsampling \cite{Zhu2019}.\\
Traditionally, the development of privacy preserving algorithms relies on formal proofs of DP (or its variants) prior to implementation. Yet, the adoption of DP in recent years has fostered interest in validation methods that can check DP for a given algorithm retrospectively.  To this end, a range of verification methods have been proposed for  standard DP (see e.g. \cite{Reed2010, Barthe2016b, Hsu2017, StatDP, DP-Finder, Kifer2019, CheckDP}). While there exist some works on the validation of approximate DP \cite{Barthe2014,Barthe2016,Liu2019,Barthe2020}, methods to study the RDP claims of an algorithm are rare.
In \cite{Sato2019} a program logic that can verify relaxations of DP, including RDP, is proposed. Its use, however,  requires access to the algorithm's code and structure, which might not always be available. This lack of access is prevalent in settings where algorithm designers and companies want to disclose as little (proprietary) information as possible. As a consequence, methods that study privacy claims in a black-box setting have gained more interest as of late \cite{Liu2019, DP-Sniper, Dette2022}. In light of
these developments, we  adopt a black-box setup in this work. Our aim is to devise, to the best of our knowledge  for the first time, estimation and inference methods for RDP in a black-box scenario.

Inspired by prior work on validating standard DP \cite{DP-Finder, DP-Sniper, Dette2022}, we base our approach on (empirical) lower bounds for RDP. We will further discuss how these help ascertain the privacy parameter for RDP in Section II. The construction of these lower bounds by statistical techniques is discussed in Section III. Theoretical underpinnings are detailed in the Appendix. We validate the performance of our method with experiments described in Section IV. 
We close with a discussion of related works and some concluding remarks.

\section{Preliminaries and Problem Statement}

\noindent In this section, we state the relevant privacy model and main objectives of this work. We also introduce  the key notion of a statistical lower bound and its use in privacy quantification.

\subsection{Privacy Definitions} \label{Sec_2_1}
\noindent The privacy definitions in this work are concerned with a generic, randomized algorithms $A$ which, given an input database $x$, produces a random output $A(x)$. Consequently, the output of $A(x)$ follows a \textit{probability distribution}. We study two types of distribution: discrete and continuous ones. Suppose that $\mathcal{X}$ is a \noindent finite, non-empty set. Then any probability measure $P$ on  $\mathcal{X}$ has a discrete probability density $p: \mathcal{X} \to [0,1]$ with 
\begin{equation} \label{Eq_def_discr_dens}
    P(B) = \sum_{t \in B} p(t)
\end{equation}
for all $B \subset \mathcal{X}$. If the output $A(x)$  follows a discrete probability density $p=p_x$ on $\mathcal{X}$, for each possible input $x$, we call $A$ a \textit{discrete (randomized) algorithm}. Next, consider a probability measure $P$ on the $d$-dimensional vector space $\R^d$. Suppose that a continuous, non-negative function $p: \R^d \to \R $ exists with
\begin{equation} \label{Eq_def_cont_dens}
    P(B) = \int_B p(t) dt
\end{equation}
for any measurable subset $B \subset \R^d$. Here the integral over $p$ is taken in the standard Lebesgue sense. 
If \eqref{Eq_def_cont_dens} holds, we call $p$ the continuous probability density of $P$. Moreover, if an algorithm $A$ has outputs in $\R^d$, with each $A(x)$ following a continuous probability density $p=p_x$, we call $A$ a \textit{continuous (randomized) algorithm}. In the following, we always consider continuous densities on the entire space $\R^d$, but adaptions of our theory to continuous distributions on subspaces of $\R^d$ are possible. While most algorithms in the DP literature are either discrete or continuous,  there exist some cases, that do not fit into either category (one example is the propose-test-release method in \cite{dworkrobust}).\\
Since most of our results can be formulated for both discrete and continuous densities, we often use the notation
$$
P(B) = \int_B p(t),
$$
that should be interpreted as summation if $p$ is discrete and integration if $p$ is continuous. When we integrate over the whole space ($\mathcal{X}$ or $\R^d$ respectively) we will usually omit the integration index. Finally, if a random variable $X$ follows the distribution $p$, we write $X \sim p$ (in particular, we can write $A(x) \sim p$ for the output of a randomized algorithm). 

\noindent Randomization obstructs adversarial inference, by weakening the link between algorithmic inputs and outputs. This effect can be quantified by comparing the distributions of $A(x)$ and $A(x')$ for "similar" databases $x$ and $x'$. More precisely, we consider databases $x = (x_1, \cdots, x_m)$ and $x' = (x'_1, \cdots, x'_m)$ that are \textit{adjacent}, i.e., that differ in one and only one data point, i.e., $x_j \neq x'_j$ for some $j \in \{1, \cdots, m \}$. We denote adjacency by $x \sim_a x'$. If each data point $x_i$ is understood as the information of an individual $i$, differential privacy requires that any one individual does not notably affect the distribution of $A(x)$.  

\begin{definition}[Differential Privacy] \label{Def_DP}
Let $A$ be a randomized algorithm, that is either discrete or continuous. We call $A$ $\varepsilon$-differentially private for $\varepsilon \geq 0$, if for any two adjacent databases $x,x'$ it holds that
\begin{equation} \label{Eq_def_DP}
\sup_{t: q(t)>0} \log\left(\frac{p(t)}{q(t)}\right) \le \varepsilon,
\end{equation}
where $A(x) \sim p$ and $A(x') \sim q$.
\end{definition}

\noindent In the above definition, $\varepsilon$ is the \textit{privacy parameter} that bounds by how much the distribution of $A$ can differ on adjacent databases. Thus, larger values of $\varepsilon$ stand for higher information leakage (low privacy) and values close to $0$ for less leakage (high privacy). Let us consider a concrete example of DP: Oftentimes, DP algorithms first aggregate the information in database $x$ before privatizing it with random noise. This is achieved by a statistic $S$ and its sensitivity 
\begin{align} \label{eq_def_sensitivity}
    \triangle_S := \sup_{x \sim_a x'} \vert S(x) - S(x') \vert 
\end{align}
is used to calibrate the random noise in the privatization. Commonly encountered statistics are the sum, mean or histogram computed over $x$.

\begin{example}[Laplace Mechanism]
Let $S$ be a real-valued statistic with sensitivity $\triangle_S$ and let $Y \sim Lap(0, b)$. The Laplace Mechanism is given by $A(x) = S(x) + Y$, has density 
\begin{align*}
   p_x(t) = \frac{1}{2 b} \, \exp\big( - \vert t - S(x) \vert \, / b \big) \;, \, t \in \mathbb{R}
\end{align*}
and is $\frac{\triangle_S}{b}$-differentially private.
\end{example}

\noindent The Laplace Mechanism is a prototype for continuous algorithms and serves as the fundamental building block for more elaborate mechanisms, among them discrete ones like the Sparse Vector Technique \cite{Lyu2017} and Report Noisy Max \cite{Dwork2014}. A close relative of the Laplace Mechanism is the Gaussian Mechanism, which adds normally distributed noise to the aggregating statistic $S$. The Gaussian Mechanism serves as a subroutine for many prominent algorithms like Noisy Gradient Descent (see Section \ref{Sec_experiments}). Yet, the Gaussian Mechanism does not satisfy \eqref{Eq_def_DP} for any $\varepsilon>0$, and hence it is not covered by DP (nor are most of the methods built on it). While for Gaussian densities  $p(t)/q(t)$ is bounded on compact subsets where most of the probability is  concentrated (even exponentially so), the density ratio is unbounded in the tails of the distributions. This example shows that by requiring $p(t)/q(t)$ to be bounded for all arguments $t$, DP sets a very high bar for randomized algorithms to count as private at all - at the cost of excluding otherwise useful mechanisms. As a consequence, more inclusive privacy notions have been proposed, such as approximate DP, also termed $(\varepsilon, \delta)$-DP, which allows $p(t)/q(t) > \exp(\varepsilon)$ on a small probability set. While $(\varepsilon, \delta)$-DP covers more algorithms than standard DP, it also encompasses methods of dubious quality that completely compromise privacy with probability $\delta$. Rényi differential privacy (RDP) is more discerning in that regard, excluding mechanisms that entail a complete breakdown of privacy \cite{Mironov2017}. RDP requires, roughly speaking, that an averaged version of the density ratio $p(t)/q(t)$ lie below $\exp(\varepsilon)$, allowing for large values of $p(t)/q(t)$ to be balanced by smaller ones. The notion of "average" is formalized via the Rényi divergence.

\begin{definition}[Rényi-Divergence] \label{Def_Rényi_Divergence}
Let $p,q$ be probability densities (discrete or continuous) and $\lambda \in (1, \infty)$ a parameter. Then we define the Rényi divergence of order $\lambda$ as
\begin{equation*}
    D_\lambda(p, q):= \frac{1}{\lambda -1} \log\left(\int p(t)^\lambda q(t)^{1-\lambda} \right)~.
\end{equation*}
\end{definition}

\noindent As $\lambda$ increases, large values of $p(t)/q(t)$ begin to dominate $D_\lambda(p,q)$ and the divergence grows. In fact, the Rényi divergence increases monotonically in $\lambda$ with $D_\lambda(p,q) \leq D_{\lambda'}(p,q)$ for $\lambda \leq \lambda'$. We can consider the limit $ D_\infty(p, q):= \lim_{\lambda \to \infty} D_\lambda(p,q)$ and note that 
\begin{equation*}
    D_\infty(p, q)= \sup_{t: q(t)>0}\log\left(\frac{p(t)}{q(t)}\right)~.
\end{equation*}
Condition \eqref{Eq_def_DP} is equivalent to $D_\infty(p,q) \le \varepsilon$. Observing that $D_\lambda(p, q) \leq D_\infty(p, q)$ for any $\lambda \in (1, \infty)$, this motivates the following relaxation of DP.

\begin{definition}[Rényi Differential Privacy] \label{Def_Rényi_DP}
Let $A$ be a randomized algorithm, that is either discrete or continuous. We call $A$ $(\lambda, \varepsilon)$-Rényi differentially private for $\lambda > 1$ and $\varepsilon \geq 0$, if for all adjacent $x,x'$ it holds that
\begin{equation} \label{Eq_def_RDP}
    D_\lambda(p, q)\leq \ve,
\end{equation}
where $A(x) \sim p$ and $A(x') \sim q$.
\end{definition}

\noindent In the above definition, $\varepsilon$ is again a privacy parameter, with small values implying high privacy. Due to monotonicity of the divergence $D_\lambda(p,q)$ in $\lambda$, the condition \eqref{Eq_def_RDP} becomes more restrictive for larger $\lambda$. Nevertheless, for any
$\lambda<\infty$ RDP is substantially more inclusive than traditional DP. For example, the above-mentioned Gaussian Mechanism satisfies RDP with $\varepsilon \propto \lambda$, whereas it does not satisfy $\varepsilon$-DP for any finite $\varepsilon$.

\subsection{Problem Formulation}
\noindent Given $\lambda > 1$, if an algorithm $A$ satisfies \eqref{Eq_def_RDP} for one $\varepsilon > 0$, the same is true for any $\varepsilon' > \varepsilon$. But while $A$ is both $(\lambda, \varepsilon) $- and $(\lambda, \varepsilon')$-differentially private, $\varepsilon'$ understates the level of privacy actually achieved by $A$. Hence, it is sensible to consider the smallest $\varepsilon$ for which \eqref{Eq_def_RDP} is met. This is the \textit{optimal privacy parameter} given by
\begin{align}
     \varepsilon(\lambda) := \sup\limits_{x \sim_a x'} D_\lambda(p_x, q_{x'})  
\label{Rényi_Sup}
\end{align}
with $A(x) \sim p_x$ and $A(x') \sim q_{x'}$ (we usually drop the dependence on $\lambda$ and write $\varepsilon=\varepsilon(\lambda)$). Identity \eqref{Rényi_Sup} implies that any instance of the Rényi divergence $D_\lambda(p, q)$ computed for $A(x) \sim p$, $A(x') \sim q$ and adjacent databases $x, x'$ constitutes a lower bound for $\varepsilon$. Recent works that aim at establishing privacy bounds for RDP mathematically, derive meaningful lower bounds by calculating $D_\lambda(p, q)$ for specific databases $x$ and $x'$ \cite{Girgis2021, Chourasia2021}. Depending on the choice of adjacent databases, a qualitatively good approximation of $\varepsilon$ from below can be achieved with 
\begin{align*}
D_\lambda(p, q) \approx \varepsilon.
\end{align*}
This approach can also serve as the basis for assessing $\varepsilon$ in a black-box scenario. Here, two questions arise: (i) Which choice of adjacent databases delivers a good approximation of $\varepsilon$ and (ii) how do we determine $D_\lambda(p, q)$ in a black-box scenario? 

Regarding the selection of $x$ and $x'$, simple heuristics have proven effective in finding suitable databases in the related literature (see \cite{StatDP, DP-Sniper}). Usually, high divergences and thus lower bounds close to $\varepsilon$ can be achieved by choosing $x$ and $x'$ to be far away by some metric. Since the black-box setting allows for choosing the algorithm inputs, these methods can be readily pursued. The greater challenge and main focus of this work is to determine $D_\lambda(p, q)$ without any knowledge of the algorithm's inner workings. Given limited access to $A$, any approximation of $\varepsilon$ can only be based on algorithmic outputs.

More precisely, consider a fixed pair of adjacent databases $x,x'$ and suppose that densities $p,q$ exist with $A(x) \sim p$ and $A(x') \sim q$. Running $A(x)$ and $ A(x')$  respectively $n$-times, produces two samples of realizations $X_1,...,X_n \sim p$ and $Y_1,...,Y_n \sim q$ that are (each) independent and identically distributed (i.i.d).  These samples can be used for inference regarding the Rényi divergence $D_\lambda(p,q)$ and hence the privacy parameter $\varepsilon$ in \eqref{Rényi_Sup}. More specifically, we can construct a \textit{statistical lower bound} $\hat \ell$ for $D_\lambda(p,q)$, which holds with probability $1-\alpha$, i.e.,
\begin{equation} \label{Eq_lower_bound}
  \mathbb{P}(D_\lambda(p,q) \ge \hat \ell) \ge 1-\alpha.  
\end{equation}
Here $\alpha \in (0,1)$ is a small, user-determined value (such as $1\%$ or $5\%$) and $1-\alpha$ is called the \textit{confidence} of the lower bound. In view of  \eqref{Rényi_Sup}, $\hat \ell$ implies with confidence $1-\alpha$, that $\varepsilon \ge \hat \ell$. Furthermore, repeating this process for multiple pairs of adjacent databases allows us to get even closer to $\varepsilon$ by taking the maximum over several lower bounds. Consider, for instance, $N$ pairs of adjacent databases $(x_1,x_1'),...,(x_N,x_N')$ with output distributions $A(x_i) \sim p_{x_i}$ and $A(x_i') \sim q_{x_i'}$. Generating independent samples for each pair allows us to construct lower bounds $\hat \ell_1,...,\hat \ell_N$ for the respective divergences $D_\lambda(p_{x_1},q_{x_1'}),...,D_\lambda(p_{x_N},q_{x_N'})$. The maximum $\hat \ell_{max} := \max_{i=1,...,N} \hat \ell_i$ then provides a  lower bound for $\max_{i=1,...,N} D_\lambda(p_{x_i},q_{x_i'})$ and hence for $\varepsilon$ (recall that $\max_{i=1,...,N} D_\lambda(p_{x_i},q_{x_i'}) \approx \varepsilon$ for suitable databases). The maximum bound  $\hat \ell_{max}$ then satisfies 
\begin{equation} \label{Eq_lower_bound_combined}
    \mathbb{P}\big(\max_{i=1,...,N} D_\lambda(p_{x_i},q_{x_i'}) \ge \hat \ell_{max}\big) \ge (1-\alpha)^N,
\end{equation}
i.e., it holds with confidence $(1-\alpha)^N$. Here we have used \eqref{Eq_lower_bound} and the independence of the bounds $\hat \ell_1, ...,\hat \ell_N$. Similar steps for the approximation of the privacy parameter have also been pursued in the standard DP model \cite{DP-Sniper, Dette2022}, with a choice of databases as indicated before. As we have seen, the central building block of such procedures is the  statistical lower bound for an individual pair of densities. Thus, the main focus of this work is developing a new statistical bound $\hat{\ell}$ for the  divergence $D_\lambda(p,q)$.

\section{Statistical bounds for the Rényi divergence} \label{Sec_3}

\noindent In this section, we develop a new method to statistically quantify the Rényi divergence $D_\lambda(p,q)$ for a pair of densities $p,q$. 
We begin our discussion, in Section \ref{Subsec_3_1}, by introducing estimators for discrete and continuous densities. Subsequently, in Section \ref{Subsec_3_2}, we employ such estimates $\hat p, \hat q$ to approximate the true Rényi divergence 
$D_\lambda(p,q)$ by a carefully regularized version of the plug-in estimator $D_\lambda(\hat p, \hat q)$. 
This regularized version follows (approximately) a normal distribution, implying statistical bounds for the true Rényi divergence. Section \ref{Subsec_3_3} provides a formal validation of this construction and may be skipped in a first reading. 

\subsection{Density estimation} \label{Subsec_3_1}

\noindent Recall the definition of a probability density (discrete and continuous) introduced in Section \ref{Sec_2_1}. 
In the following, we want to approximate a density $f$ by an estimator $\hat f$, based on a sample i.i.d. observations $X_1,...,X_n \sim f$ (think of $f \in \{p,q\}$ in our previous discussion). Our estimator will be \textit{non-parametric}, that is, we only presuppose minimal knowledge about $f$ as befits a black-box setting. Notice that in gray box scenarios, where additional information about the distribution is available, parametric estimators may be more suitable. 
We study two types of estimators: The relative frequency estimator for discrete distributions and the kernel density estimator for continuous distributions (for alternatives, see,  \cite{Tsybakov}). \\
Beginning with a discrete setup, suppose that $f$ is a density on a finite set $\mathcal{X}$. We can then approximate the probability $f(t) = \mathbb{P}(X_1 =t)$ by the relative number of observations equal to $t$, i.e.,
\begin{equation} \label{Eq_def_RDE}
      \hat f(t) := \frac{|\{X_i =t: i=1,...,n\}|}{n}. 
\end{equation}
It is well-known that this relative frequency estimator (RFE) converges to $f$ at a rate of $\mathcal{O}_P(1/\sqrt{n})$ and many concentration results exist making this statement more precise (see Appendix \ref{App_A}, where we also recap stochastic Landau symbols). \\
Next, we consider the case of a continuous density $f$ living on the space $ \R^d$ for $d \in \N$. A popular estimator for $f$ is the kernel density estimator (KDE), defined for $t \in \R^d$ as
\begin{equation}
  \label{Eq_def_KDE}
    \hat f(t) := \frac{1}{n h} \sum_{i=1}^n K\Big( \frac{t-X_i}{h}\Big).
\end{equation}
In the above formula $K : \R^d \to \R$ is a \textit{kernel}, i.e., a continuous function with $\int K(t)dt =1 $. Typical choices for $K$ are the Gaussian or the Laplace kernel, that are pre-implemented in many programming languages (for details on the choice of the kernel, see Appendix \ref{App_A}). The parameter $h>0$ in \eqref{Eq_def_KDE} is called the  \textit{bandwidth} and trades-off bias against variance in the estimation of $f$. It is hence comparable to the bin-width in a histogram, where smaller bins (smaller $h$) correspond to less bias and more noise. An adequate choice of $h$ depends on both the sample size $n$ and the smoothness of the density $f$. In section \ref{Subsec_3_3} we will explore how to measure the smoothness of a function (see Definition \ref{Def_Nikolski}). For now, we simply notice that, while the smoothness of $f$ is practically unknown, many data-driven methods exist to select $h$ (such as cross-validation).\\
It can be shown that for a proper choice of $h$, $\hat f$ is a consistent estimator for $ f$, which converges almost at a rate of $\mathcal{O}_P(1/\sqrt{n})$, for well-behaved $f$. For precise convergence rates as well as concentration results, we refer the reader to  Appendix \ref{App_A}.\\

\subsection{Constructing lower bounds for $D_\lambda(p,q)$}\label{Subsec_3_2} 

\noindent We now proceed to the construction of statistical lower bounds for the Rényi divergence $D_\lambda(p,q)$.
As a central building block, we first devise an estimator for  $D_\lambda(p,q)$. \\
Suppose that two samples of i.i.d. observations $X_1,...,X_n \sim p$ and $Y_1,...,Y_n \sim q$ are given, with sample size $n \in \N$. In order to approximate the divergence $D_\lambda(p,q)$, it seems natural to use $D_\lambda(\hat p,\hat q)$, the divergence of the estimators  $\hat p= \hat p(X_1,...,X_n)$ and $\hat q= \hat q(X_1,...,X_n)$ (see Section \ref{Subsec_3_1} for a definition of these density estimators). Versions of this \textit{empirical Rényi divergence} have been  studied in the related literature, such as \cite{Wassermann},
where also optimal rates of convergence are discussed. While $D_\lambda(\hat p,\hat q)$ provides a reasonable approximation of the Rényi divergence in some scenarios, its accuracy rests on a key premise - that the two densities are bounded away from $0$. In particular, this implies that $p,q$ are only supported on a bounded  (and usually known) subset of $\R^d$. Modest as this assumption may appear, it not generally satisfied by randomized algorithms in DP. For instance, the densities of both the Laplace and the Gaussian Mechanism live on the unbounded domain $\R$ and come arbitrarily close to $0$ for large arguments. Practically, this translates into instability of the estimator $D_\lambda(\hat p,\hat q)$, as the ratio $\hat p(t)/\hat q(t)$ (occurring in the definition of $D_\lambda(\hat p,\hat q)$) becomes unstable for small values in the denominator. Similar problems have been recognized in the literature on black-box bounds for $\varepsilon$-DP (see Definition \ref{Def_DP}), where ratios of densities or (small) probabilities have to be estimated \cite{Dette2022, DP-Sniper}.\\
One way to address this problem, pursued in \cite{Dette2022}, is replacing the estimate $\hat q$ by a “floored version”, that cannot get closer to $0$ than some constant $\tau>0$. For instance, using the pointwise maximum  $\hat q(t) \lor \tau := \max(\hat q(t), \tau)$  yields the regularized estimator 
$D_\lambda(\hat p ,\hat q\lor \tau)$. If $n$ is sufficiently large, we can expect 
\begin{equation} \label{Eq_approx_1}
    D_\lambda(\hat p ,\hat q\lor \tau) \approx D_\lambda( p, q\lor \tau)
\end{equation}
and if in turn $\tau$ is sufficiently small 
\begin{equation}\label{Eq_approx_2}
    D_\lambda( p ,q\lor \tau) \approx D_\lambda( p , q).
\end{equation}
Evidently, for both \eqref{Eq_approx_1} and \eqref{Eq_approx_2} to hold simultaneously, it is necessary to strike a balance with $\tau$, moderating variance (captured by \eqref{Eq_approx_1}) and bias (captured by \eqref{Eq_approx_2}). Yet, if $\tau=\tau(n) \downarrow 0$ converges slowly enough as $n \to \infty$,  asymptotic consistency can be demonstrated, i.e.,
$$
D_\lambda(\hat p ,\hat q\lor \tau) =  D_\lambda( p , q) +o_P(1).
$$
In principle, we can use $D_\lambda(\hat p ,\hat q\lor \tau)$ not only to approximate the Rényi divergence, but also to construct statistical lower bounds for it. For this purpose, it is necessary to study the distribution of $D_\lambda(\hat p ,\hat q\lor \tau)$. A standard tool to derive the (large sample) distribution of  an estimator is given by the so-called “delta method”. Roughly speaking, the delta method states, that an approximately normal  estimator, stays approximately normal under a differentiable transformation (for a precise statement  we refer to \cite{van2000asymptotic} chapter 3). At a first glance, this tool seems promising to analyze $D_\lambda( \hat p , \hat q\lor \tau)$, since the estimators $\hat p $ and $\hat q$ can be shown to be approximately normal. However, a closer look reveals that $D_\lambda( \hat p , \hat q\lor \tau)$ is not a differentiable transform, because the maximum-function is not smooth. How can we circumvent this problem? The answer is simple: We  replace the "hard" maximum in our estimator, by a smoothed version (a “softmax”) and hence  ensure differentiability. More precisely, for any $t \in \R$ and a smoothing parameter $\beta>0$, we define the softmax of $t$ and $\tau$ as 
\begin{equation}\label{Eq_smooth_max} 
t_\tau := \beta^{-1} \log(\exp(t \beta)+\exp(\tau \beta)), \qquad t \in \R~.
\end{equation}
This function (known in the literature as 
LogSumExp), is differentiable in $t$, satisfies the flooring condition $t_\tau \ge \tau$ and approximates the maximum $t_\tau \approx \max(t,\tau)$ for sufficiently large values of  $\beta>0$ (for a detailed discussion see Appendix \ref{App_B}). In particular, we expect 
\begin{equation} \label{eq_approx_softmax_max}
     D_\lambda(\hat p,\hat q_\tau)  \approx D_\lambda(\hat p ,\hat q\lor \tau) 
\end{equation}
to hold. As a consequence of \eqref{Eq_approx_1}, \eqref{Eq_approx_2} and \eqref{eq_approx_softmax_max},   $ D_\lambda(\hat p,\hat q_\tau)$ is an estimator for $D_\lambda( p, q)$. Moreover, as the softmax is differentiable, approximate normality of the density estimators $\hat p, \hat q$ can now filter through to $D_\lambda(\hat p,\hat q_\tau)$ by the delta method.  This yields for large enough $n$ and some variance $\sigma_n^2>0$
\begin{equation}\label{Eq_approx_normality}
    \sqrt{n} \big[D_\lambda(\hat p,\hat q_\tau)-D_\lambda( p, q_\tau) \big] \approx \mathcal{N}(0,\sigma_n^2).
\end{equation}
As we will see in the next section, the formal proof of this result is quite challenging. The softmax function becomes “less and less smooth” when approaching the hard max, making a careful mathematical analysis necessary, which involves state-of-the-art concentration results for the KDE. \\
Identity \eqref{Eq_approx_normality} implies a statistical lower bound $\hat \ell = \hat \ell(\alpha)$ for  $D_\lambda( p, q_\tau)$, that holds with approximate confidence level $1-\alpha$ (for some user-determined $\alpha \in (0,1)$). Importantly, $\hat \ell $ then also constitutes a lower bound for the true Rényi divergence $D_\lambda( p, q)$, since 
$$
D_\lambda( p, q) \ge D_\lambda( p, q_\tau).
$$
 This identity follows directly by the definition of the divergence $D_\lambda$ (see Definition \ref{Def_Rényi_Divergence}) together with the fact that $q_\tau(t) \ge q(t)$ for all $t$.  In view of \eqref{Eq_approx_normality} a lower bound for $D_\lambda( p, q_\tau)$ holding with (approximate) confidence $1-\alpha$ is given by $D_\lambda(\hat p,\hat q_\tau)+\Phi^{-1}(\alpha) \sigma_n/\sqrt{n}$, where $\Phi^{-1}$ denotes the quantile function of the standard normal distribution. While this bound is not directly applicable, as $\sigma_n^2$ is unknown, we can employ a variance estimator $\hat \sigma_n^2$, giving the feasible bound
 \begin{equation} \label{Eq_def_hat_ell}
     \hat \ell(\alpha) := D_\lambda(\hat p,\hat q_\tau)+\frac{\Phi^{-1}(\alpha) \hat \sigma_n}{\sqrt{n}}.
 \end{equation}
  To complete this approach, we have to state the variance estimator $\hat \sigma_n^2$. For this purpose we define the derivative of the softmax function \eqref{Eq_smooth_max} w.r.t. $t$ as
\begin{equation} \label{Eq_def_derivative_softmax}
   \pi(t):= \frac{\exp(\beta t)}{\exp(\beta t)+\exp(\beta \tau)}~. 
\end{equation}
 We can then define 
 \begin{align} \label{Eq_variance_estiamtor}
    \hat \sigma^2_n :=\frac{(\hat \sigma^{(1)})^2+ (\hat \sigma^{(2)})^2}{\Big( (\lambda-1)\int (\hat p(t))^\lambda (\hat q_\tau(t))^{1-\lambda}) \Big)^{2}},
\end{align}
 with 
 \begin{align*}
 (\hat \sigma^{(1)})^2 := & \lambda^2\left(\int  \hat p(t)^{2\lambda-1} \hat q_\tau(t)^{2-2\lambda}\right. \\
      &\qquad \quad \left.- \Big(\int  \hat p(t)^{\lambda} \hat q_\tau(t)^{1-\lambda}  \Big)^2\right)\\
(\hat \sigma^{(2)})^2 := & (1-\lambda)^2\left( \int \pi(\hat q(t))^2 \hat q_\tau(t)^{-2\lambda} \hat q(t) \hat p(t)^{2\lambda}  \right. \\
      &\qquad \quad \left.- \Big( \int \pi(\hat q(t)) \hat q_\tau^{-\lambda}(t)\hat q(t) \hat p(t)^\lambda  \Big)^2\right).
\end{align*}
 While the shape of the variance estimator is elaborate, its calculation is not, as it is merely another integral transform of the density estimators $\hat p, \hat q$.  The true sequence of variances $(\sigma_n^2)_{n \in \N}$ is monotonically increasing in $n$ and may diverge to $\infty$. Yet, we prove in the Appendix, that $\hat \sigma^2_n$ is a consistent estimator in the sense that $|\hat \sigma_n^2-\sigma_n^2|=o_P(1)$, which validates the statistical bound \eqref{Eq_def_hat_ell}.

\subsection{Theoretical guarantees} \label{Subsec_3_3}
\noindent We now proceed to the formal validation of the statistical lower bound, discussed in the preceding section. Before stating our results, we have to introduce some mathematical notations. In the following, for $x \in \R$, let $\lfloor x \rfloor$ denote the largest integer, that is strictly smaller than $x$. Next, for a function $f: \R^d \to \R$ that is $k \in \N$ times differentiable, we define its partial derivative $\partial^v f = \partial_1^{v_1}...\partial_d^{v_d} f $ for any multi-index $v \in \N^d$ with $v_1+...+v_d\le k$. In the case of $d=1$, we sometimes consider not only derivatives, but also weak derivatives of $f$. We also denote them by $\partial^v f $, but will point out, whenever they occur. For details on weak derivatives, we refer the reader to \cite{Ziemer}.
Finally, for a vector $v$ of any dimension, we denote by $|v|$ its Euclidean norm.   With these notations in hand, we can define a smoothness class of continuous functions, that plays a central role for (continuous) density estimation. 

\begin{definition} \rm{\label{Def_Nikolski}
Let $s>0$ be a smoothness parameter and $L>0$  a constant. Then the \textit{Nikol'ski class} $\mathcal{N}(s,L)$ consists of all densities $f:\R^d \to \R$ that are $\lfloor s \rfloor$-times continuously differentiable, with their derivatives satisfying
\begin{equation} \label{Eq_Nikolski_condition}
    \left[\int \left(\partial^{v} f(u+t)-\partial^{v}f(u)\right)^2du\right]^{1/2}\leq L|t|^{s-\lfloor s \rfloor},
\end{equation}
for all $t \in \R^d$ and all multi-indices $v=(v_1,...,v_d)$ with $v_1+...+v_d=\lfloor s \rfloor$. \\
For $d=1$, we also define the \textit{weak Nikol'ski class} $ \overline{\mathcal{N}}(s,L)$, 
which includes all densities that are $\lfloor s \rfloor -1$ times
differentiable, with $\partial^{(\lfloor s \rfloor -1)} f$ absolutely continuous and its weak derivative 
$\partial^{\lfloor s \rfloor } f$  satisfying \eqref{Eq_Nikolski_condition}. 
}
\end{definition}

\noindent 
Despite its technical appearance, the Nikol'ski class is of natural interest in the approximation of functions w.r.t. integral losses \cite{Nikolskiibook}.
In statistics,   $\mathcal{N}(s,L)$ has been studied, in the context of minimax optimal density estimation (see   \cite{Tsybakov}).
Intuitively, a function $f$ is an element of $\mathcal{N}(s,L)$, if it is $s$-times differentiable, with a non-integer value  such as $s=1.1$ interpreted as $1$-times differentiability plus smoothness of the first derivative. The weak Nikol'ski class, introduced in the second part of Definition \ref{Def_Nikolski} has a similar interpretation, but is less restrictive in its assessment of smoothness. In particular, it includes for $s=1.5$ double exponential functions, such as the Laplace density, which is differentiable everywhere, except at its peak (hence it would only be in $\mathcal{N}(1,L)$ for the ordinary Nikol'ski class). This observation is important, as our subsequent theory requires $s >1$ for continuous densities.\\
We can now state the mathematical conditions for Theorem \ref{Thm_main} in the case of continuous densities.

\begin{assumption}[Continuous Densities] \label{Ass_continuous} $ $\\
\noindent (1): There exists an $s\ge 1.5$ and a sufficiently large constant $L>0$ s.t. the densities $p,q$ are elements of $\mathcal{N}(s,L)$ or $\overline{\mathcal{N}}(s,L)$. Moreover, for $v$ with $v_1+...+v_d=1$, it holds that $|\partial^v p |, |\partial^v q |\le L$.\\
(2): The parameters $h$ (bandwidth), $\tau$ (floor) and $\beta$ (softmax parameter) are chosen depending on $n$. More precisely, there exist $u,v>0$ with $h=\mathcal{O}(n^{-u}), \tau = \mathcal{O}(n^{-v}), \beta = \mathcal{O}(n^{v})$, which satisfy
$$
\max\Big(v(2\lambda+1), \frac{1}{2s}+\frac{\lambda v}{s}\Big)<u<\frac{1}{2d}+\frac{v (\lambda+1)}{d}.
$$
(3): The kernel $K$ used in the KDEs satisfies Assumption (K), specified in Appendix \ref{App_A}.
\end{assumption}

\noindent In Assumption \ref{Ass_continuous}, part (1) is a smoothness condition, which is satisfied, e.g., by the Laplace density for $s=1.5$ (weak Nikol'ski class; see Appendix \ref{App_B} for a proof) and by the Gaussian density  for any $s >0$ (ordinary Nikol'ski class). Part (2) relates the three input parameters of the floored density estimator to each other (see Section \ref{Subsec_3_2} for a discussion). It implies that the bandwidth $h$ has to be chosen smaller than in optimal density estimation, a process called “undersmoothing”.  This approach is standard, when the task is not  estimation of the density itself, but approximation of a confidence region (see for example \cite{Horowitz2001} p.3999). 
In practice, choosing an adequate bandwidth is not too difficult, as many automated selection criteria, such cross validation, exist. Turning to the remaining parameters in Assumption \ref{Ass_continuous}, the floor $\tau$  moderates the stability of the ratio $\hat p(t)/\hat q_\tau(t)$   (see our discussion of \eqref{Eq_approx_1} and \eqref{Eq_approx_2}). To balance precision with stability, our theory suggests values of $\tau$ that are slightly larger than $h$. The constant $\beta$ should be of the same order as $\tau^{-1}$ and usually fixing $\beta=\tau^{-1}$ is a reasonable choice. Finally, part (3) of the assumption, imposes some regularity conditions on $K$, including smoothness, symmetry and fast decay of its tails. We discuss these assumptions in Appendix \ref{App_B} and only notice here that (K) is satisfied by many standard kernels in the literature (such as the popular Gaussian kernel).\\
In the next step, we formulate the assumptions for the case of discrete densities, which is easier as the RFE requires less input parameters than the KDE.

\begin{assumption}[Discrete Densities] \label{Ass_discrete} $ $\\
(1): The two functions $p,q$ are discrete densities on the finite set $\mathcal{X}$.\\
(2): The parameters $\tau$ (floor) and $\beta$ (softmax parameter) satisfy $\tau = \mathcal{O}(n^{-v})$, $\beta= \mathcal{O}(n^{v})$ with 
$$
v<(2\min\{3\lambda-3,2\lambda+1\})^{-1}
$$
\end{assumption}

\noindent With these theoretical assumptions in place, we can formulate the main mathematical result of this paper: The asymptotic validity of the lower bound $\hat \ell$ with probability $1-\alpha$.

\begin{theorem} \label{Thm_main}
Suppose that either Assumption \ref{Ass_continuous} or \ref{Ass_discrete} is satisfied, and denote by $\mathcal{F}$ the class of densities from the respective assumption (Nikol'ski class or discrete densities). Then, for any $\alpha \in (0,1)$, it holds that
\begin{equation} \label{Eq_Thm_main}
    \liminf_{n\to \infty} \inf_{p,q \in \mathcal{F}} \mathbb{P}\bigg(D_\lambda(p,q) \ge \hat \ell(\alpha)\bigg) \ge 1-\alpha.
\end{equation}
\end{theorem}

\noindent To interpret the above result, notice that it implies, for any fixed pair of densities $(p,q)$, that 
$$
   \liminf_{n\to \infty} \mathbb{P}\bigg(D_\lambda(p,q) \ge \hat \ell(\alpha)\bigg) \ge 1-\alpha.
$$
This means that asymptotically the lower bound $\hat \ell(\alpha)$ holds with confidence of at least $1-\alpha$, where we take the limit inferior, to guarantee well-definedness of the left side. In \eqref{Eq_Thm_main}, this result is further strengthened, as the confidence level holds even when minimizing over the entire class of densities (implying robustness w.r.t. the pair $(p,q)$).\\
The proof of Theorem 1 is challenging for multiple reasons: First, it requires proving asymptotic normality of $D_\lambda(\hat p, \hat q_\tau)$, while $\tau \to 0$, which inflates the variance of our estimate. In contrast, related works restrict themselves to densities bounded away from $0$ to avoid the difficulties of floored estimators. Second, as the softmax function converges to the hard max, its smoothness vanishes, complicating a proof of asymptotic normality (which requires smoothness). Third, a proof of uniform validity (as in \eqref{Eq_Thm_main}) requires the application of advanced concentration results (for all densities in the class) together with a careful study of many remainder terms. Given these challenges, we have deferred the proof to the Appendix, where we also discuss further technical details.

\section{Experiments} \label{Sec_experiments}
\noindent We evaluate our methods by applying them to various algorithms from the privacy literature. We also demonstrate that our lower bounds capture the privacy enhancements that have been recently studied through the lens of RDP.
\subsection{Setting}
\noindent We consider databases stemming from the 10-dimensional unit cube, that is $x \in [0,1]^m$, with $m = 10$ and each individual providing a data point $x_i \in [0,1]$. As discussed in Section II, choosing two adjacent databases that are "far apart" by some metric will yield tight lower bounds. In the unit cube, the furthest $x,x'$ with $x \sim_a x'$ can be apart w.r.t. the $\ell_{1}$-metric is 1. This distance is for example kept by the databases 
\begin{align} \label{data_bases}
x = (1, 0, \cdots 0) \quad \textnormal{and} \quad x' = (0, 0, \cdots, 0) 
\end{align}
and we will maintain this specific choice of $x$ and $x'$ for the remainder of this section.  The choice of databases $x$ and $x'$ in \eqref{data_bases} determines the divergence $D_{\lambda}(p,q)$ (with $A(x) \sim p$, $A(x') \sim q$), which for all studied algorithms is either equal to the optimal privacy parameter $\varepsilon$ or close to it (for a definition of the optimal privacy parameter see \eqref{Rényi_Sup}). We draw on our construction in \eqref{Eq_def_hat_ell} to infer empirical lower bounds and investigate how these compare to $D_{\lambda}(p,q)$ for each algorithm $A$ and $\lambda \in \{2,5,7\}$.

We implement our method in \texttt{R} and note that the treatment of discrete algorithms is straightforward, the computation of \eqref{Eq_def_hat_ell} being mainly grounded on relative frequencies  and summation. Regarding continuous algorithms, we employ packages and methods in \texttt{R} that are specifically designed for KDE. More concretely, we use the package "Kernsmooth" and its "bkde" function for KDEs and its "dpik" function to obtain the underlying bandwidths (we exponentiate these bandwidths with a factor $1.1$ to account for undersmoothing, see Section III). The bkde method evaluates a kernel density estimator over an even grid (in our simulations the grid consists of 1000 equidistant points) and we compute the Riemann sum over these evaluations to approximate the integrals in \eqref{Eq_def_hat_ell} and \eqref{Eq_variance_estiamtor}. 

According to our theory, parameters that influence the performance of our procedure are the prespecified confidence level, sample size, floor and smoothness parameter. We choose $\alpha = 0.05$, $n = 5 \times 10^6$, $\tau = 10^{-5}$ and $\beta = 10^5$ and maintain this selection of parameters across all algorithms studied in this section.

\subsection{Algorithms} 
\noindent \textbf{Additive Noise Mechanisms}
The Laplace and Gaussian Mechanism add random noise to the output of a statistic and constitute basic methods for privatization (see Section \ref{Sec_2_1}). For the underlying statistic $S$, we choose the sum. The additive noise mechanism is then given by
\begin{align*}
    A(x) = S(x) + Y = \sum\limits_{i = 1}^m x_i + Y
\end{align*}
where $Y \sim Lap(0, b)$ and $Y \sim \mathcal{N}(0, b^2)$ for the Laplace and Gaussian Mechanism respectively. $b$ determines the variance of the noise added to $S(x)$, with higher values translating into stronger output perturbation. We choose $b = 5$ for both Laplace and Gaussian noise. Note that on the unit cube $\triangle_S = 1$ holds (for a definition of the sensitivity  recall \eqref{eq_def_sensitivity}) and that the privacy parameter is
\begin{align*}
    \varepsilon(\lambda) =& \frac{1}{\lambda - 1} \log \Big\{ \frac{\lambda}{2 \lambda-1}  \exp\Big(\frac{\lambda -1}{b}\Big) \\&+ \frac{\lambda - 1}{2 \lambda-1}  \exp\Big(-\frac{\lambda}{b}\Big) \Big\}
\end{align*}
for the Laplace and $\varepsilon(\lambda) = \frac{\lambda}{2 b^2}$ for the Gaussian Mechanism. Our choice of databases $x, x'$ in \eqref{data_bases} delivers the privacy parameter with $\varepsilon(\lambda) = D_{\lambda}(p,q)$ for $p \sim A(x)$ and $q \sim A(x')$.\\ 

\noindent \textbf{Poisson Subsampling} We can enhance the privacy guarantee of our additive noise mechanisms by running them on a random subset of database $x$. More precisely, a mechanism $\Gamma$ is introduced that, given a generic database $x$ of size $m$, calls on independent random variables $Z_1, \cdots, Z_m$ with $Z_i \sim Ber(\gamma)$ and returns
\begin{align} \label{subset}
\{ x_i : Z_i = 1, 1 \leq i \leq m \}.
\end{align}
Each $Z_i$ follows a Bernoulli distribution and assumes $1$ with probability $\gamma$. Accordingly, $\Gamma$ includes each data point $x_i$ with probability $\gamma$ in \eqref{subset} before passing the subset on to the additive noise algorithms. Let $A$ be either the Laplace or Gaussian Mechanism and let $\varepsilon_0(\lambda)$ be the corresponding privacy parameter. Observing the privacy bounds in \cite{Zhu2019}, the enhanced privacy parameter of $A \circ \Gamma$ is given by
\begin{align*}
 \varepsilon(\lambda) =& \frac{1}{\lambda - 1} \log \Big\{ (1 - \gamma)^{\lambda-1} \, (\lambda \gamma - \gamma + 1) \\ +& \sum\limits_{j = 2}^{\lambda} \binom{\lambda}{j} (1 -\gamma)^{\lambda - j} \gamma^j \exp \big( (j-1) \, \varepsilon_0(\lambda)\big) \Big\}.
\end{align*}
As before, we fix $b = 5$ for the additive noise algorithms $A$ and choose $\gamma = 0.5$ for the subsample mechanism $\Gamma$. Note that for both Gaussian and Laplace noise this Poisson subsampling reduces the original privacy parameter $\varepsilon_0(\lambda)$ by more than 70\% for each $\lambda \in \{2,5,7\}$, amounting to a considerable increase in privacy. \\

\noindent \textbf{Randomized Response}
A commonly encountered algorithm to privatize discrete data is Randomized Response (RR). It is particularly suited for the local differential privacy model (LDP) where each user $i$ privatizes her locally held data $x_i$ before submitting it to an (untrusted) data collector. Binary Randomized Response for instance serves as a subroutine in RAPPOR \cite{Erlingsson2014}, which aims to provide privacy guarantees compliant with the LDP model. We can simulate the LDP model and binary RR in our setting by assuming that each individual $i$ randomizes its data point $x_i \in \{0,1\}$ via the mechanism
\begin{align*}
R_i(x_i) =
\begin{cases}
\, x_i \quad\quad\quad & \textnormal{with probability} \quad \frac{\exp(\varepsilon_0)}{1 + \exp(\varepsilon_0)} \\
\, 1-x_i \quad & \textnormal{with probability} \quad \frac{1}{1 + \exp(\varepsilon_0)}
\end{cases}
\end{align*}
where $\varepsilon_0 \geq 0$. Overall, the binary RR algorithm provides a vector containing the locally privatized data of all individuals given by
\begin{align*}
A(x) = (R_1(x_1), \cdots, R_m(x_m))
\end{align*}
for a database $x = (x_1, \cdots, x_m)$. A data collector can then perform her analysis and calculations over the $R_i(x_i)$ instead of the $x_i$. Importantly, the privacy parameter in the standard DP model is $\varepsilon_0$ for the entire algorithm $A$ and local mechanism $R_i$. In the RDP model, the privacy parameter for $A$ is given by 
\begin{align*}
\varepsilon(\lambda) = & \frac{1}{\lambda - 1} \log \Big\{ \Big( \frac{\exp(\varepsilon_0)}{\exp(\varepsilon_0) + 1}\Big)^{\lambda} \, \Big( \frac{1}{\exp(\varepsilon_0) + 1}\Big)^{1 -\lambda} \\
& + \Big( \frac{\exp(\varepsilon_0)}{\exp(\varepsilon_0) + 1}\Big)^{1 - \lambda} \, \Big( \frac{1}{\exp(\varepsilon_0) + 1}\Big)^{\lambda}  \Big\}.
\end{align*}
As before, we choose $\varepsilon_0 = 1.5$ for our simulations. \\

\noindent \textbf{Shuffle Model} One can further boost privacy in the LDP model by introducing an additional layer of anonymity. More precisely, the locally privatized data $R_i(x_i)$ is sent to a shuffler $\Pi$ that randomly permutes their order before passing them on to a data collector. Given data $y = (y_1, \cdots, y_m)$ the shuffler is defined by
\begin{align*}
\Pi(y) = \Pi(y_1, \cdots, y_m) = (y_{\pi(1)}, \cdots, y_{\pi(m)})
\end{align*}
where $\pi: \{1, \cdots, m \} \to \{1, \cdots, m \}$ is a random permutation chosen by $\Pi$. Applied to the binary RR algorithm $A$ discussed before, our Shuffled Randomized Response mechanism is given by $\Pi \circ A $ and 
\begin{align*}
\Pi \circ A(x) = \big( R_{\pi(1)}(x_{\pi(1)}), \cdots, R_{\pi(m)}(x_{\pi(m)}) \big).
\end{align*}
Intuitively, the shuffling operation increases privacy by further obscuring the association of individual $i$ and the privatized data $R_i(x_i)$. In \cite{Girgis2021} a general lower bound for discrete algorithms in the Shuffle Model is derived by calculating the Rényi divergence on databases as in \eqref{data_bases} for the Shuffled RR mechanism $\Pi \circ A$. To be more exact, we have
\begin{align*} \label{RR_Shuffled}
    \varepsilon(\lambda) & \geq D_{\lambda}(p,q) =  \frac{1}{\lambda - 1} \log \bigg\{ (1 + \binom{\lambda}{2} \, \frac{(\exp(\varepsilon_0) - 1)^2}{m \, \exp(\varepsilon_0)}  \\ +& \sum\limits_{j = 3}^{\lambda} \binom{\lambda}{j} \bigg( \frac{\exp(2 \varepsilon_0) - 1}{m \, \exp(\varepsilon_0)} \bigg)^j \, \mathbb{E} \bigg[  \Big( Z - \frac{m}{\exp(\varepsilon_0) + 1} \Big)^j   \bigg] \bigg\}.
\end{align*}
Here, the expected value is computed for the Binomial random variable $Z$ with
\begin{align*}
Z \sim Bin \Big(m, \frac{1}{\exp(\varepsilon_0) + 1}\Big).
\end{align*}
For our choice of databases in \eqref{data_bases}, $\varepsilon_0 = 1.5$ and $\lambda \in \{2, 5, 7 \}$ the shuffling mechanism $\Pi$ reduces the Rényi divergence by more than 60\%, illustrating the efficacy of the Shuffle Model. \\

\noindent \textbf{Noisy Gradient Descent}
Given a database $x$, consider the empirical risk minimization problem
\begin{align*}
\theta^* = {\arg \min}_{\theta \in \Theta } \mathcal{L}_x(\theta) \quad \textnormal{with} \quad  \mathcal{L}_x(\theta) = \frac{1}{m} \sum\limits_{i = 1}^m l(\theta, x_i)
\end{align*}
where $\theta^* \in \Theta$ is the optimal parameter of interest, $\Theta$ a closed and convex set and $l$ a loss function that relates parameter $\theta$ and data points $x_i$. This optimization problem can be tackled by finding an estimate $\hat{\theta}$ that closely approximates $\theta^*$. Iterative learning algorithms can be deployed to obtain $\hat \theta$ and the Noisy Gradient Descent algorithm, for instance, both delivers an estimate $\hat \theta$ and addresses privacy concerns by repeatedly adding Gaussian noise throughout its run. Starting with an initial parameter value $\theta_0$ and a learning rate $\eta$, Algorithm 1 from \cite{Chourasia2021} for instance computes \vspace{2 pt}

\begin{itemize}
\item[1.] $ g(\theta_k,x) = \sum_{i = 1}^m \nabla l(\theta_k,x_i)$ \vspace{2.5 pt} 
\item[2.] $\theta_{k+1} = \rho_{\Theta} \big( \theta_k - \frac{\eta}{m} g(\theta_k, x) + \sqrt{2 \eta} \; Y \big)$ \vspace{2pt}
\end{itemize}

\noindent for each iteration $k \in \{0, \cdots, K-1 \}$ for a total number of $K$ iterations before setting $\hat{\theta} = \theta_K$. Here, $\rho_\Theta$ is the projection onto the space $\Theta$ and $Y$ is a centered Gaussian random variable. Provided that some regularity conditions regarding the sum of loss gradients $g$ and underlying loss functions $l$ hold, the noisy gradient descent algorithm $A$ satisfies RDP. In order to show the tightness of their privacy bounds, \cite{Chourasia2021} determine a lower bound for the privacy loss associated with $A$ by calculating the Rényi divergence for a specific instance of the ERM problem. More precisely and adapted to our setting, assume that $Y \sim \mathcal{N}(0, b^2)$, $l(\theta,x_i) = \frac{1}{2} (\theta - x_i)^2$, $\theta_0 = 0$ and $\Theta = \mathbb{R}$. If the databases $x,x'$ are chosen as in \eqref{data_bases}, we have
\begin{align*}
\varepsilon(\lambda) \geq D_{\lambda}(p,q) = \frac{\lambda S_{g}^2}{4 b^2 m^2} \; \frac{2 -\eta}{1 + (1-\eta)^K} \; (1 - (1 -\eta)^K)
\end{align*}
where $S_g = 1$ is the sensitivity of the total loss gradient $g$. The corresponding simulations are carried out for $\eta = 0.2$, $b = 1$ and $K = 10$ (note that, $m = 10$ due to our choice of databases in \eqref{data_bases}).

\subsection{Runtime}

We measured the average time required to produce a single lower bound $\hat \ell$. These runtimes are averaged over 10 independent runs of our lower bound and recorded for each studied algorithm in Table \ref{runtime_table}. We used a notebook with an i7-8650U CPU and 16 GB RAM to register the runtimes reported here.

\begin{table}[h]
\centering
\begin{tabular}{ |p{4cm}|p{2.5cm}|  }
 \hline
 Algorithm & Runtime in minutes \\
 \hline
 Laplace Mechanism  & 3.34    \\
 Gaussian Mechanism &  1.76  \\
 Subsampled Laplace Mechanism & 4.74   \\
 Subsampled Gaussian Mechanism & 3.33   \\
 Randomized Response &  5.14  \\
 Randomized Response Shuffled & 6.62   \\
 Noisy Gradient Descent & 11.1  \\
 \hline
\end{tabular}
\caption{Runtimes needed to create a lower bound $\hat{\ell}$ for various algorithms. Times are averaged over 10 simulation runs.}    \label{runtime_table}
\end{table}

\subsection{Violin plots}

\noindent To display our simulation results, we use \textit{violin plots} (see Figure \ref{violin_plots}. A violin plot visualizes the distribution of a data sample, by vertically displaying its density, reflected along the $y$-axis. Accordingly, where many data points are concentrated the violin is wide, and where fewer data points exist, the violin is slim. In our case, the violin plots also include the sample median (a bold dot) and the interquartile range (a vertical black line). Hence, the violin plot also incorporates the  information of a standard box-plots.\\
In Figure \ref{violin_plots}, we use violin plots to visualize the distribution of the lower bound $\hat \ell$ for the Rényi divergence $D_\lambda(p,q)$. To make interpretations easier, we display on the $y$-axis the ratio $\hat \ell/D_\lambda(p,q)$, i.e. the relative size of the lower bound compared to the true divergence. The reference value $1$, where $\hat \ell =D_\lambda(p,q)$  is highlighted by a red horizontal line. 
Each plot is based on  $1000$ independent simulation runs.

\subsection{Interpretation of results}

\noindent In order to interpret our experiment results, we first need to understand what kind of outcomes our theory (developed in Sections \ref{Subsec_3_2} and \ref{Subsec_3_3}) predicts. First, we expect the lower bound $\hat \ell$ to be close to the true Rényi divergence $D_\lambda( p, q)$,  as we have from equation \eqref{Eq_def_hat_ell} $\hat \ell \approx D_\lambda( p, q_\tau)$ and for small enough $\tau$,  that  $D_\lambda( p, q_\tau)\approx  D_\lambda( p, q)$. 

\begin{figure*}[htp]
\centering
\begin{subfigure}{0.3\textwidth}
  \includegraphics[width=\linewidth]{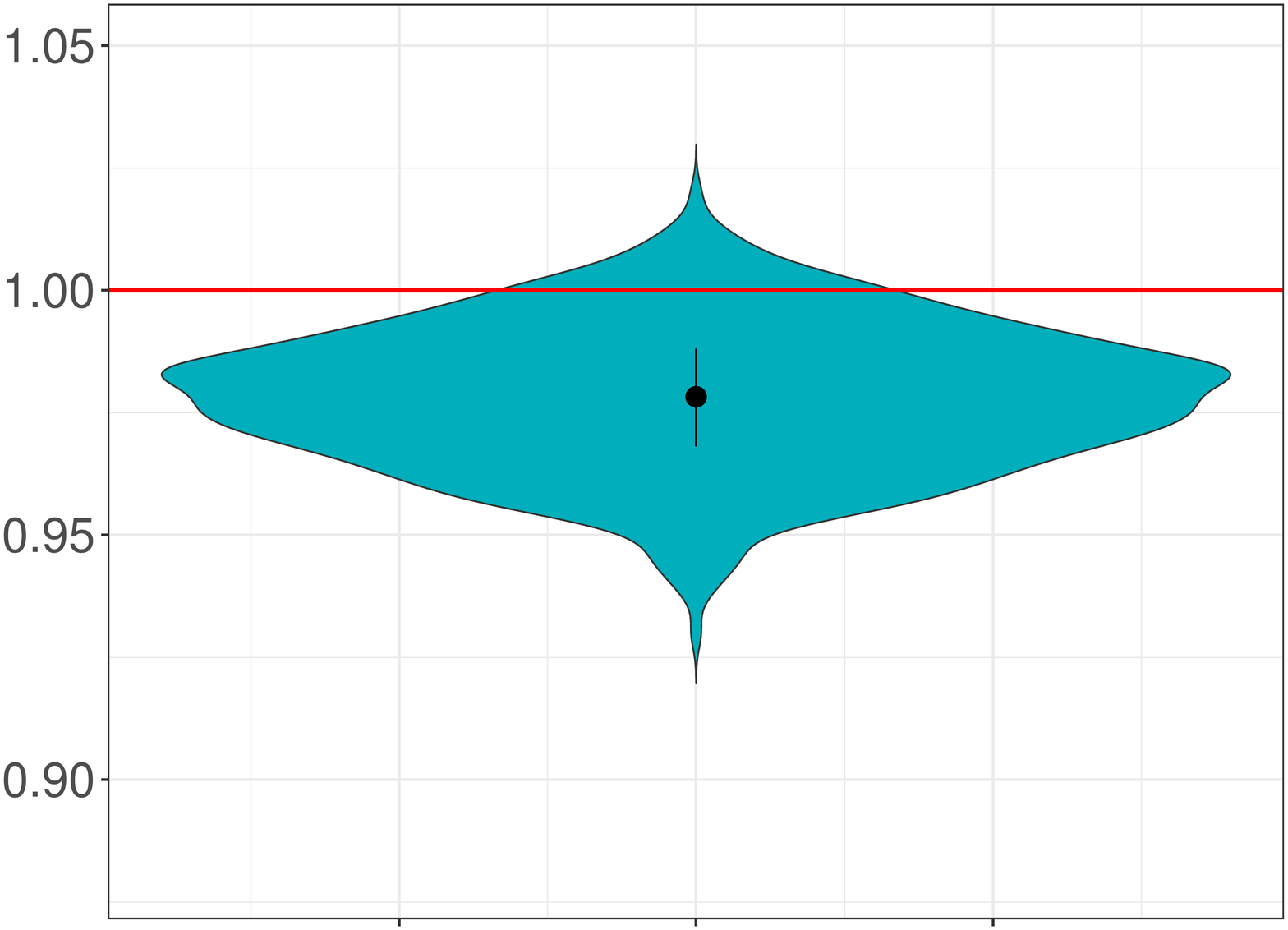}
  \caption{$\lambda=2$, $\hat\alpha=0.067$}
  \label{ngd_2}
\end{subfigure}\hfil 
\begin{subfigure}{0.3\textwidth}
  \includegraphics[width=\linewidth]{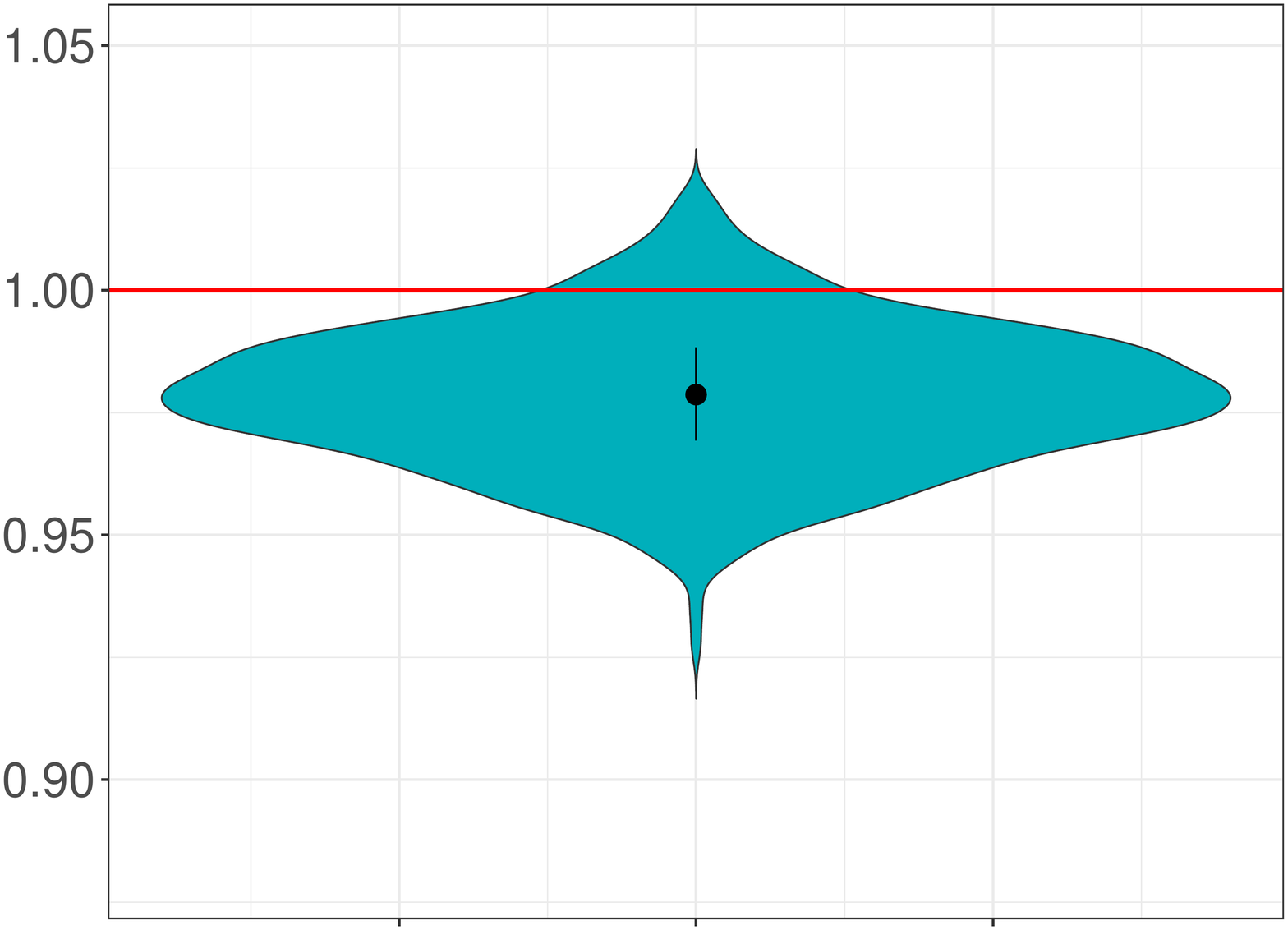}
  \caption{$\lambda=5$, $\hat\alpha=0.071$}
  \label{ngd_5}
\end{subfigure}\hfil 
\begin{subfigure}{0.3\textwidth}
  \includegraphics[width=\linewidth]{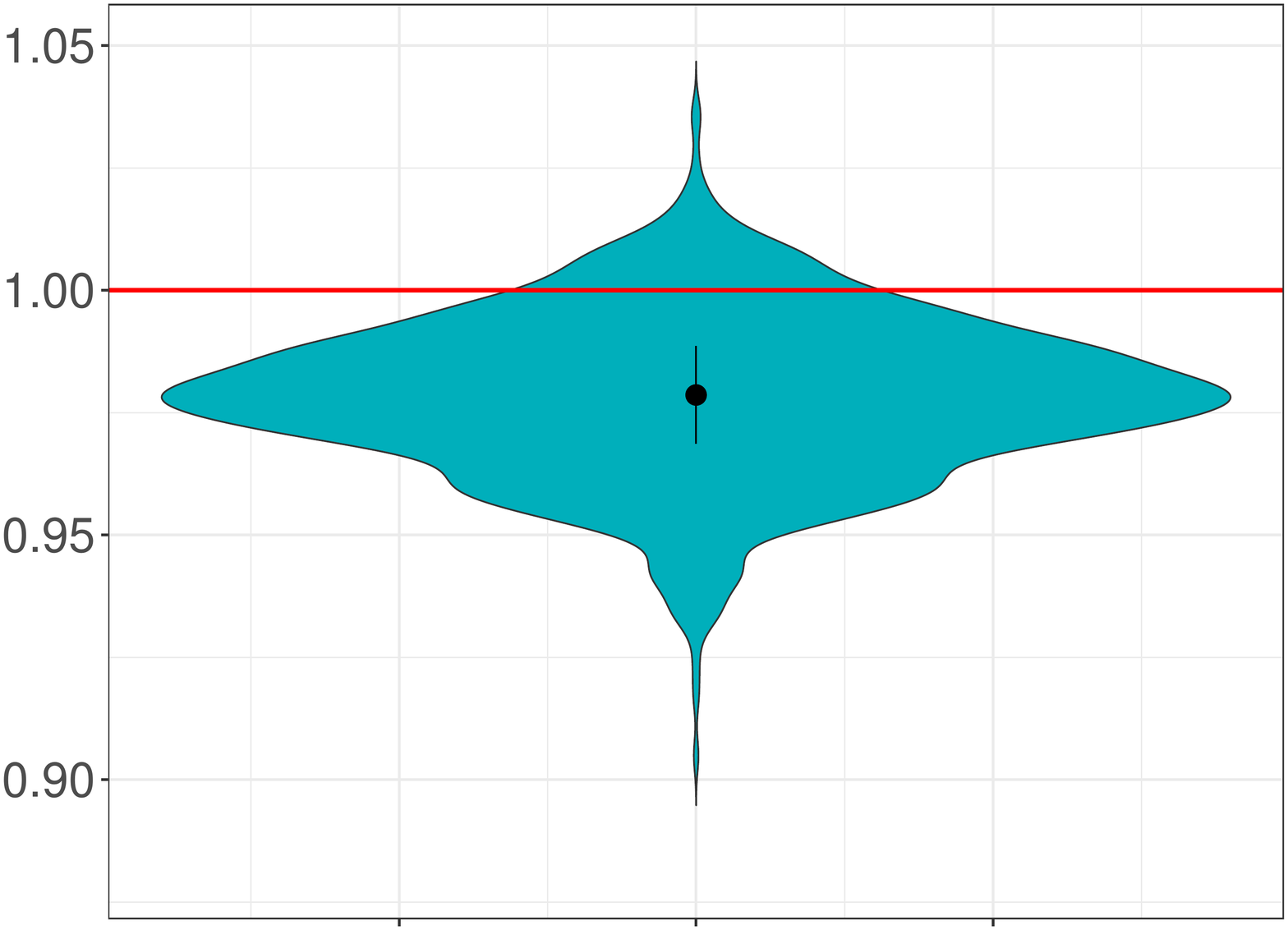}
  \caption{$\lambda=7$, $\hat\alpha=0.089$}
  \label{ngd_7}
\end{subfigure}
\captionsetup{labelformat=empty}
\caption{\romannumeral 1: Noisy Gradient Descent Algorithm}
\label{fig_ngd}
\end{figure*}
\medskip
\begin{figure*}[htp]
\centering
\begin{subfigure}{0.3\textwidth}
  \includegraphics[width=\linewidth]{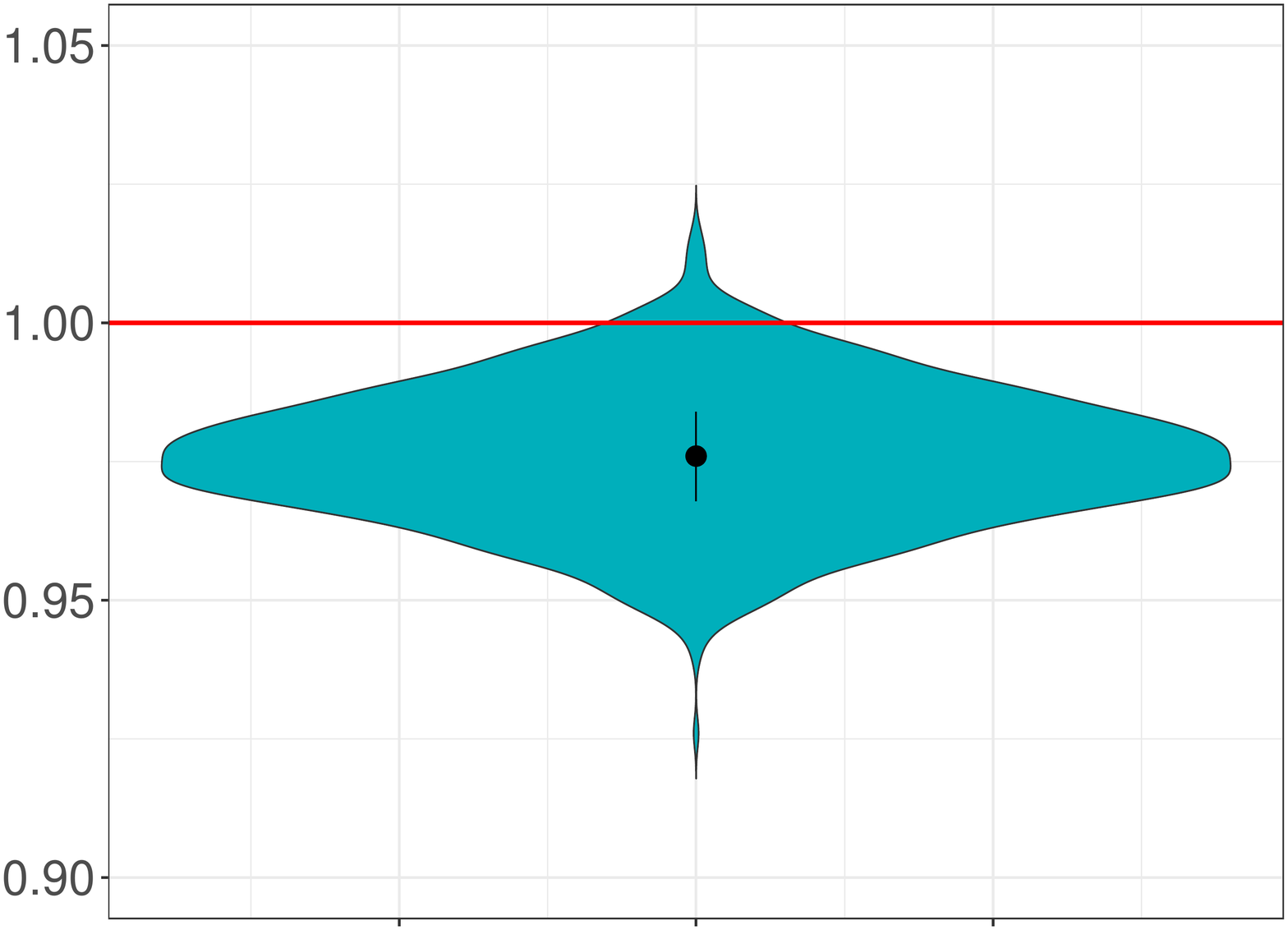}
  \caption{$\lambda=2$, $\hat\alpha=0.027$}
  \label{sub_g_2}
\end{subfigure}\hfil
\begin{subfigure}{0.3\textwidth}
  \includegraphics[width=\linewidth]{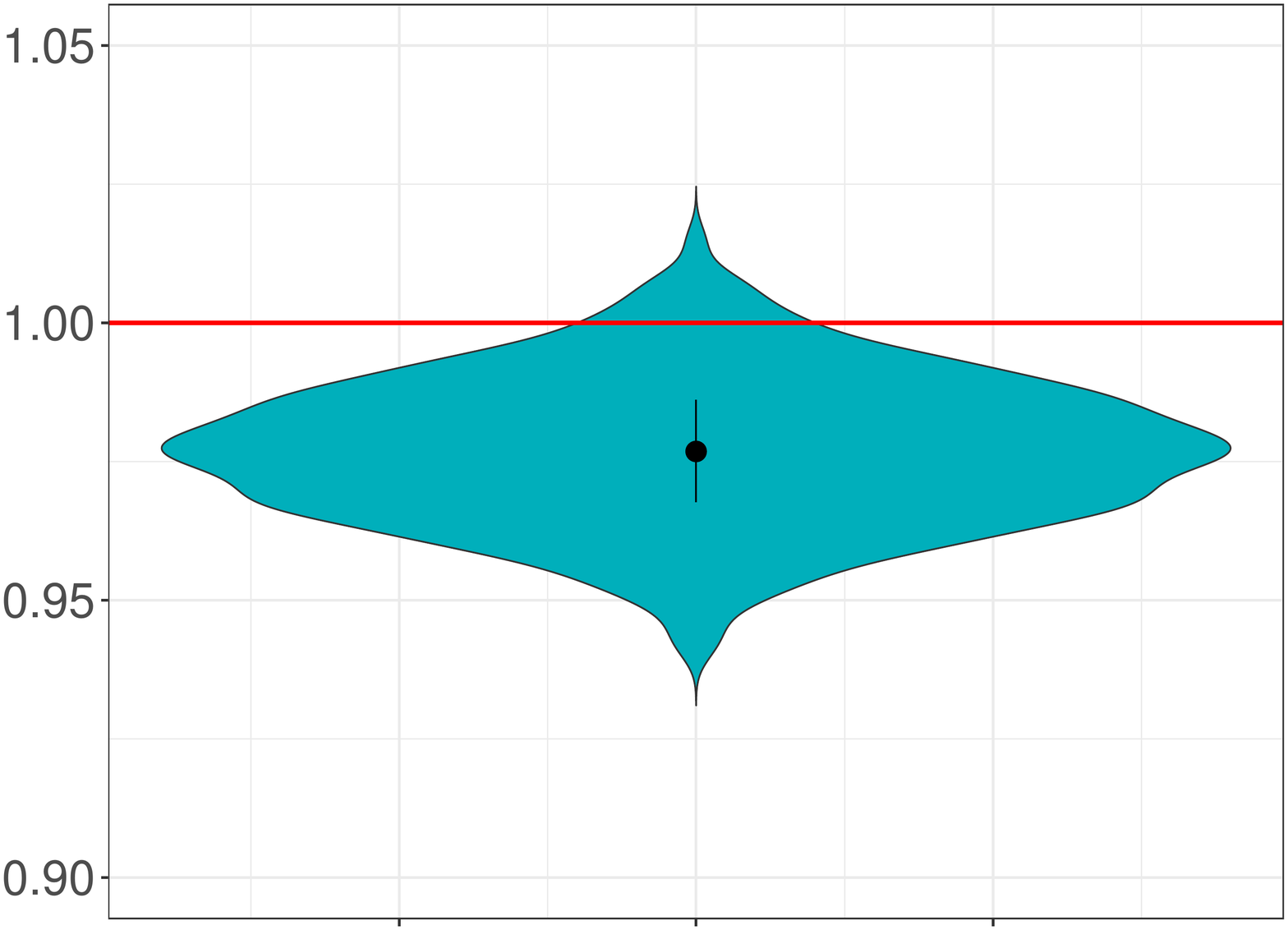}
  \caption{$\lambda=5$, $\hat\alpha=0.044$}
  \label{sub_g_5}
\end{subfigure}\hfil
\begin{subfigure}{0.3\textwidth}
  \includegraphics[width=\linewidth]{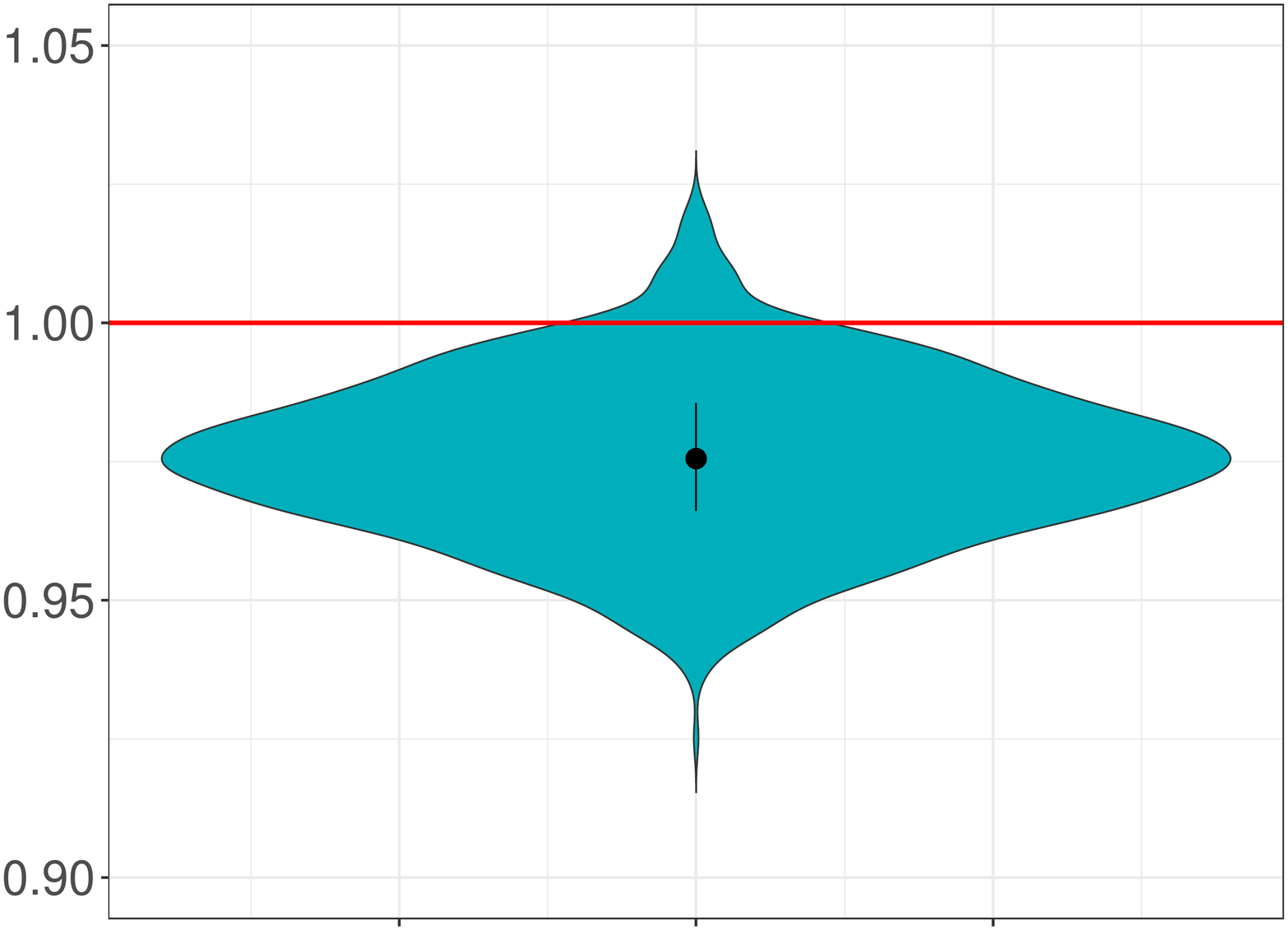}
  \caption{$\lambda=7$, $\hat\alpha=0.039$}
  \label{sub_g_7}
\end{subfigure}
\captionsetup{labelformat=empty}
\caption{\romannumeral 2: Subsampled Gaussian Mechanism}
\label{fig_sub_g}
\end{figure*}
\medskip
\begin{figure*}[htp]
\centering
\begin{subfigure}{0.3\textwidth}
  \includegraphics[width=\linewidth]{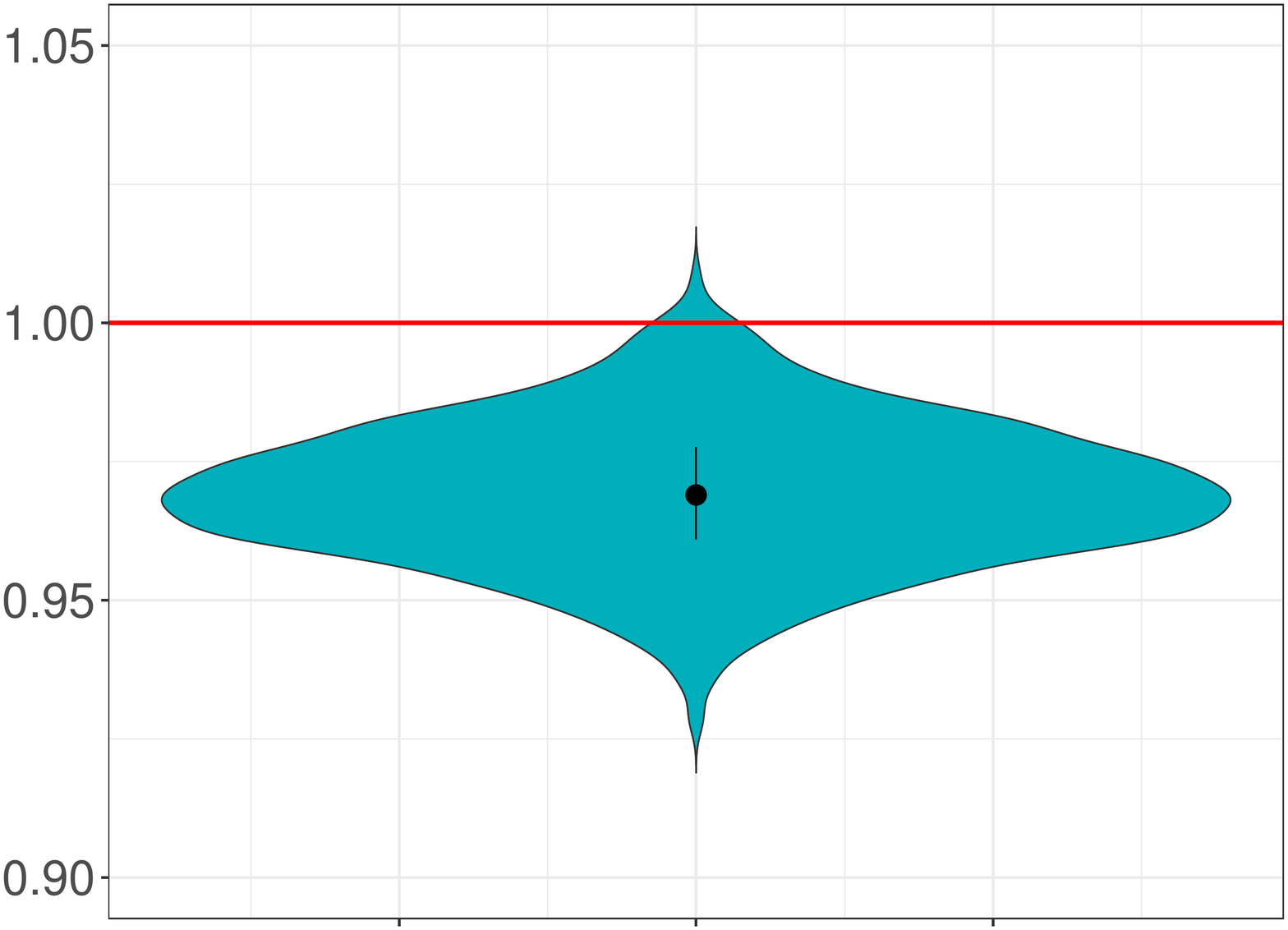}
  \caption{$\lambda=2$, $\hat\alpha=0.007$}
  \label{sub_l_2}
\end{subfigure}\hfil
\begin{subfigure}{0.3\textwidth}
  \includegraphics[width=\linewidth]{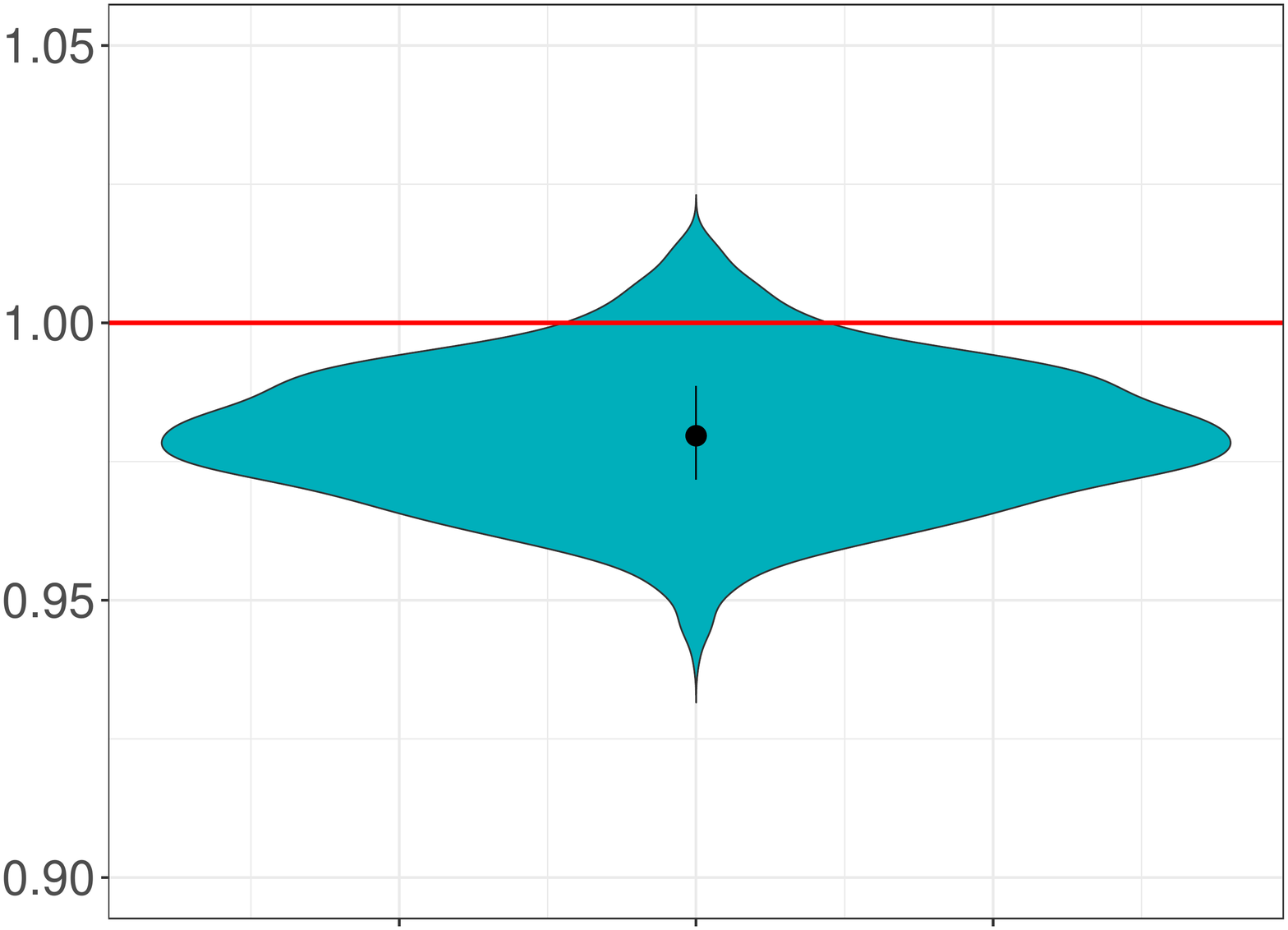}
  \caption{$\lambda=5$,$\hat\alpha=0.055$}
  \label{sub_l_5}
\end{subfigure}\hfil
\begin{subfigure}{0.3\textwidth}
  \includegraphics[width=\linewidth]{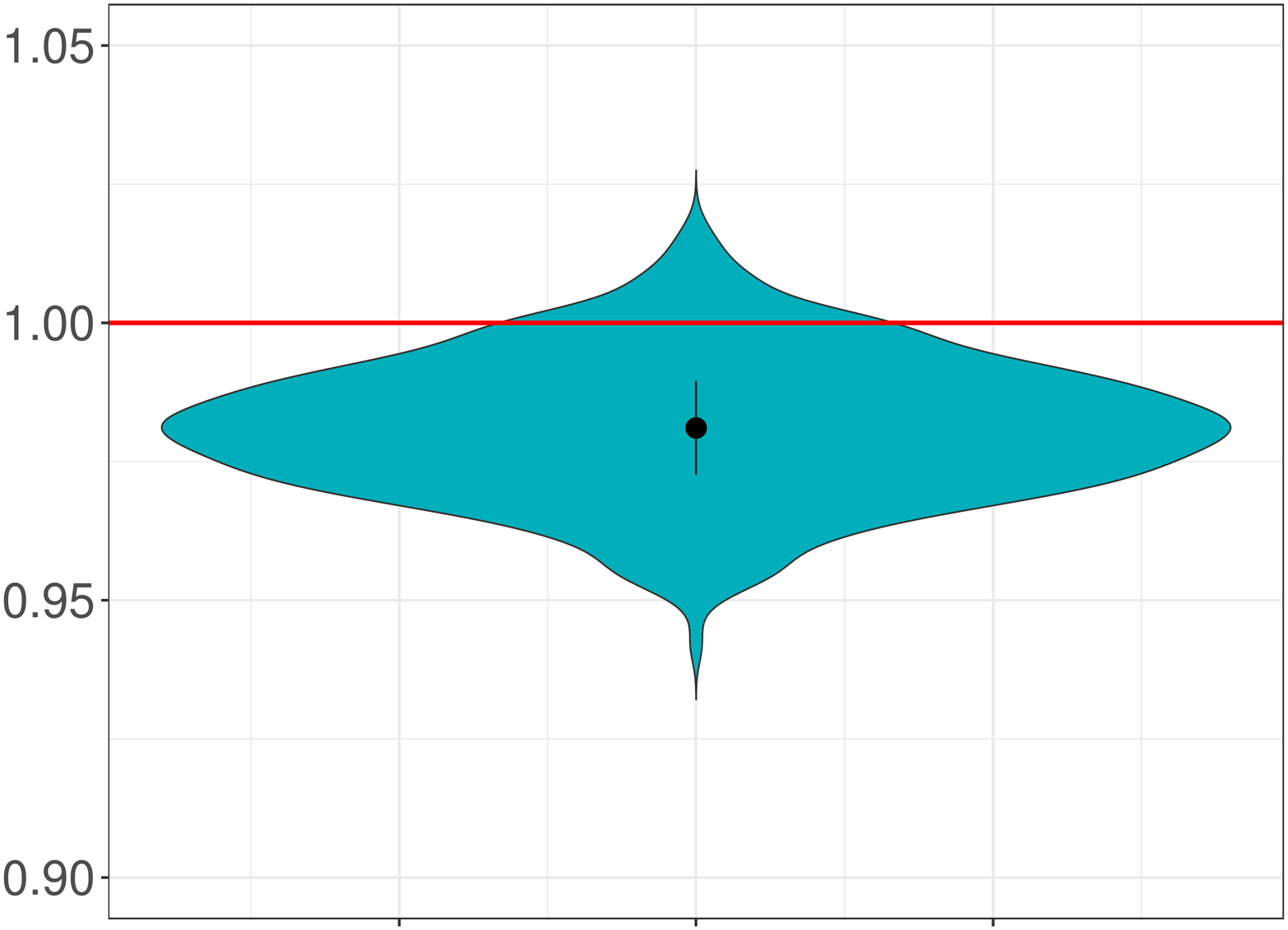}
  \caption{$\lambda=7$, $\hat\alpha=0.069$}
  \label{sub_l_7}
\end{subfigure}
\captionsetup{labelformat=empty}
\caption{\romannumeral 3: Subsampled Laplace Mechanism}
\label{fig_sub_l}
\end{figure*}
\medskip
\begin{figure*}[htp]
\centering
\begin{subfigure}{0.3\textwidth}
  \includegraphics[width=\linewidth]{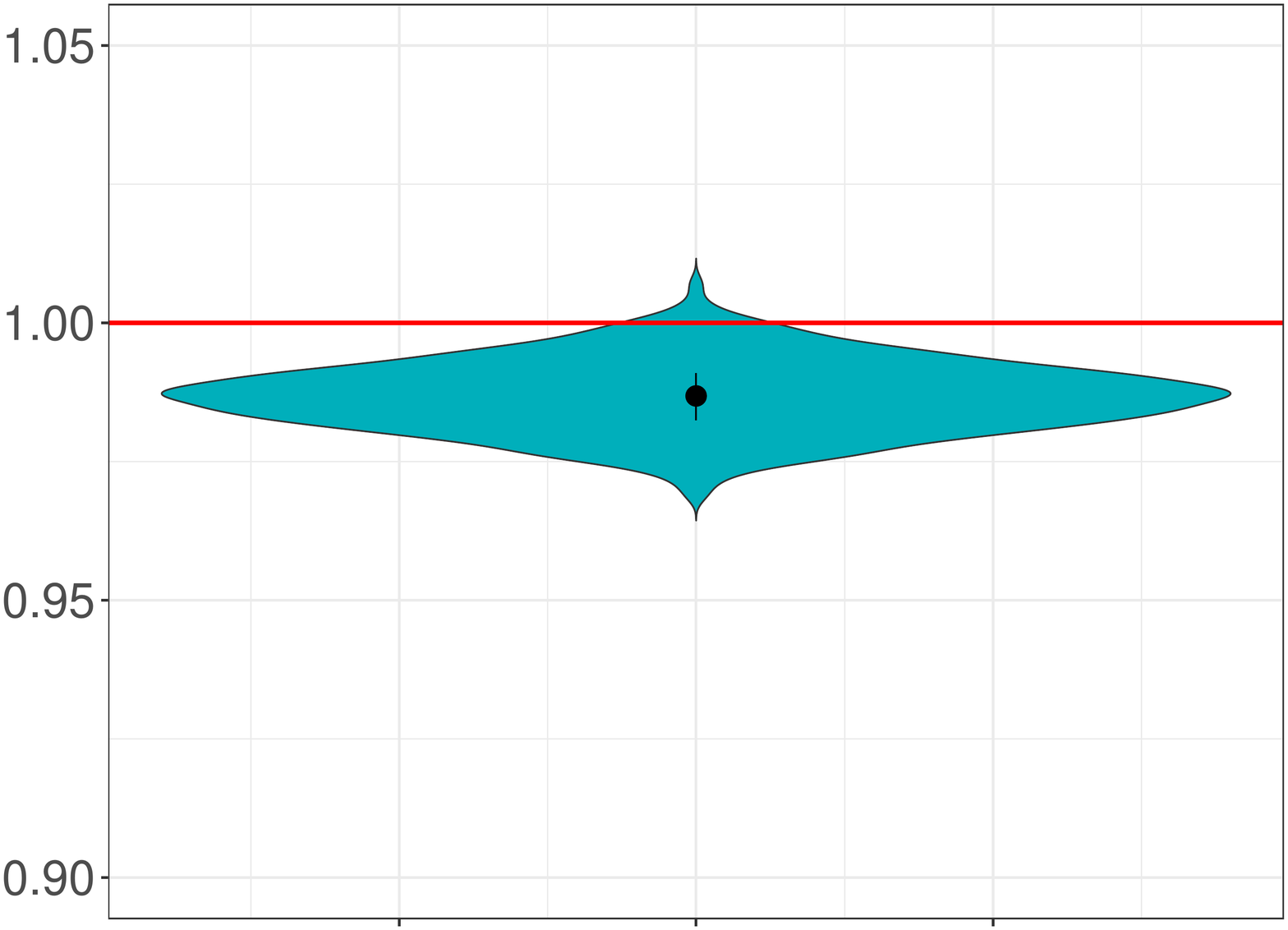}
  \caption{$\lambda=2$, $\hat\alpha=0.020$}
  \label{sum_g_2}
\end{subfigure}\hfil
\begin{subfigure}{0.3\textwidth}
  \includegraphics[width=\linewidth]{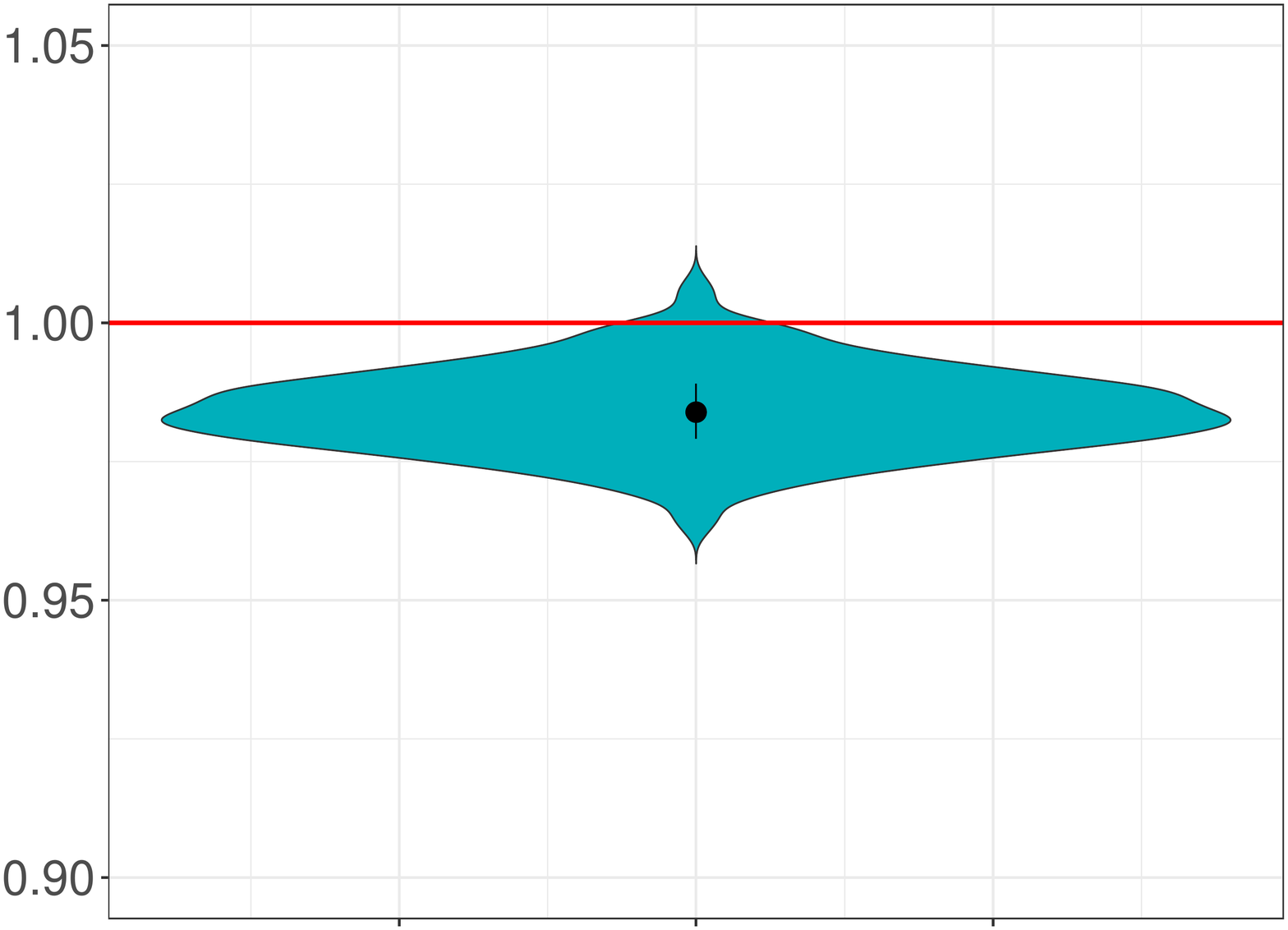}
  \caption{$\lambda=5$, $\hat\alpha=0.021$}
  \label{sum_g_5}
\end{subfigure}\hfil
\begin{subfigure}{0.3\textwidth}
  \includegraphics[width=\linewidth]{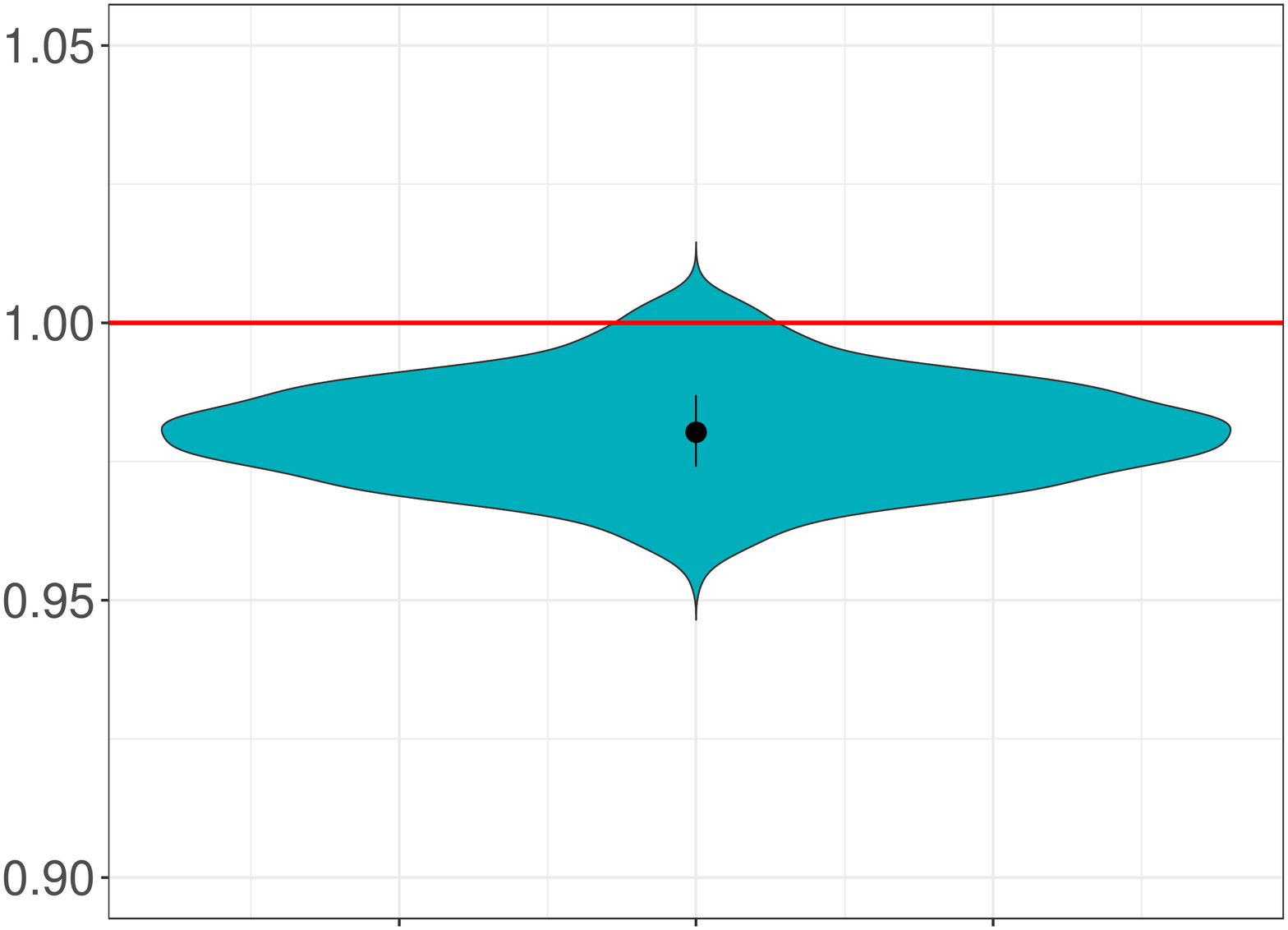}
  \caption{$\lambda=7$, $\hat\alpha=0.027$}
  \label{sum_g_7}
\end{subfigure}
\captionsetup{labelformat=empty}
\caption{\romannumeral 4: Gaussian Mechanism}
\label{fig_sum_g}
\end{figure*}
\medskip
\begin{figure*}[htp]
\centering
\begin{subfigure}{0.3\textwidth}
  \includegraphics[width=\linewidth]{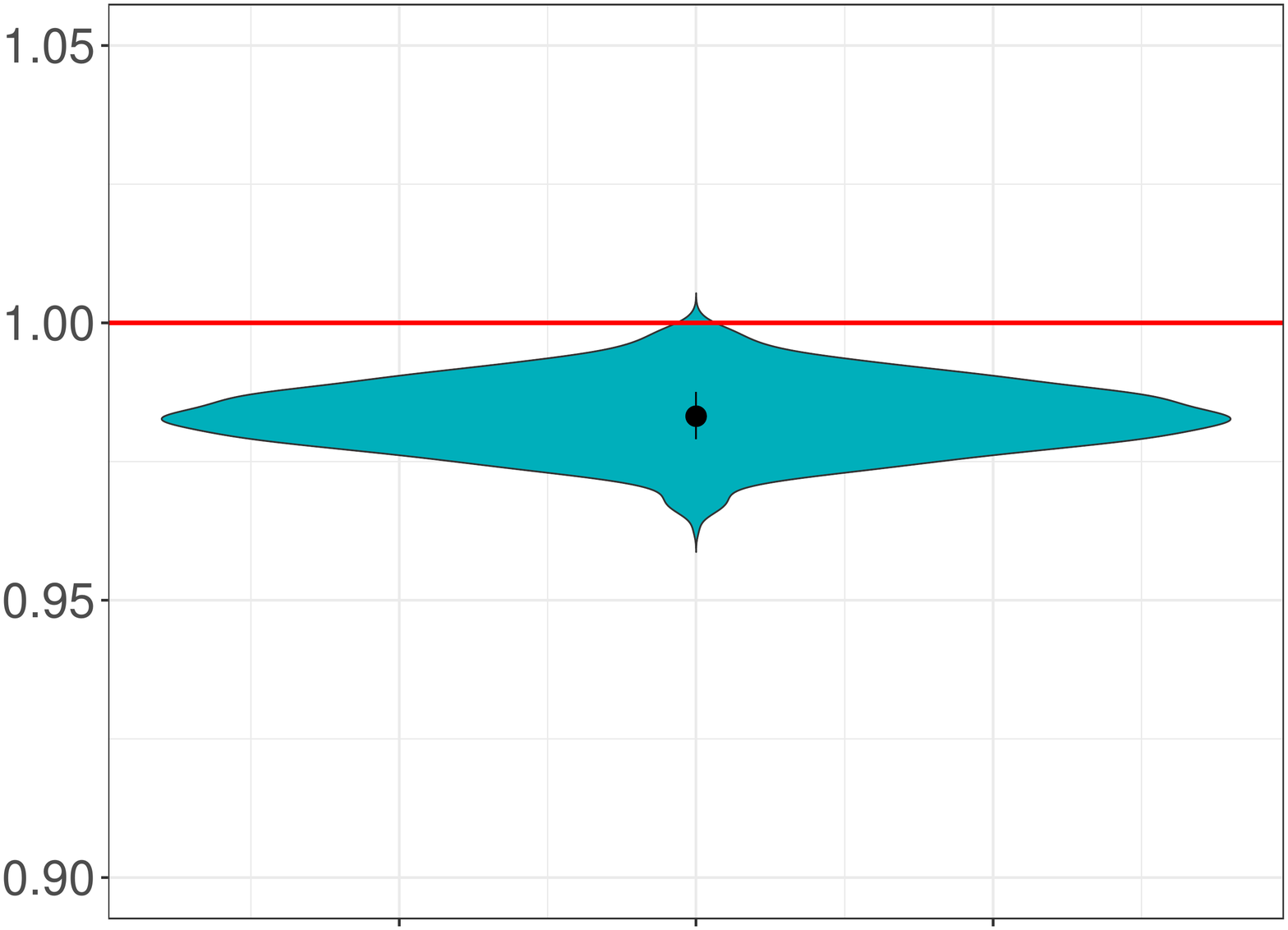}
  \caption{$\lambda=2$,  $\hat\alpha=0.001$}
  \label{sum_l_2}
\end{subfigure}\hfil
\begin{subfigure}{0.3\textwidth}
  \includegraphics[width=\linewidth]{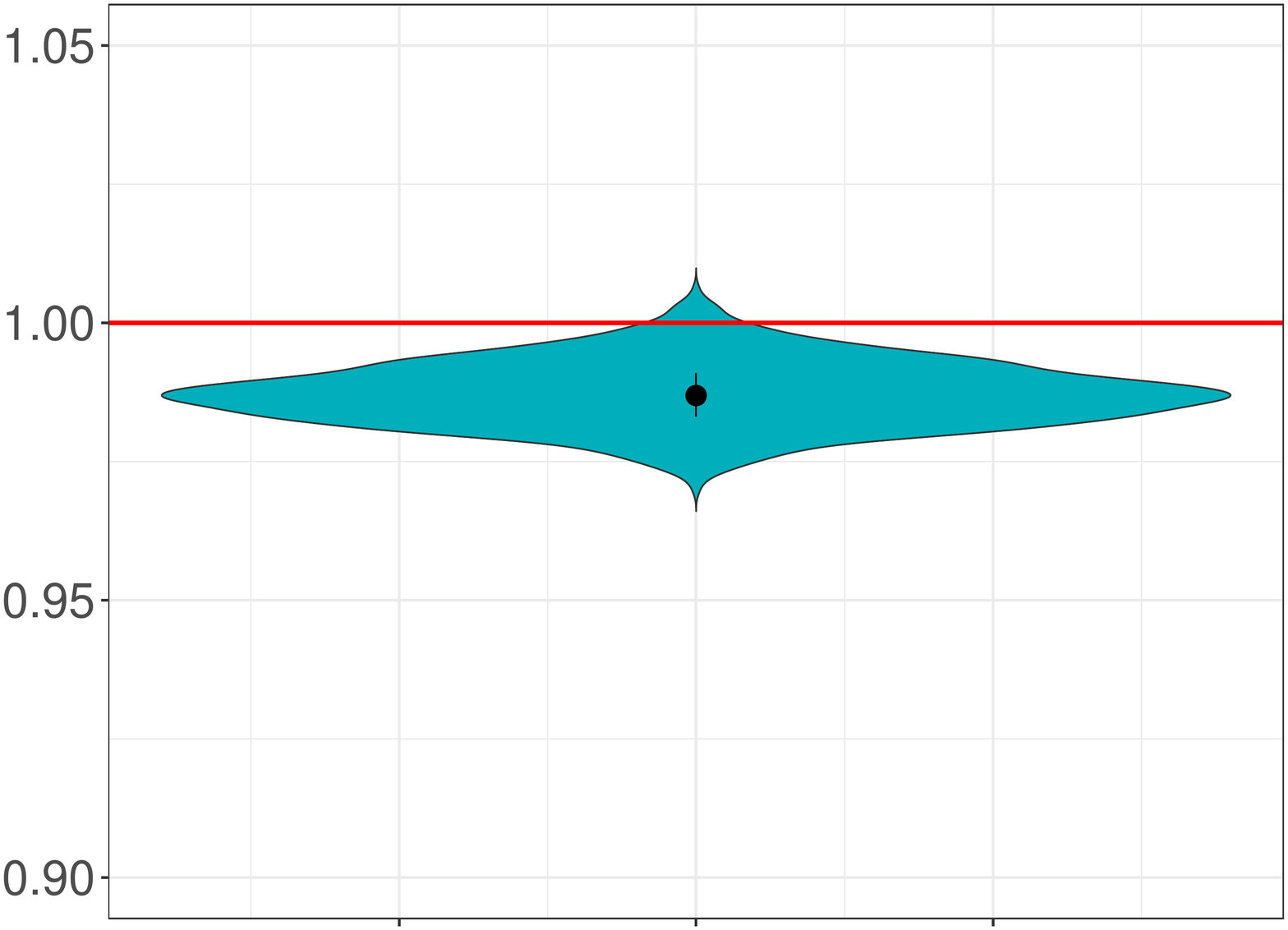}
  \caption{$\lambda=5$,  $\hat\alpha=0.017$}
  \label{sum_l_5}
\end{subfigure}\hfil
\begin{subfigure}{0.3\textwidth}
  \includegraphics[width=\linewidth]{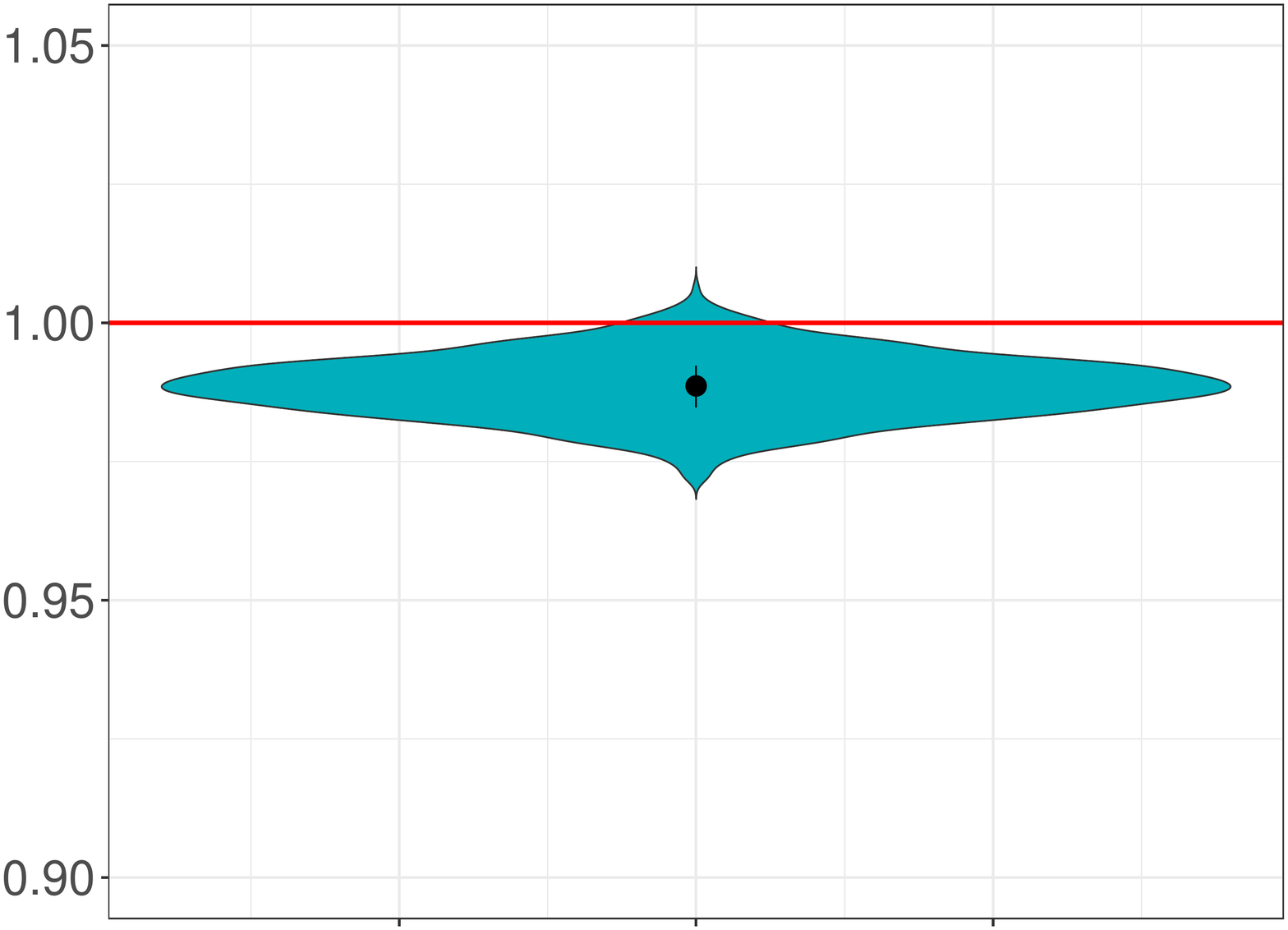}
  \caption{$\lambda=7$,  $\hat\alpha=0.020$}
  \label{sum_l_7}
\end{subfigure}
\captionsetup{labelformat=empty}
\caption{\romannumeral 5: Laplace Mechanism}
\label{fig_sum_l}
\end{figure*}
\begin{figure*}[htp]
\centering
\begin{subfigure}{0.3\textwidth}
  \includegraphics[width=\linewidth]{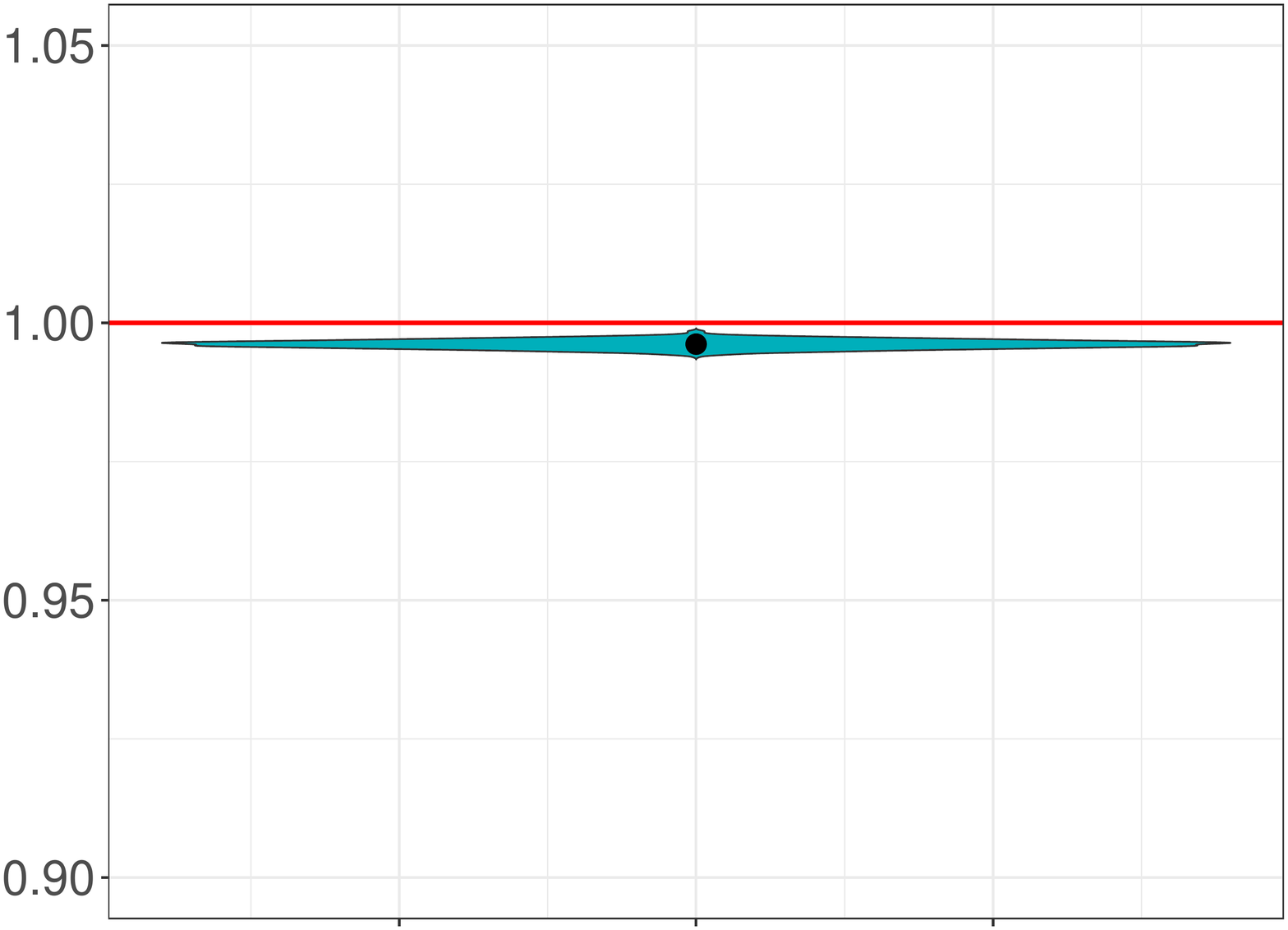}
  \caption{$\lambda=2$, $\hat\alpha=0.000$}
  \label{rr_2}
\end{subfigure}\hfil
\begin{subfigure}{0.3\textwidth}
  \includegraphics[width=\linewidth]{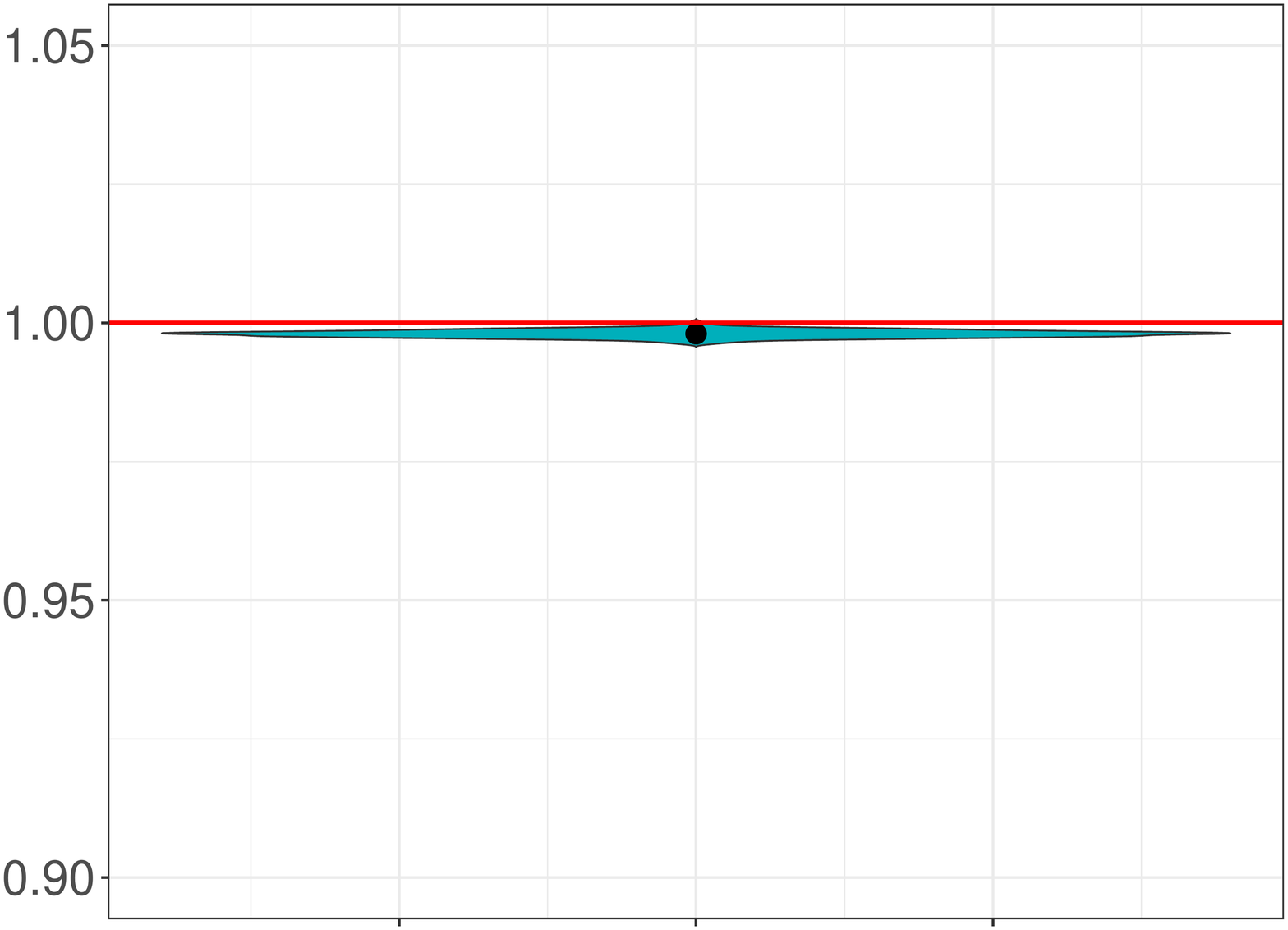}
  \caption{$\lambda=5$, $\hat\alpha=0.002$}
  \label{rr_5}
\end{subfigure}\hfil
\begin{subfigure}{0.3\textwidth}
  \includegraphics[width=\linewidth]{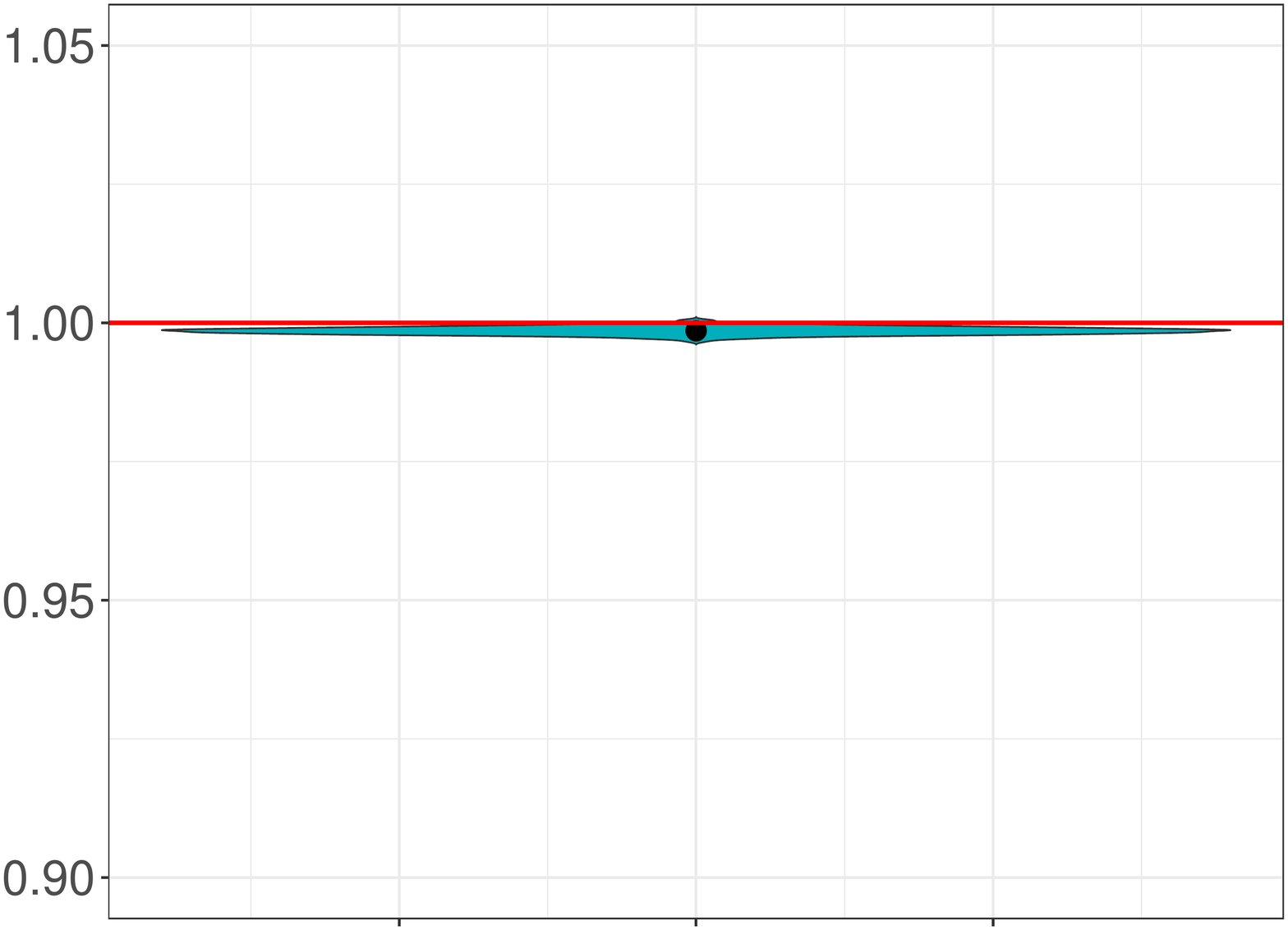}
  \caption{$\lambda=7$, $\hat\alpha=0.016$}
  \label{rr_7}
\end{subfigure}
\captionsetup{labelformat=empty}
\caption{\romannumeral 6: Randomized Response}
\label{fig_rr}
\end{figure*}
\setcounter{figure}{0}
\begin{figure*}[htp]
\centering
\begin{subfigure}{0.3\textwidth}
  \includegraphics[width=\linewidth]{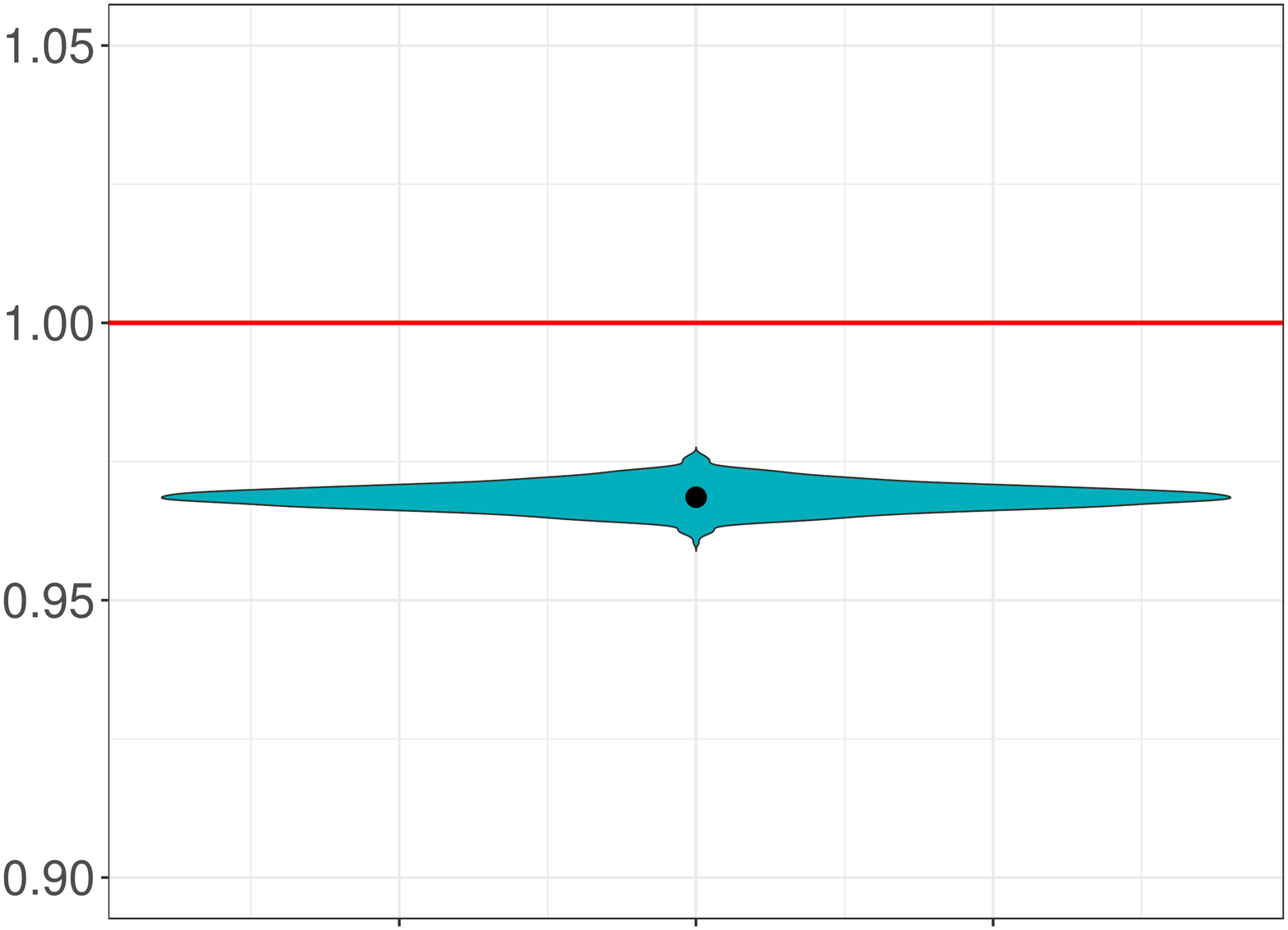}
  \caption{$\lambda=2$, $\hat\alpha=0.000$}
  \label{rrs_2}
\end{subfigure}\hfil
\begin{subfigure}{0.3\textwidth}
  \includegraphics[width=\linewidth]{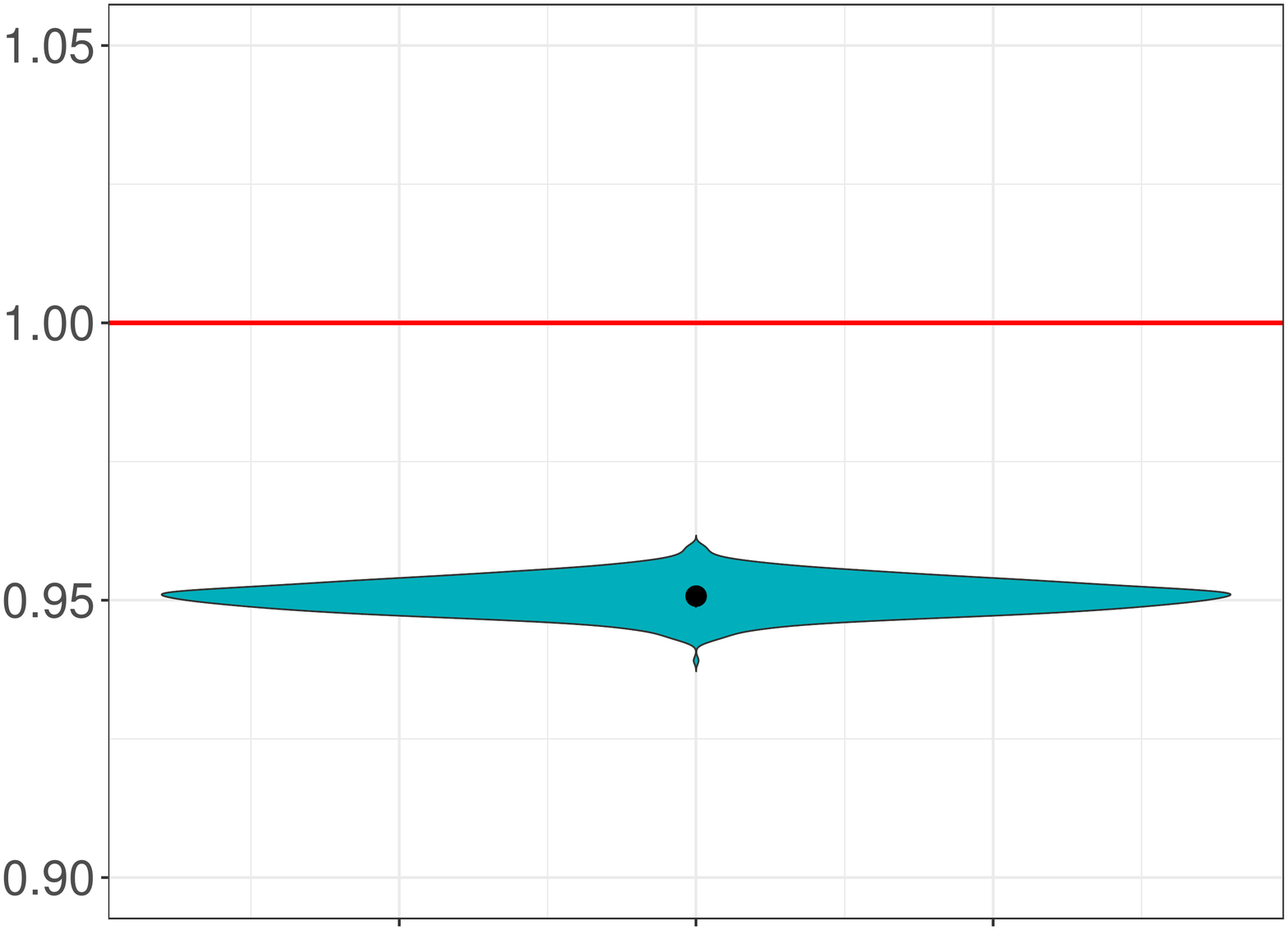}
  \caption{$\lambda=5$, $\hat\alpha=0.000$}
  \label{rrs_5}
\end{subfigure}\hfil
\setcounter{figure}{1}
\begin{subfigure}{0.3\textwidth}
  \includegraphics[width=\linewidth]{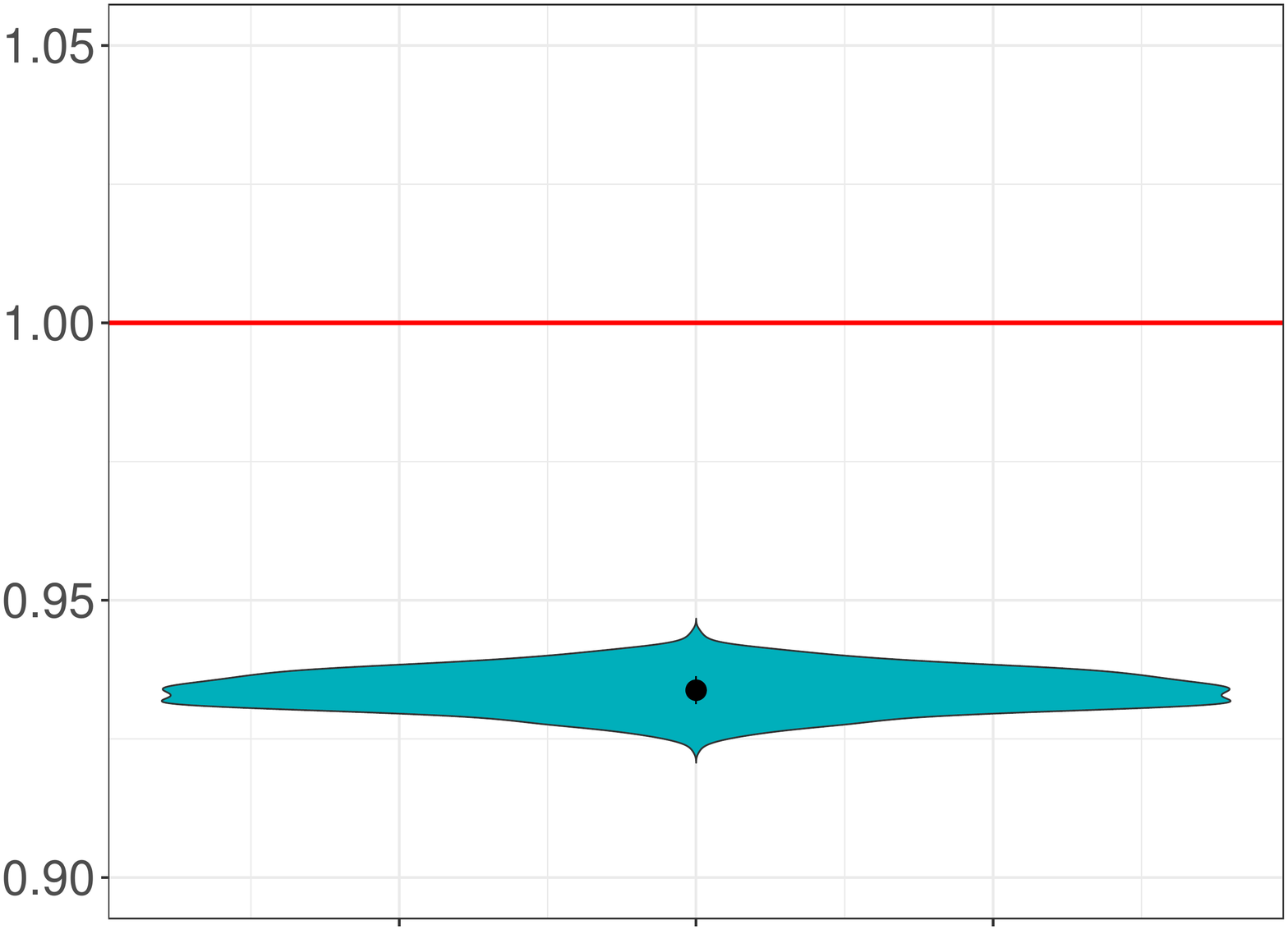}
  \caption{$\lambda=7$, $\hat\alpha=0.000$}
  \label{rrs_7}
\end{subfigure}
\captionsetup{labelformat=empty}
\caption{\romannumeral 7: Randomized Response Shuffled}\label{fig_rrs}
\captionsetup{justification=centering}
\caption*{Figure 1:  Violin plots of $\hat \ell/D_\lambda(p,q)$ for $1000$ simulation runs (each). The parameters are $\alpha = 0.05$, $n = 5 \times 10^6$, $\tau = 10^{-5}$ and 
$\beta = 10^5$ for all seven algorithms. $\hat \alpha$ refers to the empirical nominal level, that is defined as the total number of times $\,\,$\\
where $\hat \ell/D_\lambda(p,q)>1$, divided by $1000$. $\quad\qquad\qquad\qquad\qquad\qquad\qquad\qquad\qquad\qquad\qquad\qquad\qquad\qquad\qquad\qquad\qquad\,\,\,\,\,\, $}\label{violin_plots}
\end{figure*}

\noindent Translated to a violin plot this means that the displayed values of  $\hat \ell/D_\lambda(p,q)$ are reasonably close to $1$, or that the violin is wide around $1$. Moreover, for $\hat \ell$ to be a reliable approximation, its variance should not be too large, i.e., the violin should not be too long in $y$-direction. Second, since $\hat \ell$ is a statistical lower bound (see Theorem \ref{Thm_main}) we expect it to stay below $D_\lambda(p,q)$ with high probability. Translated to our violin plot this means, that  $\hat \ell/D_\lambda(p,q)$ is usually smaller than $ 1$, i.e. the main bulge of the violin is located below the red line. Of course, values greater than $1$ may occur (as our lower bound is only supposed to hold with probability $\ge 1-\alpha=95\%$) but should not constitute a higher proportion than $\alpha$ in the whole sample. Below each violin plot we therefore show $\hat \alpha$, the proportion of times where we observed overshooting values $\hat \ell/D_\lambda(p,q)>1$. We can now compare our empirical results to these theoretical standards.\\[0.7ex]
\noindent First, we find that the  lower bounds usually provide good approximations of the true divergence. In almost all cases, we observe that the median value of $\hat \ell/D_\lambda(p,q)$ is larger than $0.95$, i.e. that the lower bounds are fairly close to the ground truth. Indeed, for most algorithms the probability of observing $\hat \ell/D_\lambda(p,q)<0.95$ is slim (in many cases we did not sample a single value $<0.95$). While we observe fairly tight bounds overall, there exist noticeable differences between the algorithms. On the one hand, we see for the Randomized Response algorithm extreme concentration close to $1$ (almost no approximation error) with very little variance among the sampled bounds (the violin is very short in $y$-direction). On the other hand, for some   algorithms such as Noisy Gradient Descent, or the subsampled Gaussian Mechanism, we observe more variance among the estimated lower bounds (longer violin). This higher variance is due to the rapidly decaying densities for Gaussian algorithms, which make it difficult to approximate the density ratio $p/q$ in the Rényi divergence. Besides, the variance for all algorithms is influenced by $\lambda$, where larger values correspond to higher variance. This is due to the fact that any error of the density estimates is raised to a power of $\lambda$ in the empirical Rényi divergence. We also observe effects of $\lambda$ on the estimation bias, but only in a non-systematic way. \\[0.7ex]
\noindent Second, we observe, that $\hat \ell$ is indeed a lower bound for the true Rényi divergence with high probability (i.e., most of the violin's mass is concentrated below $1$ in each case). The empirical confidence level depends on the variance of the bounds, as well as the decay behavior of the densities. Our targeted level of at least $95\%$ is met in most cases (i.e., $1- \hat\alpha \ge 95\%$), but as before, we see differences between algorithms and for varying $\lambda$. While most of the time, the empirical confidence level is substantially higher than $95\%$ (as may be expected by \eqref{Eq_Thm_main}), we observe mild undercoverage in the case of the Noisy Gradient Descent algorithm. This effect is amplified for increasing $\lambda$, where the probability of too large values for $\hat \ell$ rises in lockstep with the variance.\\[0.7ex]
In summary, our experiments demonstrate  satisfying performance in terms of precision and coverage probabilities. Notice, that we have used identical parameters $\tau, \beta$ across all simulations. We have made this fixed choice, as in a true black-box scenario, we also cannot expect to perfectly tailor the parameters to the algorithm in hand. Yet, we want to point out, that adapting the input parameters can usually enhance performance - sometimes substantially so (see, for example, Appendix \ref{App_F}, of the online supplement, where we study the Randomized-Response-Shuffled-Algorithm for a different choice of $\tau$ and $\beta$).

\section{Related work}
Privacy validation via lower bounds has been pursued in prior work \cite{DP-Finder, DP-Sniper, Dette2022}. Lower bounds can be used to expose incorrect privacy guarantees or infer the privacy parameter $\varepsilon$. The methods in \cite{DP-Finder}, however, assume access to the algorithm's code, while the black-box methods in \cite{DP-Sniper} and \cite{Dette2022} are restricted to the standard DP model. This is also reflected in the algorithms we use to evaluate our methods. Here the only overlap with \cite{DP-Finder, DP-Sniper, Dette2022} are the Laplace Mechanism and Randomized Response. Parts of the approach in \cite{Dette2022} inform our work, since our methods also use non-parametric density estimates tailored to discrete and continuous algorithms. Yet, in contrast to prior work, we develop a method specifically designed to infer the Rényi divergence and RDP guarantees of a given algorithm.

The estimation of functionals for non-parametric density and regression estimators is a well-established subject in statistical theory \cite{fan91}. Estimation with a focus on divergences has been considered in various works  such as \cite{Wasserman2014, rubenstein2019, Poczos2011} and even statistical inference (in the sense of confidence intervals) in \cite{Wasserman2015,Moon2014}. Yet, all of these works make the common assumption of a finite (known) support of the densities. Not only does this assumption stand in tension with a black-box scenario as envisioned in this work, it also excludes most important privatizing mechanisms used in the DP literature. In particular, none of the algorithms investigated in our experiment section can be analyzed with these methods. This insight motivated our new methodology of regularized estimators and weak convergence, presented in Section \ref{Sec_3}. Analytically, our results differ from previous theory w.r.t. proofs (weighing differentiability of the softmax against approximation rates),  scope (including distributions with unbounded support) and convergence rates (instead of $\sqrt{n}$ we get a more subtle rate moderated by the decay of the densities and the choice of $\tau$).

\section{Conclusion}
\noindent We have presented methods that expand the current literature on black-box privacy assessment by targeting privacy guarantees for Rényi Differential Privacy. We provide practical estimators, lower bounds and a comprehensive theory that covers common algorithms from the DP literature as well as methods that augment their privacy. Our experiments showed tightness and reliability of the lower bounds, with reasonable runtimes. This suggests that apart from a black-box setting, our methods can also be used to complement mathematical proofs or other verification methods. Future work might include extending the methodology to an even broader class of algorithms, that are neither fully continuous nor discrete.

\section*{Acknowledgments}
This work was funded by the Deutsche Forschungsgemeinschaft (DFG, German Research Foundation) under Germany's Excellence Strategy - EXC 2092 CASA - 390781972.

\normalem

\bibliographystyle{IEEEtran}
\begin{small}
\bibliography{references}

\begin{thebibliography}{10}
\providecommand{\url}[1]{#1}
\csname url@samestyle\endcsname
\providecommand{\newblock}{\relax}
\providecommand{\bibinfo}[2]{#2}
\providecommand{\BIBentrySTDinterwordspacing}{\spaceskip=0pt\relax}
\providecommand{\BIBentryALTinterwordstretchfactor}{4}
\providecommand{\BIBentryALTinterwordspacing}{\spaceskip=\fontdimen2\font plus
\BIBentryALTinterwordstretchfactor\fontdimen3\font minus
  \fontdimen4\font\relax}
\providecommand{\BIBforeignlanguage}[2]{{%
\expandafter\ifx\csname l@#1\endcsname\relax
\typeout{** WARNING: IEEEtran.bst: No hyphenation pattern has been}%
\typeout{** loaded for the language `#1'. Using the pattern for}%
\typeout{** the default language instead.}%
\else
\language=\csname l@#1\endcsname
\fi
#2}}
\providecommand{\BIBdecl}{\relax}
\BIBdecl

\bibitem{Dwork2006}
C.~Dwork, F.~McSherry, K.~Nissim, and A.~Smith, ``Calibrating noise to
  sensitivity in private data analysis,'' in \emph{TCC'06}, 2006.

\bibitem{Erlingsson2014}
U.~Erlingsson, V.~Pihur, and A.~Korolova, ``Rappor: Randomized aggregatable
  privacy-preserving ordinal response,'' in \emph{CCS '14}, 2014.

\bibitem{Blocki2016}
J.~Blocki, A.~Datta, and J.~Bonneau, ``Differentially private password
  frequency lists,'' in \emph{{NDSS} 2016}, 2016.

\bibitem{Bolin2017}
B.~Ding, J.~Kulkarni, and S.~Yekhanin, ``Collecting telemetry data privately,''
  in \emph{NIPS'17}, 2017.

\bibitem{Abowd2018}
J.~M. Abowd, ``The {U.S.} census bureau adopts differential privacy,'' in
  \emph{Proceedings of the 24th {ACM} {SIGKDD} International Conference on
  Knowledge Discovery {\&} Data Mining, {KDD} 2018, London, UK, August 19-23,
  2018}.\hskip 1em plus 0.5em minus 0.4em\relax {ACM}, 2018, p. 2867.

\bibitem{Dwork2006_b}
C.~Dwork, K.~Kenthapadi, F.~McSherry, I.~Mironov, and M.~Naor, ``Our data,
  ourselves: Privacy via distributed noise generation,'' in
  \emph{{EUROCRYPT}'06}, 2006.

\bibitem{Mironov2017}
I.~Mironov, ``R{\'e}nyi differential privacy,'' in \emph{2017 IEEE 30th
  computer security foundations symposium (CSF)}.\hskip 1em plus 0.5em minus
  0.4em\relax IEEE, 2017, pp. 263--275.

\bibitem{Wang2019}
Y.~Wang, B.~Balle, and S.~P. Kasiviswanathan, ``Subsampled renyi differential
  privacy and analytical moments accountant,'' in \emph{{AISTATS} 2019}, 2019.

\bibitem{Feldman2021}
V.~Feldman and T.~Zrnic, ``Individual privacy accounting via a r{\'{e}}nyi
  filter,'' in \emph{NeurIPS 2021}, 2021.

\bibitem{Chourasia2021}
R.~Chourasia, J.~Ye, and R.~Shokri, ``Differential privacy dynamics of langevin
  diffusion and noisy gradient descent,'' in \emph{NeurIPS 2021}, 2021.

\bibitem{Girgis2021}
A.~M. Girgis, D.~Data, S.~N. Diggavi, A.~T. Suresh, and P.~Kairouz, ``On the
  r{\'{e}}nyi differential privacy of the shuffle model,'' in \emph{{CCS} '21},
  2021.

\bibitem{Zhu2019}
Y.~Zhu and Y.~Wang, ``Poission subsampled r{\'{e}}nyi differential privacy,''
  in \emph{{ICML}'19}, 2019.

\bibitem{Reed2010}
J.~Reed and B.~C. Pierce, ``Distance makes the types grow stronger: A calculus
  for differential privacy,'' in \emph{ICFP'10}, 2010.

\bibitem{Barthe2016b}
G.~Barthe, M.~Gaboardi, B.~Gr\'{e}goire, J.~Hsu, and P.-Y. Strub, ``Proving
  differential privacy via probabilistic couplings,'' in \emph{{LICS '16}},
  2016.

\bibitem{Hsu2017}
A.~Albarghouthi and J.~Hsu, ``Synthesizing coupling proofs of differential
  privacy,'' vol.~2, no. POPL, 2017.

\bibitem{StatDP}
Z.~Ding, Y.~Wang, G.~Wang, D.~Zhang, and D.~Kifer, ``Detecting violations of
  differential privacy,'' in \emph{CCS '18}, 2018.

\bibitem{DP-Finder}
B.~Bichsel, T.~Gehr, D.~Drachsler-Cohen, P.~Tsankov, and M.~Vechev,
  ``Dp-finder: Finding differential privacy violations by sampling and
  optimization,'' in \emph{CCS '18}, 2018.

\bibitem{Kifer2019}
Y.~Wang, Z.~Ding, G.~Wang, D.~Kifer, and D.~Zhang, ``Proving differential
  privacy with shadow execution,'' in \emph{PLDI '19}, 2019.

\bibitem{CheckDP}
Y.~Wang, Z.~Ding, D.~Kifer, and D.~Zhang, ``Checkdp: An automated and
  integrated approach for proving differential privacy or finding precise
  counterexamples,'' in \emph{CCS '20}, 2020.

\bibitem{Barthe2014}
G.~Barthe, M.~Gaboardi, E.~G. Arias, J.~Hsu, C.~Kunz, and P.~Strub, ``Proving
  differential privacy in hoare logic,'' in \emph{{CSF'14}}, 2014.

\bibitem{Barthe2016}
G.~Barthe, N.~Fong, M.~Gaboardi, B.~Gr\'{e}goire, J.~Hsu, and P.-Y. Strub,
  ``Advanced probabilistic couplings for differential privacy,'' in
  \emph{CCS'16}, 2016.

\bibitem{Liu2019}
X.~Liu and S.~Oh, ``Minimax optimal estimation of approximate differential
  privacy on neighboring databases,'' in \emph{NeurIPS '19}, 2019.

\bibitem{Barthe2020}
G.~Barthe, R.~Chadha, V.~Jagannath, A.~P. Sistla, and M.~Viswanathan,
  ``Deciding differential privacy for programs with finite inputs and
  outputs,'' in \emph{LICS '20}, 2020.

\bibitem{Sato2019}
T.~Sato, G.~Barthe, M.~Gaboardi, J.~Hsu, and S.~Katsumata, ``Approximate span
  liftings: Compositional semantics for relaxations of differential privacy,''
  in \emph{{LICS}'19}, 2019.

\bibitem{DP-Sniper}
B.~Bichsel, S.~Steffen, I.~Bogunovic, and M.~T. Vechev, ``Dp-sniper: Black-box
  discovery of differential privacy violations using classifiers,'' in
  \emph{{SP}'21}, 2021.

\bibitem{Dette2022}
{\"{O}}.~Askin, T.~Kutta, and H.~Dette, ``Statistical quantification of
  differential privacy: {A} local approach,'' in \emph{{SP} '22}, 2022.

\bibitem{dworkrobust}
\BIBentryALTinterwordspacing
C.~Dwork and J.~Lei, ``Differential privacy and robust statistics,'' in
  \emph{Proceedings of the Forty-First Annual ACM Symposium on Theory of
  Computing}, ser. STOC '09.\hskip 1em plus 0.5em minus 0.4em\relax New York,
  NY, USA: Association for Computing Machinery, 2009, p. 371–380. [Online].
  Available: \url{https://doi.org/10.1145/1536414.1536466}
\BIBentrySTDinterwordspacing

\bibitem{Lyu2017}
M.~Lyu, D.~Su, and N.~Li, ``Understanding the sparse vector technique for
  differential privacy,'' \emph{Proc. {VLDB} Endow.}, 2017.

\bibitem{Dwork2014}
C.~Dwork and A.~Roth, ``The algorithmic foundations of differential privacy,''
  \emph{Found. Trends Theor. Comput. Sci.}, 2014.

\bibitem{Tsybakov}
A.~B. Tsybakov, ``Introduction to nonparametric estimation.''\hskip 1em plus
  0.5em minus 0.4em\relax Springer, 2009.

\bibitem{Wassermann}
A.~Krishnamurthy, K.~Kandasamy, B.~Poczos, and L.~Wasserman, ``Nonparametric
  estimation of renyi divergence and friends,'' in \emph{International
  Conference on Machine Learning}, July 2014.

\bibitem{van2000asymptotic}
A.~van~der Vaart, \emph{Asymptotic Statistics}, ser. Asymptotic
  Statistics.\hskip 1em plus 0.5em minus 0.4em\relax Cambridge University
  Press, 2000.

\bibitem{Ziemer}
W.~P. Ziemer, \emph{Weakly differentiable functions: Sobolev spaces and
  functions of bounded variation}.\hskip 1em plus 0.5em minus 0.4em\relax
  Springer, 1989.

\bibitem{Nikolskiibook}
S.~M. Nikolski and J.~M. Danskin, \emph{Approximation of functions of several
  variables and imbedding theorems}.\hskip 1em plus 0.5em minus 0.4em\relax
  Springer, 1975.

\bibitem{Horowitz2001}
J.~J. Heckman and E.~Leamer, ``Handbook of econometrics, volume 5.''\hskip 1em
  plus 0.5em minus 0.4em\relax Elsevier Science B.V., 2001.

\bibitem{fan91}
J.~Fan, ``{On the Estimation of Quadratic Functionals},'' \emph{The Annals of
  Statistics}, vol.~19, no.~3, pp. 1273 -- 1294, 1991.

\bibitem{Wasserman2014}
A.~Krishnamurthy, K.~Kandasamy, B.~Poczos, and L.~Wasserman, ``Nonparametric
  estimation of renyi divergence and friends,'' in \emph{International
  Conference on Machine Learning}.\hskip 1em plus 0.5em minus 0.4em\relax PMLR,
  2014, pp. 919--927.

\bibitem{rubenstein2019}
P.~Rubenstein, O.~Bousquet, J.~Djolonga, C.~Riquelme, and I.~O. Tolstikhin,
  ``Practical and consistent estimation of f-divergences,'' \emph{Advances in
  Neural Information Processing Systems}, vol.~32, 2019.

\bibitem{Poczos2011}
B.~Poczos and J.~Schneider, ``On the estimation of alpha-divergences.''
  \emph{Journal of Machine Learning Research - Proceedings Track}, vol.~15, pp.
  609--617, 01 2011.

\bibitem{Wasserman2015}
K.~Kandasamy, A.~Krishnamurthy, B.~Poczos, L.~Wasserman, and j.~m. robins,
  ``Nonparametric von mises estimators for entropies, divergences and mutual
  informations,'' in \emph{Advances in Neural Information Processing Systems},
  C.~Cortes, N.~Lawrence, D.~Lee, M.~Sugiyama, and R.~Garnett, Eds.,
  vol.~28.\hskip 1em plus 0.5em minus 0.4em\relax Curran Associates, Inc.,
  2015.

\bibitem{Moon2014}
K.~Moon and A.~Hero, ``Multivariate f-divergence estimation with confidence,''
  in \emph{Advances in Neural Information Processing Systems}, Z.~Ghahramani,
  M.~Welling, C.~Cortes, N.~Lawrence, and K.~Weinberger, Eds., vol.~27.\hskip
  1em plus 0.5em minus 0.4em\relax Curran Associates, Inc., 2014.

\bibitem{kim2019uniform}
J.~Kim, J.~Shin, A.~Rinaldo, and L.~Wasserman, ``Uniform convergence rate of
  the kernel density estimator adaptive to intrinsic volume dimension,'' in
  \emph{International Conference on Machine Learning}.\hskip 1em plus 0.5em
  minus 0.4em\relax PMLR, 2019, pp. 3398--3407.

\bibitem{rigollet2009optimal}
P.~Rigollet and R.~Vert, ``Optimal rates for plug-in estimators of density
  level sets,'' \emph{Bernoulli}, vol.~15, no.~4, pp. 1154--1178, 2009.

\end{thebibliography}
\end{small}

\appendix

\noindent The appendix is dedicated to technical details of our methodology, as well as the proofs of our theoretical results. Throughout the appendix, we will denote by $C,C'$ generic, positive constants, that are independent of $n$ and may change from one equation to the next. Furthermore, we use the common notion of an $L^p$-norm for a real valued function $f: \R^d \to \R$, defined as
$$
\|f\|_p := \Big(\int |f(t)|^p dt \Big)^{1/p}
$$
as well as $\|f\|_\infty := \sup_t |f(t)|$ for the supremum norm. \\
We also want to point out that any references to Sections \ref{App_E} and \ref{App_F} refer to the online supplementary material.
\section{Paper}
\subsection{Details on density estimation} \label{App_A}

\noindent In Section \ref{Subsec_3_1}, we have introduced the discrete density estimator RFE and the continuous  density estimator KDE. We now gather some additional results on their convergence behavior, as well as the order requirements for the kernel $K$. \\
In order to state these results, we use stochastic Landau notations: Suppose a sequence of real valued random variables $(Z_n)_{n \in \N}$ and a sequence $(b_n)_{n \in \N}$ of positive real numbers is given. We then write $Z_n = \mathcal{O}_P(b_n)$, if 
$$
\lim_{C \to \infty} \limsup_{n \to \infty}\mathbb{P}(|Z_n/b_n| > C) =0
$$
and $Z_n = o_P(b_n)$, if for any fixed $C>0$
$$
\lim_{n \to \infty}\mathbb{P}(|Z_n/b_n| > C) =0.
$$
The interpretation is similar as for standard Landau symbols, where $Z_n = \mathcal{O}_P(b_n)$ (roughly) means that $Z_n$ is with high probability bounded by $b_n$ and $Z_n = o_P(b_n)$ means that $Z_n$ becomes with high probability negligible compared to $b_n$. \\
We can now analyze the density estimators.
Beginning with the RFE (defined in \eqref{Eq_def_RDE}), we have the following two results, the first proving its weak convergence and the second one specifying concentration of $\hat f$ around the true density $f$. 

\begin{lemma} \label{Lem_density_discrete}
Let $\mathcal{X}$ be a finite, non-empty set, $f: \mathcal{X}\to [0,1]$ a discrete density and $\hat f$ the RFE based on i.i.d. data $X_1,...,X_n \sim f$. Then
\begin{itemize}
    \item[i)] $\max_{t \in \mathcal{X}} |\hat f(t)-f(t)| =\mathcal{O}_P(1/\sqrt{n})$
    \item[ii)] There exists a positive constant $C>0$, only depending on $|\mathcal{X}|$ s.t.  
    $$
    \mathbb{P}(\max_{t \in \mathcal{X}} |\hat f(t)-f(t)|\ge \sqrt{\log(n)/n}) \ge 1-C/n.
    $$
\end{itemize}
\end{lemma}
\noindent The first result follows by an application of the union bound and Markov's inequality, while the second one follows by the union bound together with  Hoeffding's inequality.\\
Next, we consider the case of continuous density estimation. We formulate an additional assumption for the kernel $K$:
\begin{itemize}
    \item[(K)] The kernel $K: \R^d \to \R$ is symmetric, Lipschitz continuous, satisfies $\int K(t) dt=1$, $\int |K(t)| |t|^2<\infty$ and for any monomial $m(t)=t_1^{v_1}\cdot...\cdot t_d^{v_d}$ with $1 \le v_1+...+v_d \le \lfloor s \rfloor$ that $\int K(t) m(t) dt =0$ (“Kernel of order $\lfloor s \rfloor$“). Furthermore, there exists a polynomial $p: \R^d \to \R$ and a Lipschitz continuous function $\phi: \R \to \R$, s.t. $K(t)= \phi(p(t))$.
\end{itemize}
The above assumption is satisfied, e.g., by the Gaussian kernel ($K$ is the density of the standard normal) for order $\lfloor s \rfloor =1$. Another example in $d=1$ for $\lfloor s \rfloor =2$ is the Silverman kernel  $K(t)= \exp(-|t|/2)/2 \sin(|t|/\sqrt{2}+\pi/4)$. Kernels in multivariate settings can be obtained by taking the product of one dimensional kernels. For details on the construction of higher order kernels, see \cite{Tsybakov}. We can now formulate an analogue to Lemma \ref{Lem_density_discrete} for the continuous case.
\begin{lemma} \label{Lem_density_continuous}
Let $f: \R^d\to [0,1]$ be a continuous density in  $\mathcal{N}(s,L)$ (or  $ \overline{\mathcal{N}}(s,L)$) for $s \ge 1$, with $\sup_t|\partial^v f(t)| \le L$ for any multi-index $v$ with $v_1+...+v_d=1$. Moreover, let $\hat f$ be the KDE based on i.i.d. data $X_1,...,X_n \sim f$ satisfying (K). Then, for some constant, $C=C(L, s)$ it holds that
\begin{itemize}
    \item[i)]  $ \E\|\hat f-\E\hat f\|_2^2  \le \frac{C }{n h^d}$
    \item[ii)] $\| f-\E\hat f\|_2^2  \le C h^{2 s},$
    \item[iii)] If $nh^d \ge 1$
    $$
    \mathbb{P}\Big( \|\hat f-f\|_\infty \ge \sqrt{\frac{C\log(n)}{nh^d}}+C h \Big)\ \ge 1-1/n.
    $$
\end{itemize}
\end{lemma}
\noindent The first two parts of this Lemma follow by calculations analogous to Proposition 1.5 in \cite{Tsybakov}. The concentration result iii) follows by a bias-variance decomposition, with the variance part bounded via Theorem 12 in \cite{kim2019uniform}
(an investigation of their Lemma 11 shows that the constant $C$ can be chosen independent of $f$). The bias part follows by standard methods (see \cite{Tsybakov, rigollet2009optimal}).

\subsection{Properties of the softmax function}\label{App_B}
\noindent Throughout this paper, we have used the softmax function, to smoothly floor our density estimates. For ease of reference, we gather in this section some key properties of the softmax.  We start by recalling its definition for a floor $\tau>0$ and parameter $\beta>0$ as
\begin{equation*}
t_\tau = \beta^{-1} \log(\exp(t \beta)+\exp(\tau \beta)), \qquad t \in \R~.
\end{equation*}
The softmax function provides an approximation of the maximum from above, in the sense that for all $t \in \R, \tau, \beta>0$
\begin{equation*}
    t_\tau\geq \max(t, \tau)
\end{equation*}
and 
\begin{equation*}
     t_\tau - \max(t, \tau) \le \frac{2}{\beta}.
\end{equation*}
The derivative of the softmax w.r.t. to $t$ is given by the function 
$$
   \pi(t):= \frac{\exp(\beta t)}{\exp(\beta t)+\exp(\beta \tau)}~,
$$
which is obviously bounded by $1$. Thus, the softmax function is Lipschitz continuous with constant $1$. One consequence of this, that we will use repeatedly in the below proofs, is that the distance between a floored density and its floored KDE, is bounded by the distance of the density and its KDE, i.e.,
$$
|\hat q_\tau(t)-q_\tau(t)| \le |\hat q(t)-q(t)|.
$$
In particular, Lemmas \ref{Lem_density_discrete} and \ref{Lem_density_continuous} provide convergence rates as well as concentration results for the floored estimators.
\subsection{The Laplace density - an example of a weak Nikol'ski function}
\noindent In this section, we  prove that a Laplace density $p^{Lap} $ is an element of the weak Nikol'ski class $\overline{\mathcal{N}}(1.5,L)$. 
\begin{example}
Let $p^{Lap} $ denote a Laplace density 
\begin{equation*}
    p^{Lap} (t)= \frac{1}{2b}\exp(-\frac{|t-\mu|}{b})
\end{equation*}
with mean $\mu \in \R$ and variance parameter $b>0$. Then, for $L=L(b)$ sufficiently large, it follows that
\begin{equation*}
    p^{Lap} \in \overline{\mathcal{N}}(1.5,L).
\end{equation*}
\end{example}
\begin{proof}
Without loss of generality, we assume $\mu = 0$ (since the mean has no influence on the smoothness of $p^{Lap}$). Next, we notice that $p^{Lap}$ is Lipschitz continuous with constant $L = 1/(2b^2)$ and hence it is almost everywhere differentiable (Rademacher's theorem). This pointwise derivative $\partial^1 p^{Lap}$ is equal to its weak derivative and a simple calculation shows that
$$
\partial^1 p^{Lap}(u) = \frac{-sgn(u)}{2b^2}\exp\Big(-\frac{|u|}{b}\Big).
$$
We are left to prove the Nikol'ski condition \eqref{Eq_Nikolski_condition} for $s=1.5$. Due to symmetry, we can assume that $t\geq 0$. Moreover, we only consider $t \le 1$, as the case $t>1$ is much simpler. 
\begin{align*}
   &\frac{1}{2b^2}\Big[\int (sgn(u+t)\exp(-|u+t|/b)
   \\&\quad- sgn(u)\exp(-|u|/b))^2du\Big]^{1/2}
   \\&=\frac{1}{2b^2}\left[\int_{-\infty}^{-t} \left(\exp(u/b)(1-\exp(t/b))\right)^2du\right.
   \\&+\int_{-t}^{0} \left(\exp(-(u+t)/b)+\exp(u/b)\right)^2du
   \\&+\left.\int_{0}^{+\infty}\left(\exp(-(u+t)/b)-\exp(-u/b)\right)^2du\right]^{1/2}
   \\&=\frac{1}{2b^2}\left[-2b \exp(-t/b)+2t\exp(-t/b)+2b\right]^{1/2}
   \\&=\frac{\sqrt{2}}{2b^2}\exp(-t/(2b))\left[(t-b+\exp(t/b)b)\right]^{1/2}
\end{align*}
   Now, applying the mean value theorem on $g(t)=\exp(t/b)b$ for some $\xi\in (0,t)$, we have 
    \begin{align*}
        &\frac{\sqrt{2}}{2b^2}\exp(-t/(2b))\left[(t+\exp(t/b)b-\exp(0/b)b)\right]^{1/2}
        \\&=\frac{\sqrt{2}}{2b^2}\exp(-t/(2b))\left[t(1+ \exp(\xi/b))\right]^{1/2}
        \\&\leq \frac{\sqrt{2}}{2b^2}(\exp(-t/(2b))+1
        )|t|^{1/2}
        \\&\leq L|t|^{1/2}
    \end{align*}
    where (using that $t \in [0,1]$) we have defined
    \begin{equation} \label{Eq_def_Lb}
    L(b):=\frac{\sqrt{2}}{b^2}.
    \end{equation}
\end{proof}
\noindent In view of \eqref{Eq_def_Lb}, note that if $b$ is close to zero (small variance), the smoothness of $p^{Lap}$ decreases substantially.
\subsection{Proof of Theorem \ref{Thm_main}}\label{App_D}

\noindent The proof of this theorem consists of four steps: First, we show that for a derivation of an asymptotic lower bound for $D_\lambda( p , q)$, it suffices to give a lower bound for $D_\lambda( p , q_\tau)$. In the following steps, we consider the large sample behavior of the statistic $\sqrt{n}(D_\lambda(\hat p , \hat q_\tau)-D_\lambda( p , q_\tau))$. We demonstrate in the second step, that this object is asymptotically equal to a sum of independent random variables, that are shown to converge to a normal distribution in step three. Finally, in step four, we show that the variance estimator $\hat \sigma_n^2$ is asymptotically consistent in an appropriate sense. Convergence rates for technical remainders are gathered and proved in Appendix \ref{App_E}.\\
\textbf{Step 1:} By definition of the softmax function in \eqref{Eq_smooth_max}, it holds that $y_\tau \ge \max(y, \tau).$ In particular, we have $p(t)/q_\tau(t) \le p(t)/q(t)$ for all arguments $t$, which implies by Definition \ref{Def_Rényi_Divergence} of the Rényi divergence $D_\lambda( p , q) \ge D_\lambda( p , q_\tau)$. Hence, any lower bound for $D_\lambda( p , q_\tau)$ also lower bounds $D_\lambda( p , q)$.\\
\textbf{Step 2:} We first notice, that 
\begin{equation} \label{Eq_decomposition_hat_D}
\sqrt{n}(D_\lambda( \hat p , \hat q_\tau)-D_\lambda( p , q_\tau)) = \frac{\sqrt{n} \hat T}{(\lambda-1)\int \frac{p(t)^\lambda}{q_\tau(t)^{\lambda-1}}dt } +R_1.
\end{equation}
Here $R_1$ is a remainder term that is asymptotically negligible (see Appendix \ref{App_E}). Hence, to show weak convergence, we can focus on $\hat T$ in the following, which is defined as  
\begin{equation} \label{Eq_def_hat_T}
    \hat T :=  \int \frac{\hat p(t)^\lambda}{\hat q_\tau(t)^{\lambda-1}}dt-\int \frac{p(t)^\lambda}{q_\tau(t)^{\lambda-1}}dt~.
\end{equation}
By simple calculations, we can derive the decomposition $\hat T= \hat T_1+\hat T_2+\hat T_3 $, where
\begin{align*}
     \hat T_1 &= \int (\hat p(t)^\lambda-p(t)^\lambda) q_\tau(t)^{1-\lambda}dt\\
     \hat T_2 &=\int (\hat q_\tau(t)^{1-\lambda}- q_\tau(t)^{1-\lambda}) p(t)^\lambda dt\\
       \hat T_3 &=\int \left(\hat p(t)^\lambda-p(t)^\lambda\right)\left( q_\tau(t)^{1-\lambda}-\hat q_\tau(t)^{1-\lambda}\right) dt.
  \end{align*}
  In the following we show, that the terms $\hat T_1, \hat T_2$  are asymptotically normal, while $\hat T_3$ is negligible.\\
  For an appropriate value  $\chi_1$ between $\hat p(t)$ and $p(t)$ (using the mean value theorem),  we can rewrite
  \begin{align*}
      \hat T_1&= \int \lambda \chi_1^{\lambda-1} (\hat p(t)-p(t)) q_\tau(t)^{1-\lambda}dt = \breve T_1 + R_2, 
      \end{align*}
      where      
      \begin{align*}
      \breve T_1 := &\int \lambda p(t)^{\lambda-1} (\hat p(t)-p(t)) q_\tau(t)^{1-\lambda}dt,\\ 
      R_{2} := & \int \lambda (\chi_1^{\lambda-1}-p(t)^{\lambda-1}) (\hat p(t)-p(t))q_\tau(t)^{1-\lambda}dt.    
      \end{align*}
      We can show that $R_2=o_P(1/\sqrt{n})$, i.e., that it is asymptotically negligible (see App \ref{App_E}).
      Similarly, (again using the mean value theorem), we can deduce that $\hat T_2 = \breve T_2+ R_3+R_4$, where
    \begin{align*}
      \breve T_2:=&\Big(\int (1-\lambda)(\hat q(t)- q(t))\pi( q(t))\\&\times q_\tau(t)^{-\lambda}p(t)^\lambda dt\Big)\\
      R_{3}:=&\int (1-\lambda)(\chi_2^{-\lambda}-q_\tau(t)^{-\lambda})\\&\times(\hat q_\tau(t)- q_\tau(t))p(t)^\lambda dt
      \\
      R_4:=&\int (1-\lambda)(\hat q(t)- q(t))(\pi(\chi_3)-\pi( q(t)))\\&\times q_\tau(t)^{-\lambda}p(t)^\lambda dt.
    \end{align*}
    Above $\chi_2$ is a number between $\hat q_\tau(t)$ and $q_\tau(t)$ and $\chi_3$ between $\hat q(t)$ and $q(t)$. Notice that here we have employed differentiability of the softmax function (in the mean value theorem), which introduces the derivative of the softmax function $\pi$, (defined in (16)) into the formula. In Section \ref{App_E}, we establish $R_3, R_4 =o_P(1/\sqrt{n})$.\\
    Finally, we consider $\hat T_3$, which can be bounded by an application of  Cauchy-Schwarz by
    \begin{align} \label{Eq_bound_T3}
          \hat T_3 \le \norm{\hat p^\lambda -  p^\lambda}_2  \|\hat q_\tau^{1-\lambda} - q_\tau^{1-\lambda}\|_2. 
      \end{align}
      Recall that the map $t \mapsto t^\lambda$ is Lipschitz continuous on any compact interval $[0,r] \subset \R$, yielding $\norm{\hat p^\lambda -  p^\lambda}_2 \le C \norm{\hat p - \hat p}_2$ for some constant $C=C(L, s, \lambda)$ with probability converging to $1$. Here, we have used that any function $p$ in the smoothness class is uniformly bounded by some constant $C'(L,s)$ (which can be shown by basic calculations). Furthermore, since $\hat p$ is uniformly close to $p$ with probability going to $1$ (see Lemma \ref{Lem_density_continuous} part iii)), we have with probability converging to $1$ $0 \le \hat p(t), p(t) \le r$ for any $r>C'(L,s)$. We can use analogue arguments to bound  $\|\hat q_\tau^{1-\lambda} - q_\tau^{1-\lambda}\|_2$: The function $t \mapsto t^{1-\lambda}$ is Lipschitz continuous on the interval $[\tau, \infty)$ with Lipschitz constant $\tau^{-\lambda}$. Again, using boundedness of $q$ and uniform concentration of $\hat q$, implies with probability going to $1$ that $\|\hat q_\tau^{1-\lambda} - q_\tau^{1-\lambda}\|_2 \le \tau^{-\lambda}\|\hat q_\tau - q_\tau\|_2 \le \tau^{-\lambda}\|\hat q - q\|_2 $. Here we have used the softmax function is also Lipschitz with constant $1$. Together, with \eqref{Eq_bound_T3} these considerations imply (with probability going to $1)$
      $$
      \hat T_3 \le C \tau^{-\lambda}\|\hat g - g\|_2 \|\hat p - p\|_2. 
      $$
      The right side is now of order $\tau^{-\lambda}n^{-1}h^{d}=o_P(1/\sqrt{n})$, where we have used Lemma \ref{Lem_density_continuous}, together with the parameter choices from Assumption \ref{Ass_continuous}. \\
      Our derivations thus far imply that $\hat T = \breve T_1 + \breve T_2 +o_P(1/\sqrt{n})$, where $\breve T_1, \breve T_2$ are each sums of i.i.d. random variables (to see this, recall the definition of the KDE in Section \ref{Subsec_3_1}) and independent of each other. Notice, however, that $\breve T_i$ (for $i=1,2$) is not centered, as the KDE is not unbiased. Still, we can show that $\breve T_i = \tilde T_i + o(1/\sqrt{n})$ (proof in Section \ref{App_E}), where
      \begin{align*}
          \tilde T_1 := & \int  (\hat p(t)-\E \hat p(t))\lambda p(t)^{\lambda-1} q_\tau(t)^{1-\lambda}dt,\\
      \tilde T_2 := &  \int(\hat q(t)-\E \hat q(t))\pi( q(t))(1-\lambda)q_\tau(t)^{-\lambda}p(t)^\lambda dt.
      \end{align*}
are centered versions of $\breve T_1$ and $\breve T_2$ respectively. In the next step, we show asymptotic normality of $\tilde T_1, \tilde T_2$, by virtue of a Berry-Esseen argument.\\
\textbf{Step 3:} In order to apply the Theorem of Berry-Esseen to $\tilde T_i$, we have to calculate its (large sample) variance and bound its absolute third moment. Recall that by construction $\E \tilde T_i =0$ already holds.\\
First, notice that by definition of the KDE
$
    \tilde T_i = \frac{1}{n}\sum_{j=1}^n Z_{i,j},
$
where 
\begin{align*}
          Z_{i,1}:= &  \frac{\lambda}{h^d}\int \Big(K(\frac{t-X_i}{h})-\E K(\frac{t-X_i}{h})\Big)\\
          & \qquad\qquad p(t)^{\lambda-1}q_\tau(t)^{1-\lambda}dt\\[1ex]
          Z_{i,2}:= &  \frac{1-\lambda}{h^d}\int \Big(K(\frac{t-Y_i}{h})-\E K(\frac{t-Y_i}{h})\Big)\\[1ex]
          & \qquad \quad \pi( q(t))q_\tau(t)^{-\lambda}p(t)^\lambda dt          ~.
\end{align*}
Some tedious calculations (displayed in Section \ref{App_E}) now show that $\E Z_{i,j}^2 = (\sigma^{(i)})^2 + o(1)$, where the $o$-Term vanishes uniformly in $i,j$ and is only dependent on $s,L, \tau$. Here $( \sigma^{(i)})^2$ is defined as
\begin{align*}
(\sigma^{(1)})^2 :=&  \lambda^2\left(\int  p(z)^{2\lambda-1} q_\tau(z)^{2-2\lambda}dz\right.\\&-\left.\Big(\int  p(z)^{\lambda} q_\tau(z)^{1-\lambda} dz \Big)^2\right)\\
(\sigma^{(2)})^2 := & (1-\lambda)^2\left( \int \pi(q(z))^2 q_\tau(z)^{-2\lambda} q(z) p(z)^{2\lambda} dz \right. \\
      &\qquad \quad \left.- \Big( \int \pi(q(z)) q_\tau^{-\lambda}(z)q(z) p(z)^\lambda dz \Big)^2\right).
\end{align*}
Notice that $(\sigma^{(i)})^2$ is deterministic, but still depends on $n$ via $\tau$ and $\beta$ (inside the definition of $\pi$) and as $\tau \to 0$ the variance gets larger. Similar but simpler calculations than for the variance (displayed in Section \ref{App_E}) show that with some large enough constant $C=C(L, s) $
$$
\E |Z_{i,j}|^3 \le C \tau^{-3\lambda}
$$
for $i=1,2$. By the parameter choice in Assumption \ref{Ass_continuous}, we have $\tau^{-3 \lambda} =o(\sqrt{n})$ and as a consequence $\E |Z_{i,j}|^3/(\sqrt{n}(\sigma^{(i)})^{3/2})=o(1)$. Hence, the Berry-Esseen theorem yields that 
\begin{equation} \label{Eq_weak_conv_Rényi}
    \sup_t |\tilde F_i(t)- \Phi(t)|=o(1)
\end{equation}
where $\tilde F_i$ is the cumulative distribution function of  $\sqrt{n}\tilde T_i/(\sigma^{(i)})^2$  and $\Phi$ of the standard normal. Notice that this convergence holds (by our derivations) uniformly over all $p,q$ from the density class. The limiting distributions for $i=1,2$ are independent, as $\tilde T_1, \tilde T_2$ are independent. So, using  \eqref{Eq_decomposition_hat_D} and the fact that $\hat T = \tilde T_1 + \tilde T_2 +o_P(1/\sqrt{n})$, this implies 
\begin{equation} \label{Eq_convergence_Rényi_standardized}
 \sup_t|\tilde F(t)-\Phi(t)|=o(1)
\end{equation}
 where $\tilde F$ is the distribution function of $\sqrt{n/\sigma_n^2}(D_\lambda( \hat p , \hat q_\tau)-D_\lambda( p , q_\tau))$ and 
 \begin{align} \label{Eq_asymptotic_variance}
     \sigma^2_n :=\frac{ (\sigma^{(1)})^2+  (\sigma^{(2)})^2}{\Big( (\lambda-1)\int ( p(t))^\lambda ( q_\tau(t))^{1-\lambda}) dt\Big)^{2}}.
\end{align}
Notice that $\sigma_n^2$ is also deterministic, but still depends on $n$ through $\tau$ and $\beta$. According to Slutsky's Theorem, convergence in \eqref{Eq_convergence_Rényi_standardized} still holds, if we replace $\sigma_n^2$ by an estimator $\hat \sigma_n^2$, which satisfies $|\sigma_n^2-\hat \sigma_n^2|=o_P(1)$. To show this result is the objective of our last step.\\
\textbf{Step 4:} Recall the definition of the variance estimator $\hat \sigma_n^2$ from \eqref{Eq_variance_estiamtor}. In order to establish $|\sigma_n^2-\hat \sigma_n^2|=o_P(1)$, it suffices to show that $|(\sigma^{(i)})^2- (\hat\sigma^{(i)})^2|=o_P(1)$ (numerator of the variances), as well as $|D_\lambda( p , q_\tau)) -D_\lambda( \hat p , \hat q_\tau)) |=o_P(1)$ (denominator of the variances). For parsimony of presentation, we restrict ourselves to proving  $|(\sigma^{(1)})^2-(\hat\sigma^{(1)})^2|=o_P(1)$. For simplicity of notation, we now define the function $L(v,w) := \pi(w)^2 w_\tau^{-2\lambda} wv$ on all $x,y \ge 0$.
\begin{align*}
     \Big|\int& L(p(t),q(t)) p(t)^{2\lambda-1} dt  - \int L(\hat p(t),\hat q(t)) \hat p(t)^{2\lambda-1} dt\Big|.
\end{align*}
In a first step, we notice that 
\begin{align*}
    \Big| &\int L(p(t),q(t)) (p(t)^{2\lambda-1}-\hat p(t)^{2\lambda-1}) dt \Big| \\&\leq C \|p- \hat p\|_2 \tau^{-2 \lambda}.
\end{align*}
The right side is $o_P(1)$ using Lemma \ref{Lem_density_continuous} (parts i) and ii)) together with our parameter choices from Assumption \ref{Ass_continuous}. Hence, it suffices to establish that 
\begin{align}\label{eq_lip}
     \int &\big|L(p(t),q(t))  -L(\hat p(t),\hat q(t))\big| \hat p(t)^{2\lambda-1} dt =o_P(1).
\end{align}
It is not hard to show that $L$ is Lipschitz-continuous on any bounded set $M\subset \R_{\geq 0}$ with 
\begin{equation*}
    |L(v,w)-L(\tilde v, \tilde w)| \le C \tau^{-2\lambda-1} (|v-\tilde v|+|w- \tilde w|),
\end{equation*}
where $C=C(M,\lambda).$ Here we have used that the product of Lipschitz continuous, bounded functions $f_1,...,f_k$ with Lipschitz constants $l_1,...,l_k$ is again Lipschitz, with constant  $\sum_{j=1}^k l_j \prod_{i \neq j} \sup_t |f_i(t)|$ (this follows by a simple induction).\\
To apply Lipschitz continuity of $L$ to \eqref{eq_lip}, we  notice that  $p(t),q(t)$ are bounded by some universal constant $C'=C'(L,s)$ and hence   $\hat p(t),\hat q(t)$ are bounded by $C'+1$ with probability converging to $1$ according to Lemma \ref{Lem_density_continuous} part iii) (uniform approximation of $p(t),q(t)$ by $\hat p(t),\hat q(t)$).
Consequently, we have for \eqref{eq_lip} that (with probability converging to $1$)
\begin{align} \label{Eq_Jensen}
    \int& \big| L(p(z), q(z)) - L(\hat p(z), \hat q(z)) \big| \hat p(z)^{2\lambda -1}dz \\&
    \le  C \tau^{-2\lambda-1} \nonumber  \left(\int |\hat p(z)-p(z)|\hat p(z) dz \right.\\&+\left.  \int |\hat q(z)-q(z)| \hat p(z) dz .\right) \nonumber
\end{align}
 Now, using Jensen's inequality we can upper bound each of the integrals on the right side of \eqref{Eq_Jensen}. For instance, focusing on the first one, we get with probability converging to $1$
\begin{align*}
    \int |\hat p(z)-p(z)|\hat p(z) dz &\le \Big\{\int |\hat p(z)-p(z)|^2 \hat p(z) dz  \Big\}^{1/2} \\&\le (C'+1) \|\hat p-p\|_2.
\end{align*}
Here we have again used boundedness of $\hat p(t)$. Similarly, we get $  \int |\hat q(z)-q(z)| \hat p(z) dz \le (C'+1) \|\hat q-q\|_2$ (with probability converging to $1$) and hence for the difference in \eqref{Eq_Jensen} the rate $\mathcal{O}_P((\tau^{-2\lambda-1}) [\|\hat q-q\|_2+\|\hat p-p\|_2])$. This product is of order $o_P(1)$, using the convergence rates of the KDE in Lemma \ref{Lem_density_continuous} parts i) and ii), together with the parameter choices in Assumption \ref{Ass_continuous}. This concludes the proof for $|\sigma_n^2-\hat \sigma_n^2|=o_P(1)$, which implies by Slutsky's theorem and \eqref{Eq_convergence_Rényi_standardized}, that
\begin{equation} \label{Eq_convergence_Rényi_standardized}
 \sup_t| F_{\hat \sigma_n^2}(t)-\Phi(t)|=o(1)
\end{equation}
where $F_{\hat \sigma_n^2}$ is the distribution function of $\sqrt{n/\hat \sigma_n^2}(D_\lambda( \hat p , \hat q_\tau)-D_\lambda( p , q_\tau))$. Accordingly, if we denote by $q_{1-\alpha}$ the upper $\alpha$-quantile of the standard normal distribution, we get
$$
\mathbb{P}(\sqrt{n/\hat \sigma_n^2}(D_\lambda( \hat p , \hat q_\tau)-D_\lambda( p , q_\tau)) \le q_{1-\alpha}) = 1-\alpha +o(1),
$$
  for any $\alpha \in (0,1)$, with the vanishing term on the right independent of $p$ and $q$. Rewriting this yields
  $$
\mathbb{P}(D_\lambda( p , q_\tau) >q_\alpha\sqrt{\hat \sigma_n^2/n} +D_\lambda( \hat p , \hat q_\tau)) = 1-\alpha +o(1),
$$
  where  $q_\alpha = -q_{1-\alpha}$ is the lower $\alpha$-quantile of the standard normal. Since we have deterministically $D_\lambda( p , q) \ge D_\lambda( p , q_\tau)$ (see step 1) and our remainder $o(1)$ vanishes uniformly over the entire function class, this implies 
  $$
  \liminf_{n \to \infty} \inf_{p,q}\mathbb{P}(D_\lambda( p , q) >q_\alpha\sqrt{\hat \sigma_n^2/n} +D_\lambda( \hat p , \hat q_\tau)) \ge 1-\alpha,
  $$
  which is \eqref{Eq_Thm_main}. This concludes the proof of Theorem \ref{Thm_main}.

  \subsection{Convergence rates for remainders} \label{App_E}
  
\noindent In the following, we prove bounds for all remainder terms, which occurred in the course of Section \ref{App_D}.\\
  \textbf{$\mathbf{R_1}$:} Recall Definition \ref{Def_Rényi_Divergence} of the Rényi divergence.  To derive \eqref{Eq_decomposition_hat_D}, we apply a mean value theorem to the logarithm (in the Rényi divergence), which yields
  $$
  \sqrt{n}(D_\lambda( \hat p , \hat q_\tau)-D_\lambda( p , q_\tau)) = \frac{\sqrt{n} \hat T}{(\lambda-1)\chi_4 } 
  $$
  for some value $\chi_4$ between   $\int \frac{\hat p(t)^\lambda}{\hat q_\tau(t)^{\lambda-1}}dt$ and $\int \frac{p(t)^\lambda}{q_\tau(t)^{\lambda-1}}dt$. Recall that $\hat T$ is defined in \eqref{Eq_def_hat_T}. Now, defining 
  $$
  R_1 := \frac{\sqrt{n} \hat T}{(\lambda-1)\chi_4 } -\frac{\sqrt{n} \hat T}{(\lambda-1)\int \frac{p(t)^\lambda}{q_\tau(t)^{\lambda-1}}dt }
  $$
  we show that $R_1$ is asymptotically negligible, i.e., 
  $$
  R_1 = o_P\bigg( \frac{\sqrt{n} \hat T}{\int \frac{p(t)^\lambda}{q_\tau(t)^{\lambda-1}}dt } \bigg).
  $$
  This holds, if $|\chi_4- \int \frac{p(t)^\lambda}{q_\tau(t)^{\lambda-1}}dt|=o_P(1)$, or equivalently, if 
  $$
  \Big|  \int \frac{\hat p(t)^\lambda}{\hat q_\tau(t)^{\lambda-1}}-  \frac{p(t)^\lambda}{q_\tau(t)^{\lambda-1}}dt \Big|=o_P(1).
  $$
  We can upper bound the left side by the sum $S_1+S_2$, where
  \begin{align*}
      S_1 := & \Big|\int  (\hat p(t)^\lambda- p(t)^\lambda)\hat q_\tau(t)^{1-\lambda}dt\Big| \\
      S_2 :=& \Big|\int p(t)^\lambda (\hat q_\tau(t)^{1-\lambda} -q_\tau(t)^{1-\lambda}) dt \Big|.
  \end{align*}
  We now demonstrate that $S_1 =o_P(1)$ (the proof for $S_2$ works by similar strategies). Using the mean value theorem for $t \mapsto t^\lambda$, we have
  \begin{align*}
      S_1&=\Big|\int  \lambda (\hat p(t)- p(t)) \chi_5^{\lambda-1} \hat q_\tau(t)^{1-\lambda}dt\Big|
      \\&\leq  \Big|\int  \lambda (\hat p(t)- \E \hat p(t)) \chi_5^{\lambda-1} \hat q_\tau(t)^{1-\lambda}dt\Big|
      \\&\quad +\Big|\int  \lambda (\E \hat p(t)- p(t)) \chi_5^{\lambda-1} \hat q_\tau(t)^{1-\lambda}dt\Big|
      \\&=:A+B~,
  \end{align*}
  where $\chi_5$ is between $\hat p(t)$ and $p(t)$. We start by upper bounding $A$. First, notice that we can decompose $A$ into two parts given by
  \begin{align*}
      &\Big|\int  \lambda (\hat p(t)- \E \hat p(t)) \chi_5^{\lambda-1} \hat q_\tau(t)^{1-\lambda}dt\Big|
      \\&\leq \lambda \Big| \sum_{i=1}^n\frac{1}{nh^d} \int \Big(K(\frac{t-X_i}{h})-\E K(\frac{t-X_i}{h})\Big)\\&\quad\times (\chi_5^{\lambda-1}-p(t)^{\lambda-1}) \hat q_\tau(t)^{1-\lambda}dt\Big|\\&+\lambda\Big| \sum_{i=1}^n\frac{1}{nh^d} \int \Big(K(\frac{t-X_i}{h})-\E K(\frac{t-X_i}{h})\Big)\\&\quad \times p(t)^{\lambda-1} \hat q_\tau(t)^{1-\lambda}dt\Big|=:A_1+A_2
  \end{align*}
Let $\delta>0$ be sufficiently small (it is specified later). Then, we obtain with Chebyshev's inequality and the independece of $X_i$ ($i=1,\hdots, n$) that
  \begin{align*}
      & \quad \,\, \PR(A_2>n^{-\delta})\\&\leq \frac{C \E \Big( \int\frac{1}{h^d} K(\frac{t-X_i}{h}) p(t)^{\lambda-1} \hat q_\tau(t)^{1-\lambda}dt\Big)^2}{n n^{-2\delta}}
      \\&= \frac{C\tau^{2-2\lambda}}{n^{1-2\delta}}  \E \int \int\frac{1}{h^{2d}} K(\frac{t-X_i}{h})K(\frac{y-X_i}{h})\\&\quad \times p(t)^{\lambda-1} p(y)^{\lambda-1}dtdy
      \\&= \frac{C\tau^{2-2\lambda}}{n^{1-2\delta}}  \int \int \int\frac{1}{h^{2d}} K(\frac{t-z}{h})K(\frac{y-z}{h})\\&\quad \times p(t)^{\lambda-1} p(y)^{\lambda-1}p(z)dtdydz
      \\&= \frac{C\tau^{2-2\lambda}}{n^{1-2\delta}}  \int \int \int\frac{1}{h^{2d}} K(t)K(y)p(th+z)^{\lambda-1} \\&\quad\times p(yh+z)^{\lambda-1}   p(z)dtdydz
      \\&\leq \frac{C'\tau^{2-2\lambda}}{n^{1-2\delta}}  \int\int K(t)K(y)dtdy=o(1)~.
  \end{align*}
  In the above calculations, we have exploited Assumption \ref{Ass_continuous} (e.g., using boundedness of the densities $p,q$ or the fact that $K$ is a kernel). The final equality holds by Assumption \ref{Ass_continuous}, part (2) for a sufficiently small choice of $\delta$, which then implies $A=o_P(n^{-\delta})$.\\ Similarly, using the Cauchy-Schwarz inequality, we have  \begin{align*}
      \E A_1 &\leq C \sum_{i=1}^n \E \frac{1}{nh^d} \int \Big|K(\frac{t-X_i}{h})-\E K( \frac{t-X_i}{h})\Big|\\&\quad\times |\hat p(t)-p(t)|q_\tau^{1-\lambda}(t)dt
      \\&\leq  \frac{C}{h^d} \tau^{1-\lambda} \E \Big[\left( \int |K(\frac{t-X_i}{h})-\E K( \frac{t-X_i}{h})|^2dt\right)^{1/2}
      \\&\quad\times\norm{\hat{p}-p}_2\Big]
      \\&\leq \frac{C}{h^d} \tau^{1-\lambda} \E \left[  \int |K(\frac{t-X_i}{h})-\E K( \frac{t-X_i}{h})|^2dt\right]^{1/2} \\&\quad\times\left(\E \norm{\hat{p}-p}_2^2\right)^{1/2}=o(1)
  \end{align*}
 In the last step, we have used Lemma \ref{Lem_density_continuous} parts i) and ii), together with the convergence rate of $\tau$, given in Assumption \ref{Ass_continuous} part (2).  Markov's inequality implies that $A_2=o_P(1)$ if $\E |A_2|=\E A_2=o(1)$. This concludes the proof that $A=o_P(1)$. We are left to show that $B=o_P(1)$. In analogy to $A$, we can decompose $B$ into two parts
  \begin{align*}
      B&=\Big|\int  \lambda (\E \hat p(t)- p(t)) \chi_5^{\lambda-1} \hat q_\tau(t)^{1-\lambda}dt\Big|
      \\&\leq \Big|\int  \lambda (\E \hat p(t)- p(t)) (p(t)^{\lambda-1}) \hat q_\tau(t)^{1-\lambda}dt\Big|\\&\quad+\Big|\int  \lambda (\E \hat p(t)- p(t)) (\chi_5^{\lambda-1}-p(t)^{\lambda-1}) \hat q_\tau(t)^{1-\lambda}dt\Big|
    \\&=:B_1+B_2~.
  \end{align*}
 With similar arguments as in the proof of $A_2=o_P(1)$, we can demonstrate that $B_2=o_P(1)$ (this proof is omitted for sake of brevity). In contrast, for $B_1$ we need a different strategy. This is due to the fact that $p(t)^{\lambda-1}$ is not necessarily integrable (for $\lambda <2$). Hence, we will rewrite $B_1$ appropriately to obtain that $B_1=o_P(1)$~. First we consider the case, where $\lambda\in (1,2)$. Pulling the absolute value into the integral yields
 \begin{equation} \label{Eq_boundB_1}
     B_1\leq \lambda \int |\E \hat p(t)- p(t)|p(t)^{\lambda-1}\hat q_\tau(t)^{1-\lambda}dt~.
 \end{equation}
Next we decompose the integrand as follows:
\begin{align*}
    |\E \hat p(t)- p(t)|&=|\E \hat p(t)- p(t)|^{\lambda-1}|\E \hat p(t)- p(t)|^{2-\lambda}
    \\&\leq|\E \hat p(t)- p(t)|^{\lambda-1}(|\E \hat p(t)|+ p(t))^{2-\lambda}
    \\&\leq |\E \hat p(t)- p(t)|^{\lambda-1}\\&\quad\times((|\E \hat p(t)|)^{2-\lambda} + p(t)^{2-\lambda})~.
\end{align*}
In the last step we have used that $0<2-\lambda<1$, which implies subadditivity. 
Plugging this into the right side of \eqref{Eq_boundB_1} and using the triangle inequality yields the following bound for $B_1$:
\begin{align}\label{Eq_B1}
     B_1\leq & C \int  |\E \hat p(t)- p(t)|^{\lambda-1} p(t)^{\lambda-1} (\E |\hat p|)^{2-\lambda}\hat q_\tau(t)^{1-\lambda}
     \\&\quad + C \int |\E \hat p(t)- p(t))^{\lambda-1}| p(t) \hat q_\tau(t)^{1-\lambda}dt\Big|\nonumber~.
 \end{align}
  With that in hand, we can apply  Hölder's inequality (note $(\lambda-1)+(2-\lambda)=1$) on each part and obtain for the first one
  \begin{align*}
      &C\int  |\E \hat p(t)- p(t)|^{\lambda-1} p(t)^{\lambda-1} (\E |\hat p|)^{2-\lambda}\hat q_\tau(t)^{1-\lambda}dt
     \\&\leq \tau^{1-\lambda} C\int \Big| \E \hat p(t)- p(t)\Big|^{\lambda-1} p(t)^{\lambda-1} (\E |\hat p|)^{2-\lambda}dt 
    \\&\leq  \tau^{1-\lambda}C\norm{\E \hat p- p}_\infty^{\lambda-1} \norm{p(t)^{\lambda-1} (\E |\hat p|)^{2-\lambda}}_1
    \\&\leq  \tau^{1-\lambda}C\norm{\E \hat p- p}_\infty^{\lambda-1} \norm{p^{\lambda-1} (\E |\hat p|)^{2-\lambda}}_1
  \end{align*}
  The last factor on the right can bounded (according to Hölder's inequality) by
  \begin{align*}
   \norm{p^{\lambda-1} (\E |\hat p|)^{2-\lambda}}_1 &\leq \norm{p^{\lambda-1}}_{\lambda-1}\norm{(\E |\hat p|)^{2-\lambda}}_{2-\lambda}.
  \end{align*}
  On the right side $\norm{p^{\lambda-1}}_{\lambda-1}=1$ ($p$ is a density) and  $\norm{(\E |\hat p|)^{2-\lambda}}_{2-\lambda} = \norm{|K|^{2-\lambda}}_{2-\lambda}$ (it follows by assumption (K) and some easy calculations, that this norm is bounded by some constant $C$ only depending on the choice of $K$).
  For the second term in \eqref{Eq_B1}, we have with Hölder's inequality 
  \begin{align*}
      &\Big|\int (\E \hat p(t)- p(t))^{\lambda-1} p(t) \hat q_\tau(t)^{1-\lambda}dt\Big|
      \\&\quad\leq \tau^{1-\lambda}\norm{\E \hat p- p}_\infty^{\lambda-1}
  \end{align*}
  Thus, combining our above considerations, we see that
  \begin{equation*}
      B_1\leq C \norm{\E \hat p- p}_\infty^{\lambda-1}  \tau^{1-\lambda}~.
  \end{equation*}
  We are left to consider $\norm{\E\hat p -p}_\infty$.  Applying the definition of the estimator, we obtain
  \begin{align*}
      \E\hat p(t) -p(t)&=\int \frac{1}{h^d}K(\frac{t-u}{h}) p(u) du -p(t)
      \\&=\int (p(uh+t)-p(t)) K(u)du
      \\&\leq L h\int |u|K(u)du\leq C h~.
  \end{align*}
  Notice that in the final inequality we have used Assumption (K), which implies that the integral on the left is finite. Hence, we obtain  $B_1 \le C'(h\tau^{-1})^{1-\lambda} =o(1)$ as $h \tau^{-1}=o(1)$ by Assumption \ref{Ass_continuous}, part (2). For $\lambda\geq 2$, we can see that
  \begin{align*}
  B_1&\leq  \tau^{1-\lambda}C\int |\E \hat p(t)- p(t)|p(t)^{\lambda-1}
 dt
 \\&\leq C
 \tau^{1-\lambda}\norm{p(t)^{\lambda-2}}_\infty \int |\E \hat p(t)- p(t)|p(t)dt 
 \\&\leq C
 \tau^{1-\lambda}\int |\E \hat p(t)- p(t)|p(t)dt 
 \\&\leq C
 \tau^{1-\lambda}  \Big\{\int |\E \hat p(t)- p(t)|^2  p(z) dz  \Big\}^{1/2}
 \\&\leq C
 \tau^{1-\lambda} \norm{\E \hat p(t)- p(t)}_2=o(1)~.
  \end{align*}
  Taken together, all of the above considerations combined yield
   $$
  \Big|  \int \frac{\hat p(t)^\lambda}{\hat q_\tau(t)^{\lambda-1}}dt-  \frac{p(t)^\lambda}{q_\tau(t)^{\lambda-1}}dt \Big|=o_P(1).
  $$
  or in other words for $R_1$, that we have
  \begin{equation*}
      R_1=o_P\bigg( \frac{\sqrt{n} \hat T}{\int \frac{p(t)^\lambda}{q_\tau(t)^{\lambda-1}}dt } \bigg)~.
  \end{equation*}
\textbf{$\mathbf{R_2}$}: Recall that 
$$R_{2} :=  \int \lambda (\chi_1^{\lambda-1}-p(t)^{\lambda-1}) (\hat p(t)-p(t))q_\tau(t)^{1-\lambda}dt $$
  and note that $t \mapsto t^\lambda$ is Lipschitz-continuous for any $\lambda>1$ on any compact subinterval of $\R_{\ge 0}$. Furthermore, observe that all Lipschitz continuous densities (with constant $L$) are uniformly bounded by some constant $C'=C'(L)$ (this observation follows by a simple calculation, using that $\int p(t) dt =1$ for any density). Moreover, due to uniform convergence of the kernel density estimator (see Lemma \ref{Lem_density_continuous}, part iii)) it holds with probability going to $1$ that $\|\hat p \|_\infty \le 2C'$. Now, due to the Hölder's inequality, we have
        \begin{align*}
                R_{2}&\leq \lambda\norm{(\chi_1^{\lambda-1}-p^{\lambda-1})(\hat p-p)}_1 \norm{q_\tau^{1-\lambda}}_\infty
                \\&=\lambda\norm{\chi_1^{\lambda-1}-p^{\lambda-1}}_2 \norm{\hat p-p}_2 \tau^{1-\lambda}
                \\&\leq \lambda C\norm{\hat p-p}_2^2 \tau^{1-\lambda}
                \leq  \frac{\lambda C\tau^{1-\lambda}}{nh}=o_P(1/\sqrt{n})
        \end{align*}
    where the second inequality holds with probability going to $1$ (uniformly over the function class). The constant $C$ in the third line depends on $L$ (via $C'=C'(L)$) as well as $\lambda$ (via Lipschitz continuity of $t \mapsto t^\lambda$ on $[0, 2C']$).  In the final step we have used the convergence rates for the density estimator from Lemma \ref{Lem_density_continuous} (parts i) and ii)), together with the rates specified for $\tau$ and $h$ in Assumption \ref{Ass_continuous} part ii).\\
  \textbf{$\mathbf{R_3}$}:
  Recall that
  $$
   R_{3}:=\int (1-\lambda)(\chi_2^{-\lambda}-q_\tau(t)^{-\lambda})(\hat q_\tau(t)- q_\tau(t))p(t)^\lambda dt,
    $$
    where $\chi_2$ is a number between $\hat q_\tau(t)$ and $q_\tau(t)$.
    Due to the Lipschitz property of the softmax function, we have $|q_\tau(t)-\hat q_\tau(t)|\le |q(t)-\hat q(t)|$ for any $t$. Moreover, recall the Lipschitz-continuity (for $\lambda>1$) of $t \mapsto t^{-\lambda}$ with constant $\tau^{-\lambda-1}$ on the interval $[\tau, \infty)$. Now applying these results, together with Cauchy-Schwarz, we have
    \begin{align*}
           R_{3}&\leq |1-\lambda|\norm{q_\tau-\hat q_\tau}_2\norm{\hat q_{\tau}^{-\lambda}-q_{\tau}^{-\lambda}}_2\norm{p^\lambda}_\infty\\&\leq C  \tau^{-\lambda-1}  \norm{\hat q-q}^2_2 
           \\&\leq C \tau^{-\lambda-1}\frac{1}{nh^d}=o_P(1/\sqrt{n})~.
    \end{align*}
    To get the final rate we have used Assumption \ref{Ass_continuous} part ii).\\
    \textbf{$\mathbf{R_4}$}: Recall that
    \begin{align*}
        R_4:=&\int (1-\lambda)(\hat q(t)- q(t))(\pi(\chi_3)-\pi( q(t)))\\&\times q_\tau(t)^{-\lambda}p(t)^\lambda dt,
    \end{align*}
    where $\chi_3$ is a number between $\hat q(t)$ and $q(t)$.
    Similarly as before,  we have by two applications of Hölder's inequality
      \begin{align*}
          R_{4}&\leq C \tau^{-\lambda}\norm{\hat q- q}_2\norm{\pi(\chi_3)-\pi( q)}_2(\lambda-1)\norm{p^\lambda}_\infty~.
      \end{align*}
      By a simple calculation and application of the mean value theorem, it follows that
      \begin{align*}
      &|\pi(\chi_3)-\pi(q(t))|\leq |\hat p(t)-p(t)| \beta.
     \end{align*}
      Consequently, we have with probability going to $1$ (uniformly over the function class)
      \begin{align*}
      R_{4} \leq C \norm{\hat q - q }_2^2 \beta \tau^{-\lambda}\leq C \beta\frac{1}{nh^d}  \tau^{-\lambda}=o_P(1/\sqrt{n}).
      \end{align*}
      With that in hand, we can consider 
     \\\textbf{Second moments}:
    Recall the definition of $Z_{i,1}$ and note that the variance can be decomposed in
      \begin{align}\label{Eq_2Moment_Y}
      \E Z_{i,1}^2=&\E  \frac{\lambda^2}{h^{2d}}\int \int K(\frac{t-X_i}{h})K(\frac{y-X_i}{h}) \\&\times p(t)^{\lambda-1} p(y)^{\lambda-1}q_\tau(t)^{1-\lambda}q_\tau(y)^{1-\lambda}dtdy \nonumber
    \\&- \frac{\lambda^2}{h^{2d}}\int\int \E K(\frac{t-X_i}{h}) \E K(\frac{y-X_i}{h})  \nonumber \\&\times p(t)^{\lambda-1} p(y)^{\lambda-1}q_\tau(t)^{1-\lambda} q_\tau(y)^{1-\lambda} dtdy  \nonumber
  \end{align}
  For the first term, we have for the expectation that
  \begin{align*}
      & \E\frac{\lambda^2}{h^{2d}}\int \int K(\frac{t-X_i}{h})K(\frac{y-X_i}{h}) \\&\quad\times p(t)^{\lambda-1} p(y)^{\lambda-1}q_\tau(t)^{1-\lambda}q_\tau(y)^{1-\lambda}dtdy
       \\&= \frac{\lambda^2}{h^{2d}}\int \int \int K(\frac{t-z}{h})K(\frac{y-z}{h}) \\&\quad \times p(t)^{\lambda-1} p(y)^{\lambda-1}q_\tau(t)^{1-\lambda}q_\tau(y)^{1-\lambda}p(z)dtdydz
       \\&=\lambda^2\int \int \int K(t)K(y) p(th+z)^{\lambda-1} p(yh+z)^{\lambda-1}\\&\quad\times q_\tau(th+z)^{1-\lambda}q_\tau(yh+z)^{1-\lambda}p(z)dtdydz
       \\&=\lambda^2\int \int \int K(t)K(y) p(z)^{\lambda-1} p(z)^{\lambda-1}\\&\quad\times q_\tau(z)^{1-\lambda}q_\tau(z)^{1-\lambda}p(z)dtdydz+R_5
       \\&=\lambda^2\int p(z)^{2\lambda-2} q_\tau(z)^{2-2\lambda}p(z)dz+R_5
  \end{align*}
   Here the remainder $R_5$ is simply the difference of the integral in the second and third equation.
   Proving that $R_5$ can be split up into four separate parts (we replace $q(th+z)^{\lambda-1}$ by $q(z)^{\lambda-1}$, $q(yh+z)^{\lambda-1}$ by $q(z)^{\lambda-1}$ etc.). For the purpose of illustration, we confine ourselves to the first replacement. In the following we use the boundedness of all densities involved together with their Lipschitz continuity to see that
    \begin{align*}
     & \int \int \int K(t)K(y) (p(th+z)^{\lambda-1}-p(z)^{\lambda-1})p(yh+z)^{\lambda-1}\\&\quad \times \lambda^2p(th+z)^{1-\lambda} q_\tau(yh+z)^{1-\lambda}p(z)dtdydz
      \\&\leq C \tau^{2-2\lambda} \int \int  p(z) |p(th+z)^{\lambda-1}-p(z)^{\lambda-1}|  K(t) dt dz \\
      &\le   C \tau^{2-2\lambda} \sup_{t \in supp(K)} \int  p(z) |p(th+z)^{\lambda-1}-p(z)^{\lambda-1}|  dz  \\&\le C h \tau^{2-2\lambda} ~.
  \end{align*}
   By further, analogous calculations we can show that $|R_5| \le  C \tau^{1-2\lambda} h =o(1)$. 
     For the second term on the left of \eqref{Eq_2Moment_Y}, we can derive a similar expression: 
  \begin{align*}
      & \Big(\int \int \frac{\lambda}{h^d} K(\frac{t-z}{h}) p(t)^{\lambda-1} q_\tau(t)^{1-\lambda}p(z) dt dz \Big)^2 \\
      = & \Big(\int \int \lambda K(t) p(th+z)^{\lambda-1} q_\tau(th+z)^{1-\lambda}p(z) dt dz \Big)^2 \\
      = & \Big(\int \lambda  p(z)^{\lambda} q_\tau(z)^{1-\lambda} dz \Big)^2 + R_6,
  \end{align*}
   where $R_6 \le C h^2 \tau^{-2\lambda} =o(1)$ (here we have used again the Lipschitz continuity of the densities).\\
   In analogy to $Z_{i,1}$, one can decompose $Z_{i,2}$ such that
    \begin{align}\label{Eq_2moment_two}
        \E Z_{i,2}^2=&\E  \frac{(1-\lambda)^2}{h^{2d}}\int \int K(\frac{t-Y_i}{h})K(\frac{y-Y_i}{h}) \\&\times \pi( q(t))q_\tau(t)^{-\lambda}p(t)^\lambda\pi( q(y))q_\tau(y)^{-\lambda}p(y)^\lambda dtdy \nonumber
    \\&- \frac{(1-\lambda)^2}{h^{2d}}\Big(\int\int \E K(\frac{t-Y_i}{h})  \nonumber 
    \\&\qquad\times \pi( q(t))q_\tau(t)^{-\lambda}p(t)^\lambda dt \Big) ^2 \nonumber
    \end{align}
    Likewise, we can obtain by simple computations that the first term \eqref{Eq_2moment_two} can be decomposed in
    \begin{align*}
      &\E  \frac{(1-\lambda)^2}{h^{2d}}\int \int K(\frac{t-Y_i}{h})K(\frac{y-Y_i}{h}) \\&\times \pi( q(t))q_\tau(t)^{-\lambda}p(t)^\lambda\pi( q(y))q_\tau(y)^{-\lambda}p(y)^\lambda dtdy  
      \\&=  (1-\lambda)^2 \int\pi(q(z))^2q_\tau(z)^{-2\lambda}p(t)^{2\lambda}q(z)dz+R_7
    \end{align*}
    where again $R_7$ only consists of analogues replacements as for $Z_{i,1}$. Once more, one can show that the remainder can be split up in six parts and is bounded by $|R_7|\leq C\tau^{-2\lambda}h+C'\beta\tau^{-2\lambda}h=o(1)$. Similarly, one can derive for the second term in \eqref{Eq_2moment_two} that
    \begin{align*}
       & \frac{(1-\lambda)^2}{h^{2d}}\Big(\int\int \E K(\frac{t-Y_i}{h}) \pi( q(t))q_\tau(t)^{-\lambda}p(t)^\lambda dt \Big) ^2\\&=(1-\lambda)^2\Big(\int\pi( q(z))q_\tau(z)^{-\lambda}p(z)^\lambda q(z) dz \Big) ^2+R_8~,
    \end{align*}
    where $|R_8|\leq C \beta^2\tau^{-2\lambda}h^2 + C'\tau^{-2\lambda-2}h^2=o(1)$.
     \\\textbf{Third moments}:
     Here we will upper bound the third moment. Given an upper bound that is $o(\sqrt{n})$, the Berry-Esseen bound will imply the desired result Theorem \ref{Thm_main}. First, note that
\begin{align*}
         \E Z_{i,1}^3=&\frac{\lambda^3}{h^{3d}}\E\Big(\int (K(\frac{t-X_i}{h})-\E K(\frac{t-X_i}{h})) \\&\times p(t)^{\lambda-1}q_\tau(t)^{1-\lambda}dt\Big)^3
         \\\leq& \frac{\lambda^3}{h^{3d}}\int\left( \int K(\frac{t-y}{h}) p(t)^{\lambda-1}q_\tau(t)^{1-\lambda}dt\right)^3q(y)dy
        \\=& \int\left( \int K(t)\lambda p(th+y)^{\lambda-1}q_\tau(th+y)^{1-\lambda}dt\right)^3\\&\times q(y)dy
\end{align*}
Recall that $p(t)$ is bounded and $q_\tau(t)^{1-\lambda}\leq \tau^{1-\lambda}$. Therefore, we have
\begin{align*}
    \E Z_{i,1}^3\leq C\tau^{3-3\lambda} \Big(\int K(t)dt\Big)^3=C \tau^{3-3\lambda}.
\end{align*}

 \begin{figure*}[h]
\centering
\begin{subfigure}{0.3\textwidth}
  \includegraphics[width=\linewidth]{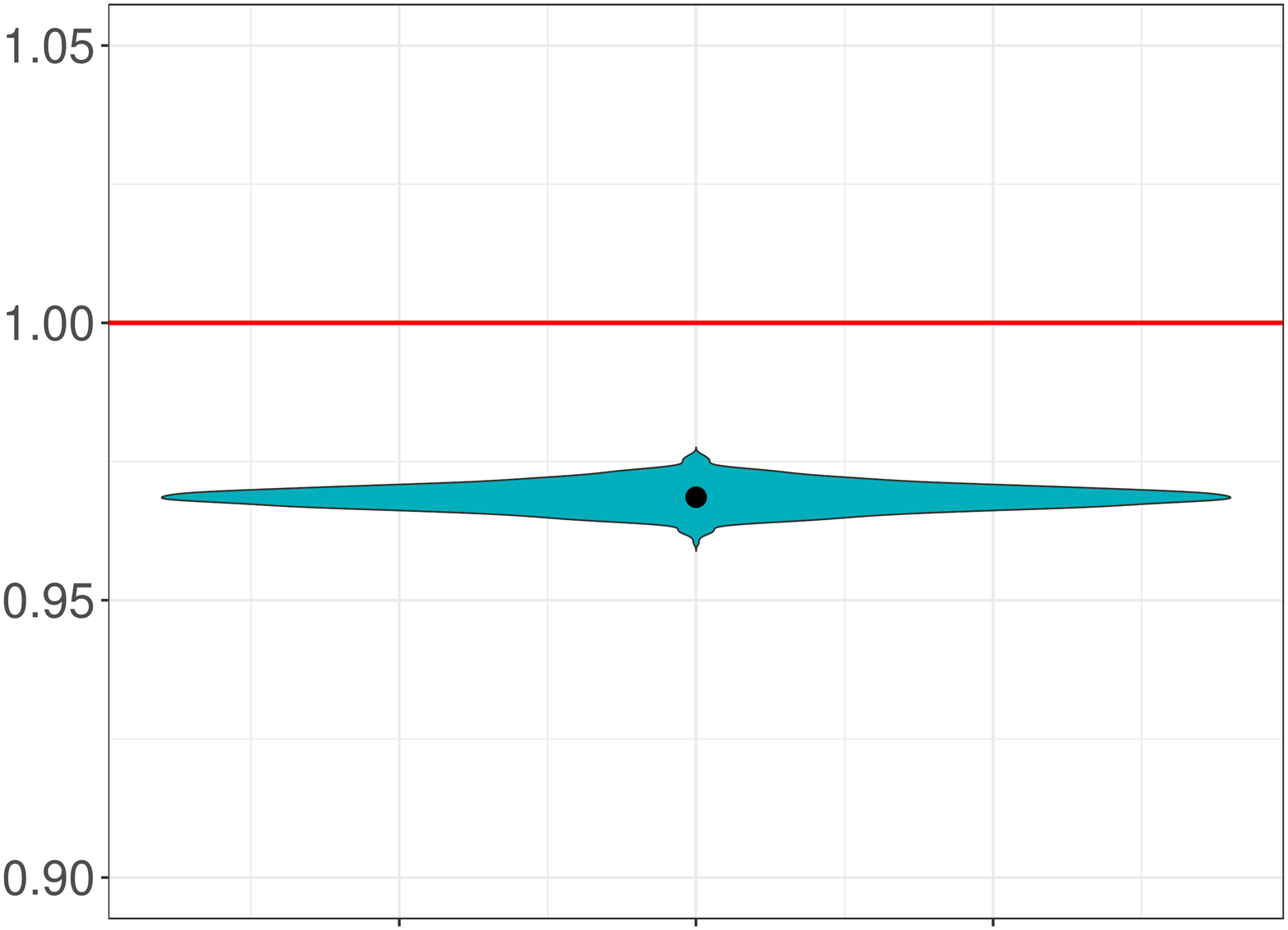}
  \caption{$\lambda=2$, $\hat\alpha=0$}
  \label{rrs_2_p}
\end{subfigure}\hfil
\begin{subfigure}{0.3\textwidth}
  \includegraphics[width=\linewidth]{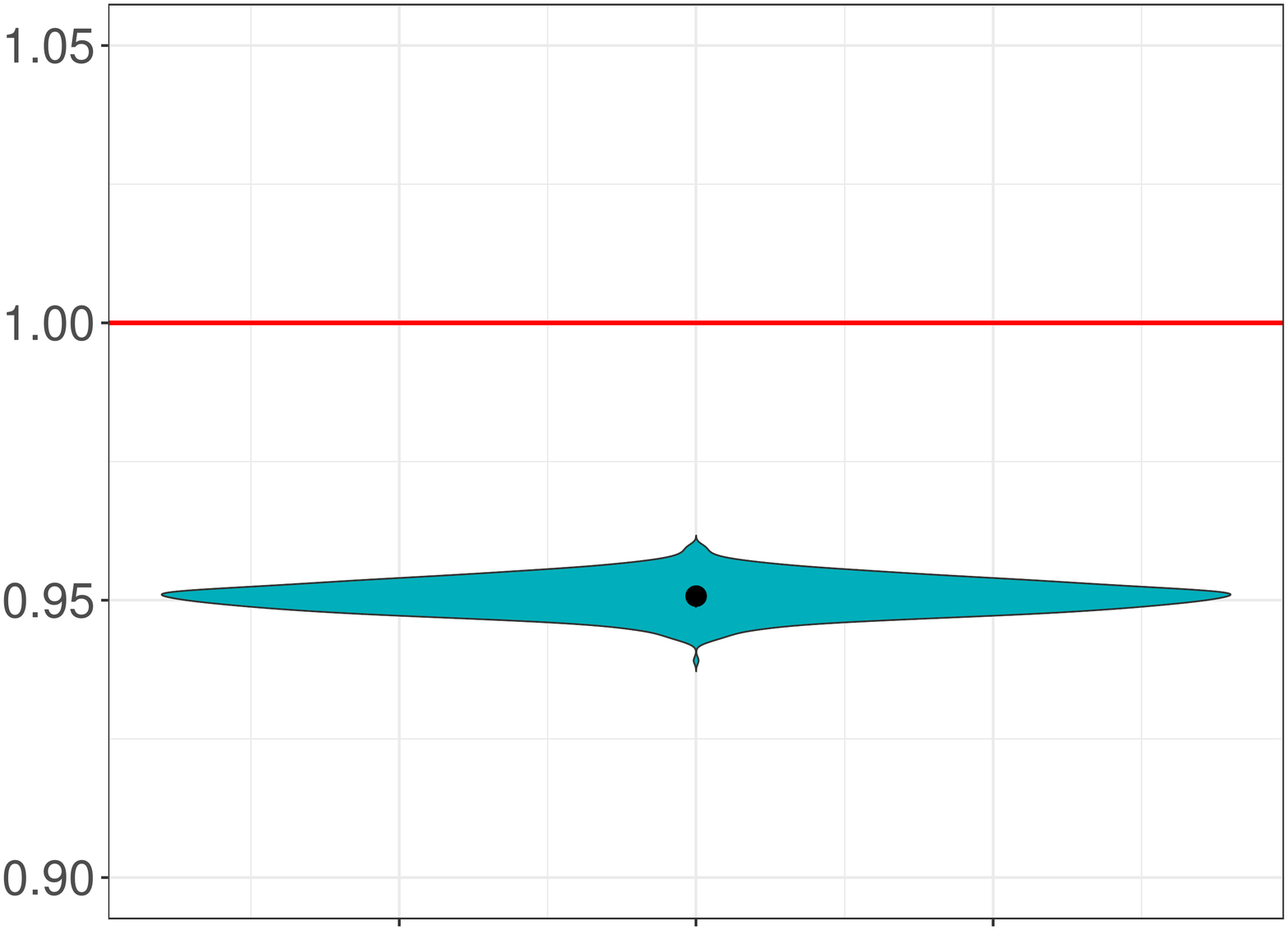}
  \caption{$\lambda=5$, $\hat\alpha=0$}
  \label{rrs_5_p}
\end{subfigure}\hfil
\begin{subfigure}{0.3\textwidth}
  \includegraphics[width=\linewidth]{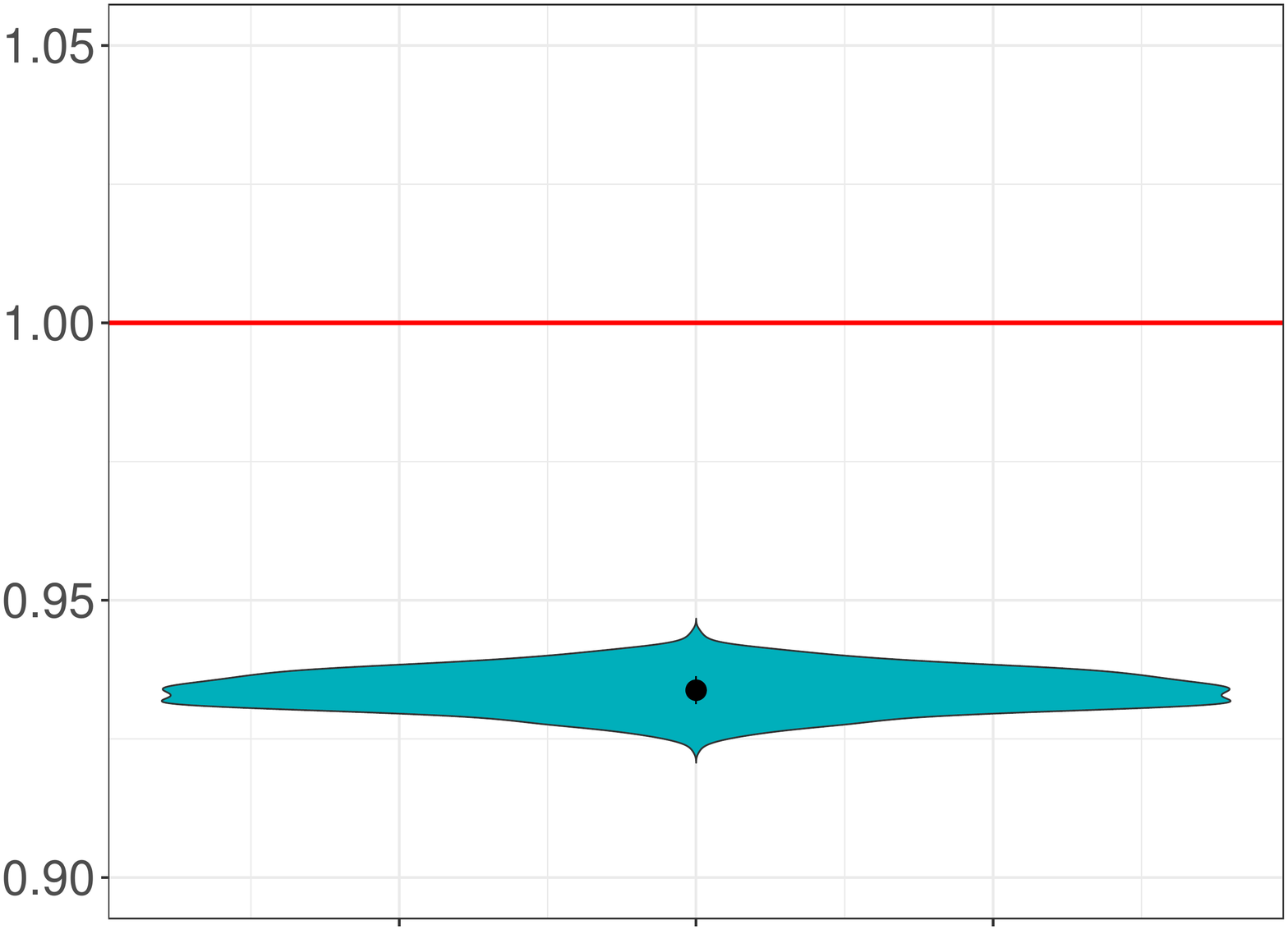}
  \caption{$\lambda=7$, $\hat\alpha=0$}
  \label{rrs_7_p}
\end{subfigure}
\captionsetup{labelformat=empty}
\caption{\romannumeral 1: Randomized-Response-Shuffled-Algorithm with $\tau= 10^{-5}, \beta=\tau^{-1}$}
\label{fig_rrs_p}
\end{figure*}
\begin{figure*}[h]
\centering
\begin{subfigure}{0.3\textwidth}
  \includegraphics[width=\linewidth]{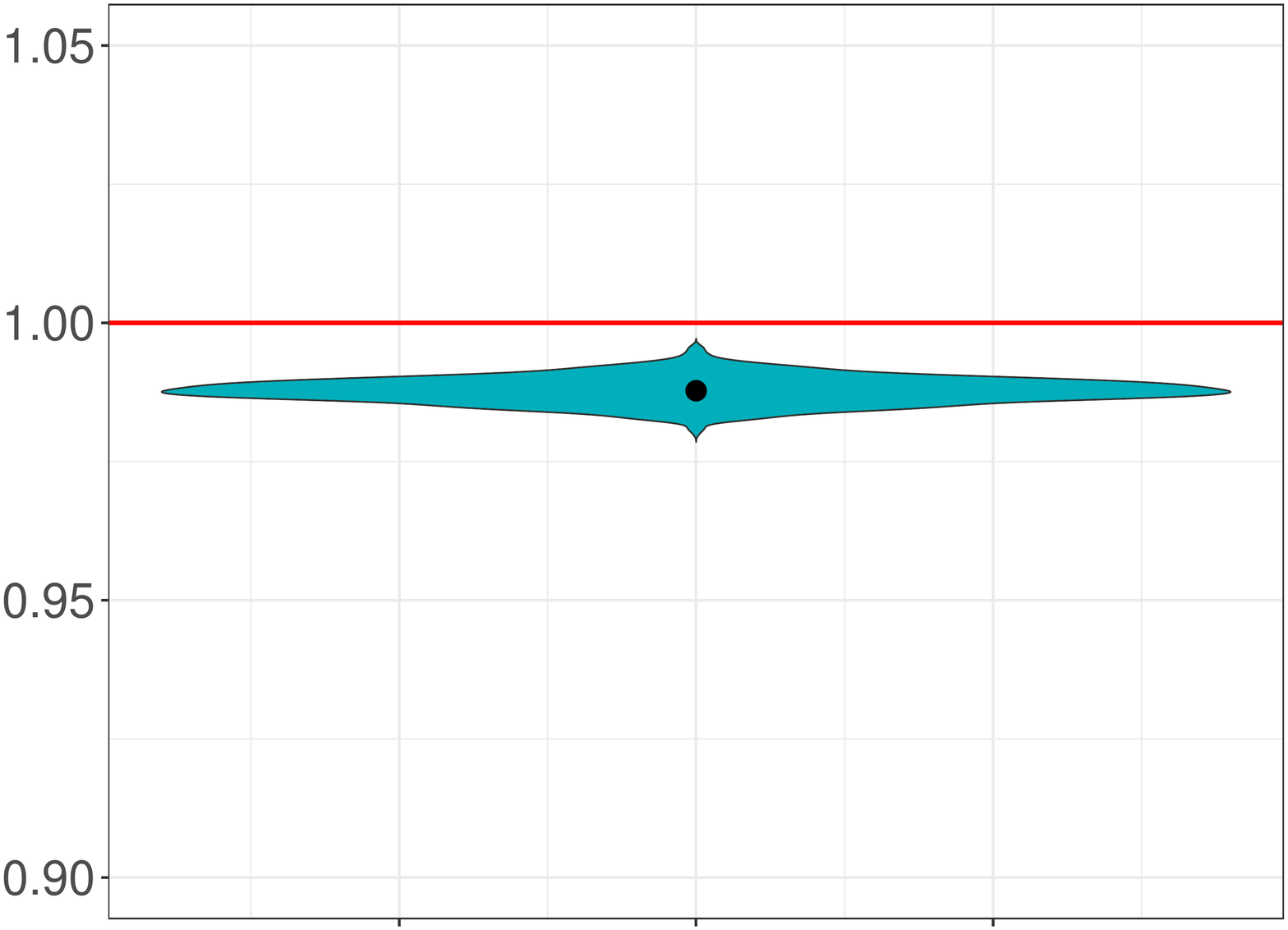}
  \caption{$\lambda=2$, $\hat\alpha=0$}
  \label{rrs_2_pp}
\end{subfigure}\hfil
\begin{subfigure}{0.3\textwidth}
  \includegraphics[width=\linewidth]{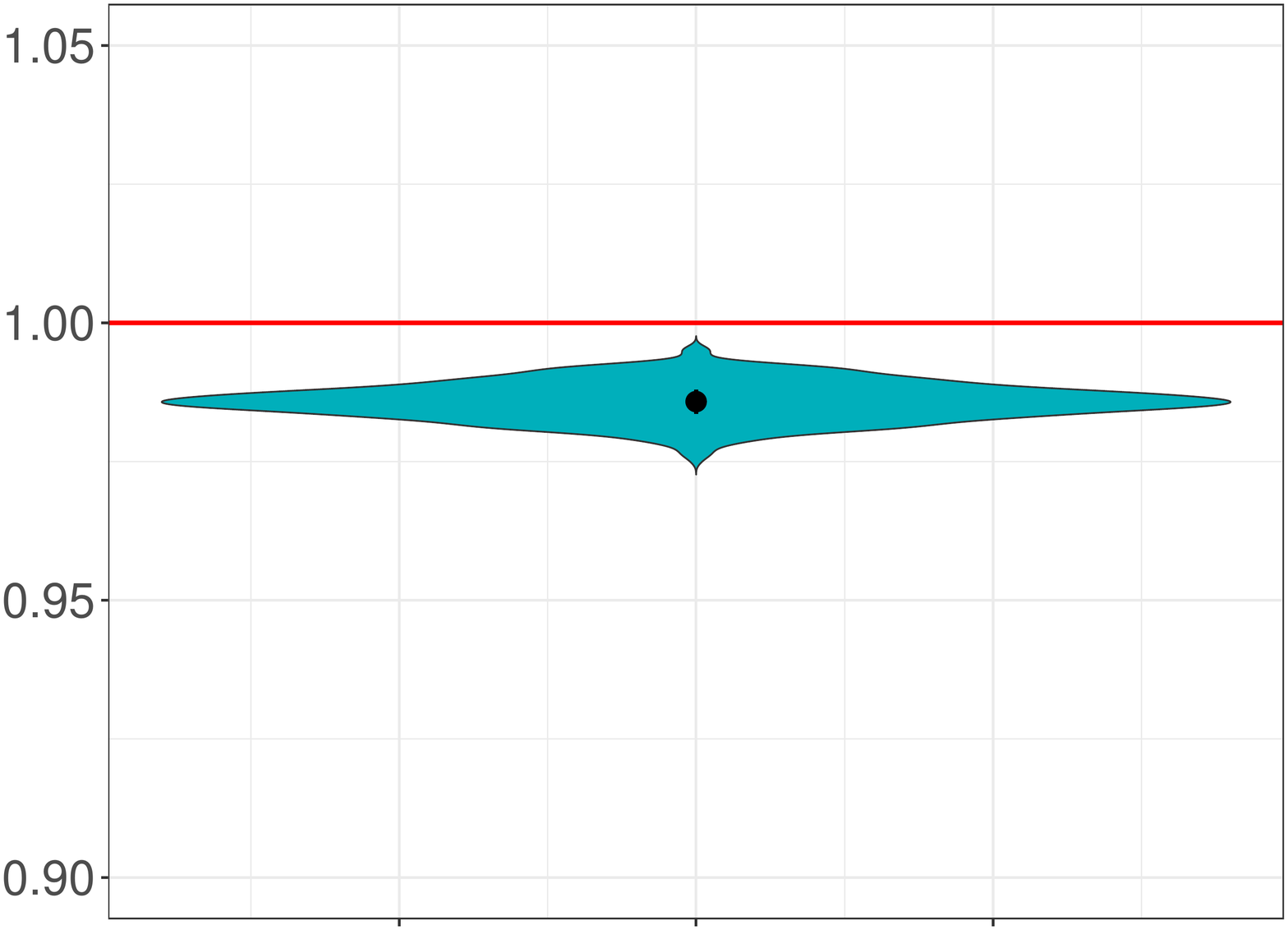}
  \caption{$\lambda=5$, $\hat\alpha=0$}
  \label{rrs_5_pp}
\end{subfigure}\hfil
\begin{subfigure}{0.3\textwidth}
  \includegraphics[width=\linewidth]{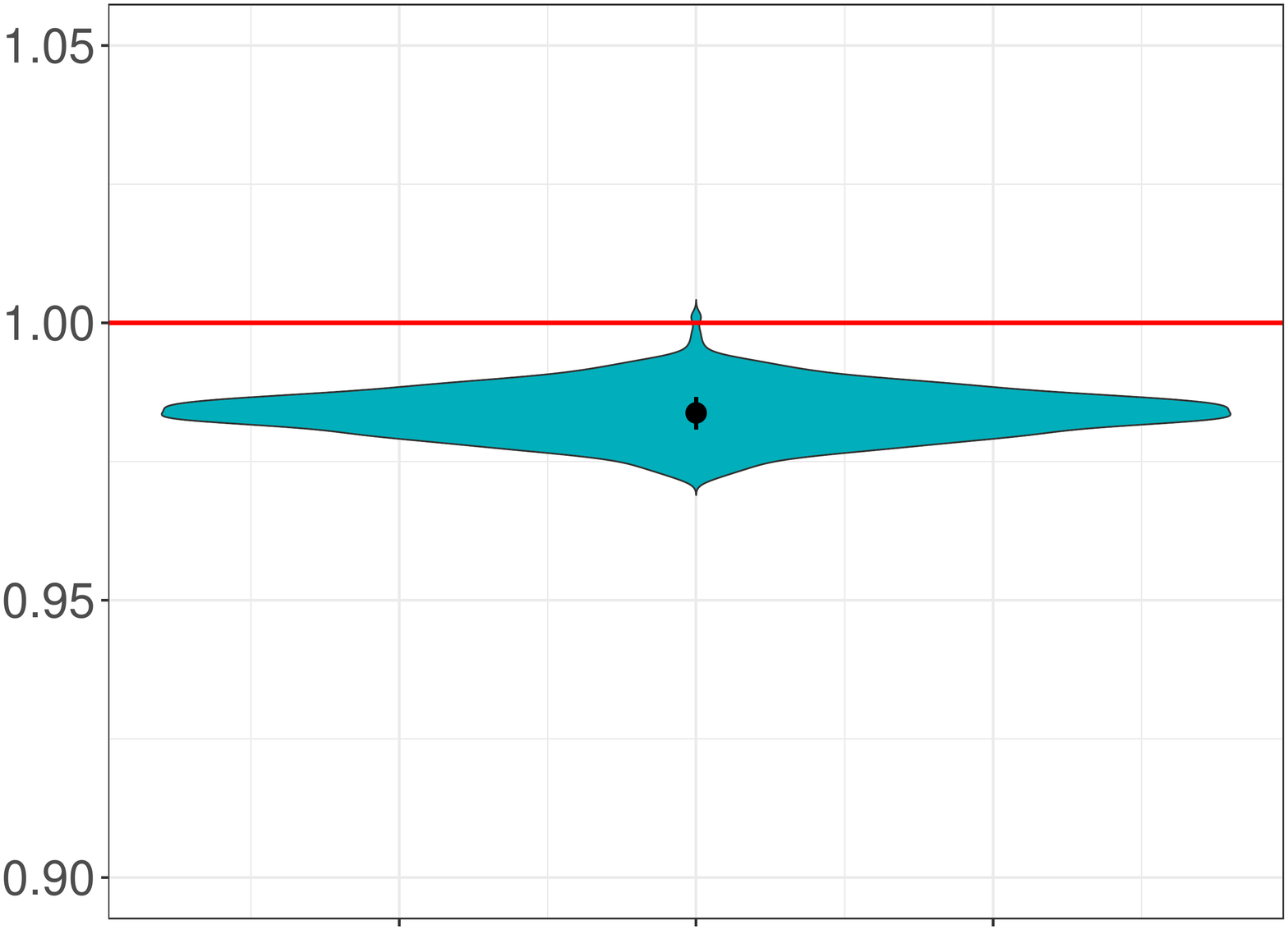}
  \caption{$\lambda=7$, $\hat\alpha=0.002$}
  \label{rrs_7_pp}
\end{subfigure}
\captionsetup{labelformat=empty}
\caption{\romannumeral 2: Randomized-Response-Shuffled-Algorithm with $\tau=5\times 10^{-6}, \beta=\tau^{-1}$}
\label{fig_rrs_pp}
\end{figure*}

Similarly, we have
 \begin{align*}
         \E Z_{i,2}^3=&\frac{1}{h^{3d}}\E\Big(\int (K(\frac{t-X_i}{h})-\E K(\frac{t-X_i}{h}))\\&\times\pi( q(t), \tau)(1-\lambda)q_\tau(t)^{-\lambda}p(t)^\lambda dt\Big)^3
        \\=& \int\Big( \int K(t)\pi( q(th+y), \tau)(1-\lambda)\\&\times q_\tau(th+y)^{-\lambda}p(th+y)^\lambda dt\Big)^3q(y)dy
        \end{align*}
   Due to $\pi(p(t))\leq 1$, we have
    \begin{align*}
    \E \tilde T_1^3&\leq C \tau^{-3\lambda}~.
    \end{align*}
    
Finally, we have that both third moments are $o(\sqrt{n})$.

\vfill\break

\subsection{Improving the parameters}\label{App_F}
   In this section, we briefly illustrate how an adapted choice of parameters (especially of $\tau,\beta$) can improve the estimation. It showcases that prior knowledge can usually improve the performance of our procedure. This insight is relevant because only rarely will a user have absolutely no prior knowledge about the algorithm in question (even though it may fall short of having the algorithm's source code). We will demonstrate this for the Randomized-Response-Shuffled Algorithm defined in Section \ref{Sec_experiments}. The first results are simulated with parameters $\tau=10^5, \beta=\tau^{-1}$ (the parameter choice used in our experiment section), while the improved ones are simulated with $\tau=5\times 10^6, \beta=\tau^{-1}$. The violin plots on this page illustrate the effect, mainly due to a reduced bias.\\

\end{document}